\documentclass[trackchanges]{aastex701}

\usepackage{amsmath}
\usepackage{textcomp}
\usepackage{enumitem}
\usepackage{xcolor}

\usepackage{subcaption}

\usepackage{tikz}
\usetikzlibrary{arrows.meta, positioning, shapes, fit}

\begin{document}

\title{Cosmology with one galaxy: An analytic formula relating $\Omega_{\rm m}$ with galaxy properties}

\author{Kito Liao}
\affiliation{Center for Cosmology and Particle Physics, Department of Physics, New York University, New York, NY, 10003, USA}
\affiliation{City University of New York, 275 Convent Ave, New York, NY , 10031, USA}
\email{kl4180@nyu.edu}

\author{Francisco Villaescusa-Navarro}
\affiliation{Center for Computational Astrophysics, Flatiron Institute, 162 5th Avenue, New York, NY, 10010, USA}
\affiliation{Department of Astrophysical Sciences, Princeton University, Peyton Hall, Princeton NJ 08544, USA}
\email{fvillaescusa@flatironinstitute.org}

\author[gname=IceSheet]{Romain Teyssier}
\affiliation{Department of Astrophysical Sciences, Princeton University, Peyton Hall, Princeton NJ 08544, USA}
\email{teyssier@princeton.edu}

\author[gname=IceSheet]{Natalí S. M. de Santi}
\affiliation{Berkeley Center for Cosmological Physics, University of California, Berkeley, 341 Campbell Hall, Berkeley, CA 94720, USA}
\affiliation{Physics Division, Lawrence Berkeley National Laboratory, 1 Cyclotron Road, Berkeley, CA 94720, USA}
\email{natalidesanti@gmail.com}


\begin{abstract}

Standard cosmological analyses typically treat galaxy formation and cosmological parameter inference as decoupled problems, relying on population-level statistics such as clustering, lensing, or halo abundances.
However, classical studies of baryon fractions in massive galaxy clusters have long suggested that gravitationally bound systems may retain cosmological information through their baryonic content. Building on this insight, we present the first analytic and physically interpretable cosmological tracer that links the matter density parameter, $\Omega_m$, directly to intrinsic galaxy-scale observables, demonstrating that cosmological information can be extracted from individual galaxies.
Using symbolic regression applied to state-of-the-art hydrodynamical simulations from the \textsc{CAMELS} project, we identify a compact functional form that robustly recovers $\Omega_m$ across multiple simulation suites (\textsc{IllustrisTNG}, \textsc{ASTRID}, \textsc{SIMBA}, and \textsc{Swift-EAGLE}), requiring only modest recalibration of a small number of coefficients. The resulting expression admits a transparent physical interpretation in terms of baryonic retention and enrichment efficiency regulated by gravitational potential depth, providing a clear explanation for why $\Omega_m$ is locally encoded in galaxy properties. Our work establishes a direct, interpretable bridge between small-scale galaxy physics and large-scale cosmology, opening a complementary pathway to cosmological inference that bypasses traditional clustering-based statistics and enables new synergies between galaxy formation theory and precision cosmology.

\end{abstract}



\section{Introduction}

Modern cosmology has traditionally relied on the statistics of the large--scale distribution of matter to infer fundamental cosmological parameters. Measurements of two--point and higher--order correlation functions, redshift--space distortions, gravitational lensing, and galaxy clustering have long provided the primary route to constrain the matter density parameter $\Omega_m$ and other cosmological quantities \citep{1980lssu.book.....P, 2001PhR...340..291B}. These methods, while powerful, fundamentally treat galaxies as tracers of an underlying continuous dark matter density field, and extract cosmological information from their collective spatial and dynamical behavior across large volumes.

The CAMELS project \citep{CAMELS:2020cof} has enabled a new generation of machine--learning-based approaches to cosmological inference by providing a controlled numerical framework in which both cosmological parameters and astrophysical prescriptions are systematically varied \footnote{\url{https://www.camel-simulations.org}}. Early studies based on the CAMELS simulation suite \citep{Villaescusa-Navarro:2022twv,Lue:2025zqk} have demonstrated that neural networks could identify non-trivial correlations between cosmological parameters and the internal properties of individual simulated galaxies, even in the presence of strong and highly nonlinear astrophysical processes.
These findings indicated that cosmological information may survive baryonic processing at the level of individual galaxies, although the physical origin and robustness of this signal remain unclear. While such correlations are not intended to compete with standard cosmological probes, they provide a novel framework for studying how cosmological information may be indirectly encoded in galaxy-scale observables within simulations.
Subsequent work \citep{Lue:2025zqk} moved beyond direct parameter regression toward understanding the internal structure of the high--dimensional galaxy population itself. In this framework, neural networks were shown to learn low--dimensional manifolds in galaxy property space, encoding joint information about cosmology and baryonic physics. These studies suggested that intrinsic galaxy properties themselves encode cosmological information and that the observed structure in galaxy feature distributions reflects the imprint of cosmology on galaxy formation physics, rather than relying solely on large-scale clustering statistics.

One of the challenges highlighted by the \emph{Cosmology with One Galaxy} framework is that, while the properties of individual galaxies have been shown to encode cosmological information, the physical origin of this connection has so far remained only partially formulated and not yet fully understood, which has limited its broader physical interpretation and acceptance. In contrast, our work approaches the inference of  $\Omega_m$ from an explicitly physical perspective. We obtain compact formulae that relate the value of  $\Omega_{\rm m}$ with the properties of individual galaxies. This approach examines how cosmology influences the thermodynamic, chemical, and structural outcomes of galaxy formation. 
As we shall show below, we identify a robust latent variable structure with clear physical interpretation, which persists across a high--dimensional parameter space spanning variations in both cosmology and astrophysical subgrid models. This structure survives extensive tests across multiple simulation suites and feedback prescriptions, indicating that the mapping between galaxy--scale physics and cosmology is not coincidental, but reflects an underlying physical connection between baryonic regulation and cosmology itself.

This paper is organized as follows. Section~\ref{sec:section2} describes the simulation data and machine--learning methodology. Section~\ref{sec:LH_result} presents the results obtained from the reduced cosmology–and–feedback variation simulation runs, and Section~\ref{sec:LH-physics} provides their physical interpretation. Section~\ref{sec:SB28tests} extends the analysis to a higher dimensional parameter space dataset, testing the analytic framework under broad variations in cosmology-and-feedback and relating the results to our existing physical interpretation. Finally, we summarize and discuss the main results of this work in Section~\ref{sec:summary}.

\section{Methods} \label{sec:section2}

\subsection{Simulations}

The Cosmology and Astrophysics with MachinE Learning Simulations project
(\textsc{CAMELS}; \citealt{CAMELS:2020cof}) is a large suite of cosmological simulations designed to study how variations in cosmology and baryonic physics impact galaxy formation and large-scale structure.
CAMELS consists of over 16,000 hydrodynamic and $N$-body simulations and provides a controlled framework for exploring the joint effects of cosmological parameters and subgrid feedback models in a setting tailored for statistical analysis and machine learning applications.
Each simulation of the first generation evolves $2 \times 256^3$ dark matter and gas resolution elements in a periodic comoving volume of $(25\,h^{-1}\mathrm{Mpc})^3$, from an initial redshift of $z=127$ down to $z=0$.

A key feature of CAMELS is the coordinated variation of both cosmology and galaxy formation physics across multiple independent simulation frameworks, enabling robust comparisons between models and the identification of physical trends that overcome differences in numerical implementation.

\subsubsection{The Latin Hypercube Dataset}

In the first part of this work, we use the \textbf{Latin Hypercube (LH)} sets of CAMELS as our primary dataset.
Each LH suite consists of 1,000 simulations in which cosmological and astrophysical parameters
are simultaneously varied using Latin-hypercube sampling to efficiently explore the
high-dimensional parameter space.
The LH sets vary over two cosmological parameters:
$\Omega_m$ and $\sigma_8$,
as well as four astrophysical parameters that
control the efficiency of supernovae (SN) and active galactic nuclei (AGN) feedback:
\begin{equation}
\Omega_m \in [0.1, 0.5], \qquad
\sigma_8 \in [0.6, 1.0],
\end{equation}
\begin{equation}
A_{\rm SN1}, A_{\rm AGN1} \in [0.25, 4.0], \qquad
A_{\rm SN2}, A_{\rm AGN2} \in [0.5, 2.0].
\end{equation}
All simulations adopt fixed background values for the remaining cosmological parameters:
\[
\Omega_b = 0.049, \quad
H_0 = 0.067\,({\rm km\,s^{-1}})/{\rm kpc},\quad
n_s = 0.9624,\quad 
\sum m_\nu = 0 \rm  eV, \quad
w = -1.
\]

We use hydrodynamic simulations from four independent galaxy formation models
implemented in the CAMELS project:

\begin{enumerate}

\item \textsc{IllustrisTNG}: run with the \textsc{AREPO} code ~\citep{2020ApJS..248...32W}. The subgrid physics models are the same as the original \textsc{IllustrisTNG} simulations ~\citep{Pillepich:2017fcc}.

\item \textsc{SIMBA}: run with the \textsc{GIZMO} code ~\citep{Hopkins:2017blv} and employs the same subgrid physics model as the
original SIMBA simulation, building on its precursor MUFASA with the addition of supermassive black hole
growth and feedback~\citep{Dave:2019yyq,Angles-Alcazar:2016krf}.

\item \textsc{ASTRID}: run with the \textsc{MP-Gadget} code
~\citep{Bird:2022ulj}. Detailed subgrid physics models are described in ~\cite{CAMELS:2023wqd}.

\item \textsc{Swift-EAGLE} galaxy formation model is a modified version of the original EAGLE simulation ~\citep{Schaye:2014tpa},
implemented in the \textsc{Swift} ~\citep{SWIFT:2023dix}
code to solve the equations of gravity and the
\textsc{SPHENIX} ~\citep{Borrow_2021} formulation of smoothed particle hydrodynamics.
Detailed subgrid physics models are described in~\cite{Lovell:2024tei}.
\end{enumerate}

\subsubsection{28-parameter TNG Dataset: TNG-SB28} \label{sec:style}

In the second part of this work, we focus on the \textsc{TNG-SB28} dataset ~\citep{CAMELS:2023wqd}, an extended version of the standard
TNG-LH suite that explores a substantially larger parameter space.
Whereas TNG-LH varies six parameters, TNG-SB28 simultaneously varies \textbf{28 parameters}.
These parameters span all major physical modules of the model:
five cosmological parameters, and 23 astrophysical parameters. For the astrophysical parameters,
two governing star formation and the interstellar medium (ISM),
two describing stellar population modeling,
ten controlling galactic wind feedback,
three regulating supermassive black hole growth,
and six associated with AGN feedback.
More information about these parameters is provided in Table \ref{tab:SB28}, where the physical meaning, sampling range, and fiducial value of each parameter are listed in detail.

\begin{table*} [h]
\centering
\caption{Summary of the 28 model parameters varied in the \textsc{TNG-SB28} dataset. \label{tab:SB28}}
\label{tab:SB28}
\begin{tabular}{lcc}
\hline
Parameter & Physical meaning & Range (sampling) \\
\hline
$\Omega_m$ &
Matter density parameter &
$[0.1,\,0.5]$ (linear; fiducial $0.3$) \\

$\sigma_8$ &
Amplitude of matter fluctuations &
$[0.6,\,1.0]$ (linear; fiducial $0.8$) \\

$A_{\rm SN1}$ &
Wind energy normalization &
$[0.9,\,14.4]$ (log; fiducial $3.6$) \\

$A_{\rm SN2}$ &
Wind velocity normalization &
$[3.7,\,14.8]$ (log; fiducial $7.4$) \\

$A_{\rm AGN1}$ &
Radio-mode AGN energy normalization &
$[0.25,\,4.0]$ (log; fiducial $1.0$) \\

$A_{\rm AGN2}$ &
AGN feedback reorientation frequency &
$[10,\,40]$ (log; fiducial $20$) \\

$\Omega_b$ &
Baryon density parameter &
$[0.029,\,0.069]$ (linear; fiducial $0.049$) \\

$H_0$ &
Hubble constant &
$[0.4711,\,0.8711]$ (linear; fiducial $0.6711$) \\

$n_s$ &
Spectral index of the initial fluctuations &
$[0.7624,\,1.1624]$ (linear; fiducial $0.9624$) \\

$t_0^\star$ &
Star-formation timescale &
$[1.135,\,4.54]\,\mathrm{Gyr}$ (log; fiducial $2.27\,\mathrm{Gyr}$) \\

$q_{\rm EOS}$ &
Effective equation-of-state interpolation factor &
$[0.1,\,0.9]$ (log; fiducial $0.3$) \\

$\alpha_{\rm IMF}$ &
High-mass IMF slope ($M > 1\,M_\odot$) &
$[-2.8,\,-1.8]$ (linear; fiducial $-2.3$) \\

$M_{\rm SNII,min}$ &
Minimum stellar mass for SN~II &
$[4,\,12]\,M_\odot$ (linear; fiducial $8\,M_\odot$) \\

$\tau_{\rm w}$ &
Thermal fraction of the galactic wind feedback energy &
$[0.025,\,0.4]$ (log; fiducial $0.1$) \\

$\mathrm{mom}_w$ &
Wind specific momentum normalization &
$[0,\,4000]\,\mathrm{km\,s^{-1}}$ (linear; fiducial $0$) \\

$\rho_{\rm recouple}$ &
Wind recoupling density factor &
$[0.005,\,0.5]$ (log; fiducial $0.05$) \\

$v_{w,\min}$ &
Minimum galactic wind velocity &
$[150,\,550]\,\mathrm{km\,s^{-1}}$ (linear; fiducial $350\,\mathrm{km\,s^{-1}}$) \\

$f_{w,Z}$ &
Wind energy reduction factor (metallicity-dependent) &
$[0.0625,\,1.0]$ (log; fiducial $0.25$) \\

$Z_{w,\rm ref}$ &
Reference metallicity for wind energy transition &
$[5\times10^{-4},\,8\times10^{-3}]$ (log; fiducial $2\times10^{-3}$) \\

$\gamma_{w,Z}$ &
Wind energy reduction exponent (metallicity transition) &
$[1,\,3]$ (linear; fiducial $2$) \\

$1-\gamma_w$ &
Wind dump factor  &
$[0.2,\,1.0]$ (linear; fiducial $0.6$) \\

$M_{\rm BH,seed}$ &
Seed black hole mass &
$[2.5\times10^{5},\,2.5\times10^{6}]\,M_\odot$ (log; fiducial $8\times10^{5}\,M_\odot$) \\

$A_{\rm BH}$ &
Black hole accretion normalization &
$[0.25,\,4.0]$ (log; fiducial $1.0$) \\

$A_{\rm Edd}$ &
Eddington accretion normalization &
$[0.1,\,10]$ (log; fiducial $1.0$) \\

$\epsilon_{f,\rm high}$ &
Black Hole feedback factor &
$[0.025,\,0.4]$ (log; fiducial $0.1$) \\

$\epsilon_r$ &
Black hole radiative efficiency &
$[0.05,\,0.8]$ (log; fiducial $0.2$) \\

$\chi_0$ &
Quasar-mode Eddington threshold &
$[6.3\times10^{-5},\,6.3\times10^{-2}]$ (log; fiducial $2\times10^{-3}$) \\

$\beta$ &
Quasar threshold power &
$[0,\,4]$ (linear; fiducial $2$) \\

\hline
\end{tabular}
\end{table*}

The 28-dimensional parameter space is sampled using a \textbf{Sobol low-discrepancy sequence} \citep{SOBOL1967}
rather than a Latin-hypercube.
This choice provides deterministic, faster, and lower-discrepancy coverage.
Parameter ranges are calibrated using a combination of physical priors and sensitivity tests,
aiming to span realistic values while avoiding edge effects and equilibrating the impact
of variations across parameters.

\subsection{Machine Learning Method -- Symbolic Regression}
\label{sec:ML-SR}

During the process of discovering the analytic form of $\Omega_{m}$ as a function of galaxy properties, we employ \textit{symbolic regression} to identify the most interpretable and accurate functional expression directly from the data. 

\begin{figure}[h]
    \centering
    \scalebox{1.0}{
    \begin{tikzpicture}[
    node distance=0.9cm and 0.9cm,
    every node/.style={font=\scriptsize, align=center},
    box/.style={rectangle, rounded corners=3pt, draw=black!80, fill=gray!5,
                minimum width=1.4cm, minimum height=0.5cm, inner sep=2pt},
    arrow/.style={-{Latex[length=1.3mm,width=1.3mm]}, thick}
]

% Nodes
\node[box] (init) {Init.};
\node[box, right=of init] (eval) {Eval.};
\node[box, right=of eval] (evo) {Select\\\& Mutate};
\node[box, right=of evo] (check) {Stop?};
\node[box, right=of check] (out) {Result};

% Arrows
\draw[arrow] (init) -- (eval);
\draw[arrow] (eval) -- (evo);
\draw[arrow] (evo) -- (check);
\draw[arrow] (check) -- node[above]{Yes} (out);
\draw[arrow] (check) |- ++(0,-0.7) -| node[below,yshift=1pt]{No} (eval.south);

    \end{tikzpicture}}

    \caption{\label{Fig:ML_flow}Symbolic regression search loop: The procedure starts with initialization (Init.), evaluates candidate expressions (Eval.), and applies selection and modification operators to generate new equations. A stopping condition is then checked; if satisfied, the algorithm terminates and outputs the resulting expression, otherwise the search continues iteratively.}
    
\end{figure}
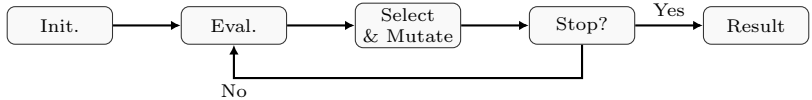

The symbolic regression algorithm\footnote{\url{https://github.com/MilesCranmer/PySR}}~\citep{cranmerDiscovering2020} follows six general steps: 
(1) \textbf{Initialization} --- a population of random candidate equations is generated, where each equation is represented as a tree structure composed of operators and variables; 
(2) \textbf{Evaluation} --- each equation is assessed using a customized error metric (e.g., Mean Squared Error (MSE) or Mean Absolute Error (MAE)) combined with a complexity penalty to balance accuracy and simplicity; 
(3) \textbf{Selection} --- the best-performing equations are chosen as parents for the next generation; 
(4) \textbf{Crossover} --- selected parent equations exchange parts of their expression trees to form new offspring; 
(5) \textbf{Mutation} --- small random alterations are applied to individual equations to introduce diversity and explore new regions of the search space; and 
(6) \textbf{Iteration} --- the population is repeatedly evaluated, selected, and updated over multiple generations until convergence criteria are met. 
The overall workflow of the symbolic regression search process is illustrated in Fig.~\ref{Fig:ML_flow}, which summarizes the iterative evolution from initialization to convergence. Throughout this process, symbolic regression explores combinations of mathematical building blocks---operators, variables, and constants---to uncover an analytic expression.

\subsubsection{Symbolic Regression Configuration}

The symbolic regression phase was designed to balance physical interpretability, numerical stability, and model expressiveness. We adopt a single \emph{exploratory} search phase, supported by a restricted but physically motivated operator set, with the goal of identifying a compact analytic relation linking galaxy properties to $\Omega_m$.

\begin{itemize}

\item \textbf{Phase 1: Exploratory Search.}  
During the exploratory phase, we set \texttt{deterministic = False} to expand the search space.
This configuration allows random initialization, crossover, and mutation operations to vary between runs, promoting diversity in the expression population and enabling the algorithm to escape local minima.
Each run may therefore yield slightly different analytic forms, increasing the likelihood of discovering novel and physically meaningful equations.

\item \textbf{Operator Design.}  
To preserve physical interpretability while allowing flexible nonlinearity, the symbolic regression search employed a restricted set of mathematical operators.
The \texttt{binary operators} were defined as
\[
\{ +,\; -,\; \times,\; /,\; \texttt{power}(x,y) \},
\]
while the \texttt{unary operators} included common analytic transformations together with a custom sigmoid activation,
\[
\{\exp,\; \ln,\; |\cdot|,\; \sqrt{\cdot},\; \sigma(x)\}, 
\qquad 
\sigma(x) = \frac{1}{1 + e^{-x}}.
\]
The inclusion of the smooth sigmoid function was essential for capturing regulated, saturating relations—particularly those linking metallicity, stellar mass fraction, and the cosmological parameter $\Omega_m$.

\item \textbf{Operator Constraints.}  
To prevent unphysical or unstable expressions, several constraints were imposed:
\begin{itemize}
    \item Exponents were restricted to the range $-3 \leq \alpha \leq 3$.
    \item Recursive application of identical unary operators (e.g., $\exp(\exp(x))$ or $\ln(\ln(x))$) was disallowed through nesting constraints.
    \item \texttt{weight\_optimize=True} was enabled to refine numerical coefficients after the symbolic structure was discovered.
\end{itemize}

\item \textbf{Training Configuration.}  
The primary training parameters were
\[
\texttt{niterations}=1500, \quad
\texttt{populations}=30, \quad
\texttt{model\_selection}=\text{"best"}, \quad
\texttt{elementwise\_loss}=(x-y)^2.
\]
A total of 1500 evolutionary generations were performed across 30 independent populations, each exploring distinct subregions of the symbolic expression space.
At the end of each run, the expression with the lowest loss was retained according to the \texttt{``best''} model-selection criterion.

\item \textbf{Loss Function and Parsimony.}  
The primary loss function was the mean squared error,
\[
\mathcal{L} = \frac{1}{N}\sum_i (y_i - \hat{y}_i)^2,
\]
where $y_i$ denotes the true simulation value of the cosmological parameter $\Omega_m$ and $\hat{y}_i$ is the corresponding value predicted by the symbolic expression, 
ensuring smooth and continuous fits suitable for cosmological parameter prediction.
To discourage overly complex models, a parsimony penalty $\lambda = 10^{-4}$ was added to the effective loss,
\[
\mathcal{L}_{\mathrm{eff}} = \mathcal{L} + \lambda N_{\mathrm{terms}},
\]
where $N_{\mathrm{terms}}$ denotes the total number of operators and constants in the expression. The effective loss $\mathcal{L_{\rm eff}}$ is used both to guide the evolutionary search and to select the final model, ensuring that the chosen expression reflects an optimal balance between predictive accuracy and analytic simplicity, rather than minimizing the \texttt{elementwise\_loss} alone.

\item \textbf{Complexity Constraints.}  
The symbolic search was bounded by 
$\texttt{maxsize}=18$ and $\texttt{maxdepth}=8$,
which restricted both the total number of terms and the depth of nested function calls.
These limits guided the evolution toward concise and interpretable analytic forms.

\item \textbf{Optimization and Parallelism.}  
To enhance exploration and prevent premature convergence, training was executed in multithreaded mode
(\texttt{parallelism="multithreading"}) with random seed \texttt{random\_state=42}.
The multithreading option of \texttt{PySR} was used, with the number of Julia threads set to 8.
Mini-batch evaluation with \texttt{batch\_size=2048} enabled stochastic assessment on random subsets of the dataset, introducing mild noise that helped avoid local minima—complementary to the nondeterministic search behavior.
Each run was limited to \texttt{timeout\_in\_seconds=1200}, ensuring convergence within practical wall-time limits.

\end{itemize}

\subsubsection{Training Procedure}

The symbolic regression model was trained using galaxy property data from the \textsc{IllustrisTNG} simulations at redshift $z=0$. From each of the 1000 simulations, we randomly sampled 30 galaxies at $z=0$, forming the training set used to identify the baseline analytic relation. To account for redshift evolution, additional multiplicative redshift-dependent factors were subsequently fitted using galaxy samples drawn at higher redshifts ($z = 0.5$, $1.0$, $1.5$, $2.0$, and $3.0$). This procedure isolates the discovery of the underlying functional structure at $z=0$ from the modeling of its redshift dependence, allowing the evolution of the model parameters to be captured in a controlled and interpretable manner.

\subsubsection{Model Evaluation Metrics}

The performance of the analytic $\Omega_m$ predictor is quantified using: the coefficient of determination ($R^2$), and the accuracy.

\begin{itemize}
\item \textbf{Coefficient of determination ($R^2$).}  
The $R^2$ score measures how well the predicted values reproduce the variance of the true targets,
\begin{equation}
R^2 = 1 - 
\frac{\displaystyle\sum_{i=1}^{N}
\left(y_i - \hat{y}_i\right)^2}
{\displaystyle\sum_{i=1}^{N} 
\left(y_i - \langle y \rangle\right)^2}\,,
\label{eq:R2_def}
\end{equation}
where $N$ is the total number of samples, $y_i$ and $\hat{y}_i$ denote the true and predicted values of $\Omega_m$, respectively. $\langle y \rangle$ denotes the mean (average) of the true values $y_i$ over all samples.

\item \textbf{Accuracy.}  
The accuracy is defined as the root--mean--square deviation between predictions and true values,
\begin{equation}
\mathrm{Accuracy} = 
\sqrt{\frac{1}{N}\sum_{i=1}^{N}
\left(y_i - \hat{y}_i\right)^2}\,.
\label{eq:accuracy_def}
\end{equation}
This metric quantifies the absolute prediction error in the same physical units as $\Omega_m$, with smaller values indicating higher accuracy.

\end{itemize}

\subsection{Dimensionless Parameter Construction}
\label{sec:dimensionless_par_construction}

In the studies,
we construct dimensionless parameters that encapsulates the joint dependence of baryonic, kinematic, and metallicity properties.
The formulation of these dimensionless quantities combine key physical variables to eliminate dimensional dependencies, in our case, serve as a better candidate for $\Omega_{m}$ analytic equation searching.

The raw galaxy catalogs from the CAMELS suites— \textsc{IllustrisTNG}, \textsc{SIMBA}, \textsc{ASTRID}, and \textsc{Swift-EAGLE}— each supply a consistent collection of fundamental physical quantities, ensuring consistency across simulation models:
\[
\{ M_{\rm gas},\, M_\star,\, M_{\rm BH},\, M_{\rm total},\, 
V_{\max},\, V_{\rm disp},\, Z_{\rm gas},\, Z_{\rm star},\, 
{\rm SFR},\, \text{spin},\, V_{\rm pec},\, 
R_\star,\, R_{\rm total},\, R_{\max}\}
\]
In addition, the IllustrisTNG suite uniquely provides synthetic photometric magnitudes,
\(
\{ U_{\rm mag},\, K_{\rm mag},\, g_{\rm mag} \},
\)
which are not directly available in the SIMBA, ASTRID, or Swift-EAGLE catalogs. 
A detailed description of these quantities and their physical definitions is provided in Table~\ref{tab:galaxy_properties}.

\begin{table*}[h]
\centering
\caption{ Galaxy properties extracted from the \textsc{CAMELS} simulations. Halos and subhalos are identified using \textsc{SUBFIND} \citep{Springel:2000qu}. Galaxies are defined as subhalos containing more than 20 star particles.}
\label{tab:galaxy_properties}
\begin{tabular}{l l p{10cm}}
\hline
Symbol & Property & Description \\
\hline
$M_g$ & Gas mass & Total gas mass of the galaxy, including contributions from the circumgalactic medium \\

$M_{\rm BH}$ & Black hole mass & Total mass of the central supermassive black hole \\

$M_\star$ & Stellar mass & Total stellar mass of the galaxy \\

$M_t$ & Total subhalo mass & Sum of dark matter, gas, stars, and black holes bound to the subhalo \\

$V_{\max}$ & Maximum circular velocity & Maximum value of the circular velocity profile \\

$\sigma_v$ & Velocity dispersion & Velocity dispersion of all particles within the galaxy subhalo \\

$Z_g$ & Gas metallicity & Mass-weighted gas-phase metallicity \\

$Z_\star$ & Stellar metallicity & Mass-weighted stellar metallicity \\

${\rm SFR}$ & Star-formation rate & Instantaneous star-formation rate of the galaxy \\

$J$ & Subhalo angular momentum & Magnitude of the galaxy subhalo spin vector \\

$V$ & Peculiar velocity & Magnitude of the galaxy subhalo peculiar velocity \\

$R_\star$ & Stellar half-mass radius & Radius enclosing half of the stellar mass \\

$R_t$ & Total half-mass radius & Radius enclosing half of the total subhalo mass \\

$R_{\max}$ & Radius of $V_{\max}$ & Radius at which the circular velocity reaches $V_{\max}$ \\

$U_{\rm mag}$ & $U$-band magnitude & Absolute galaxy magnitude in the $U$ band \\

$K_{\rm mag}$ & $K$-band magnitude & Absolute galaxy magnitude in the $K$ band \\

$g_{\rm mag}$ & $g$-band magnitude & Absolute galaxy magnitude in the $g$ band \\
\hline
\end{tabular}
\end{table*}

We further define the core dimensionless combinations being used in the symbolic regression process. Here, we only present the ones that are relevant to our project goals; one could, in principle, define more combinations with the given raw galaxy properties.
The following quantities are derived from the raw galaxy catalogs and used in subsequent analysis:
\begin{itemize}
    \item \textbf{Metallicities in Solar Units:}  
    Metal abundances are normalized to the solar value $Z_\odot$ \citep{2009ARA&A..47..481A}, with 
    $Z_{\rm gas,\,solar} = Z_{\rm gas}/Z_\odot$ and 
    $Z_{\star} = Z_{\rm star}/Z_\odot$.

    \item \textbf{Effective Dynamical Velocity:}  
    Rotational and dispersion components are combined as 
    $V_{\text{eff}} = \sqrt{V_{\max}^2 + \kappa\,V_{\rm disp}^2}$ with $\kappa = 2.0$, 
    capturing both disk- and bulge-dominated systems within a unified kinematic scaling.

    \item \textbf{Baryonic mass:}  
    $M_{\rm b} = M_{\star} + \,M_{\rm gas}$.

    \item \textbf{Dimensionless Mass Fractions:}  
    The relative contributions of each component are expressed as 
    $f_{\rm gas} = M_{\rm gas}/M_{\rm b}$, 
    $f_\star = M_{\star}/M_{\rm b}$, and 
    $f_{\rm BH} = M_{\rm BH}/M_{\rm b}$.

\end{itemize}
With the above dimensionless quantities, we define two dimensionless combinations: 
\begin{itemize}
    \item  \textbf{$S_{\text{core}}$:}
    represents a dimensionless baryonic 
    binding scale linking galaxy-scale dynamics to cosmological baryon content
    \[
    S_{\text{core}} 
    = \frac{\Omega_b^{2}\,V_{\text{eff}}^{3}}
           {G\,H_0\,M_{\rm b}}.
    \]
In the LH simulation suite, $\Omega_b$ is not varied and therefore cannot be inferred by the symbolic regression. We include it in the construction of the dimensionless variable as part of a physically motivated normalization, where a single factor of $\Omega_b$ reflects the cosmic baryon fraction.
The precise dependence on $\Omega_b$ is not fixed \textit{a priori}. Empirically, we find that including an additional factor of $\Omega_b$ leads to more stable and compact expressions, resulting in an effective quadratic scaling.
In Section~\ref{sec:SB28:result}, we analyze the TNG-SB28 suite, where $\Omega_b$ varies, and recover the same scaling. This indicates that the $\Omega_b^2$ dependence is not an artifact of the normalization, but reflects a robust feature of the relation between galaxy properties and cosmology.
    \item \textbf{Compactness Parameter:}  
    The spatial concentration of baryons is quantified as
    \[
    R_{\text{compact}} = \frac{R_{\star}}{R_{\rm total}}.
    \]

\end{itemize}

\section{Results} 
\label{sec:LH_result}

In this section, we present the analytic predictor selected from the symbolic regression analysis. Subsection~\ref{sec:LH-redshift_zero_result} presents the results at redshift $z=0$, and Subsection~\ref{sec:LH-higher_redshift} provides the fitted redshift evolution of the analytic predictor at higher redshifts.

\subsection{Analytic predictor for \(\Omega_{m}\) at redshift $z=0$ }
\label{sec:LH-redshift_zero_result}

We present the analytic form of the cosmological predictor trained on the \textsc{IllustrisTNG} dataset at redshift $z=0$:

\begin{equation}
\Omega_{m,\,z=0}^{\mathrm{sim}}
=
\ln\!\left[
  \sigma\!\left(
    \frac{Z_\star \, S_{\mathrm{core}}}{f_\star\,  k_{\mathrm{sim}}}
  \right)
  + c_0^{\mathrm{sim}}
\right]
- \frac{a_0^{\mathrm{sim}}}{R_{\mathrm{compact}}}\
=
\ln\!\left[
  \sigma\!\left(
    \frac{Z_\star}{M_\star}
    \frac{\Omega_b V_{\mathrm{eff}}^{3}}{G H_0}
    \frac{\Omega_b}{k_{\mathrm{sim}}}
  \right)
  + c_0^{\mathrm{sim}}
\right]
- \frac{a_0^{\mathrm{sim}}}{R_{\mathrm{compact}}}\,.
\label{z=0,eq}
\end{equation}
Where $\sigma$ function is 
$\sigma(x) = 1/(1 + e^{-x})$.
The subscript ``\text{sim}'' denotes simulation-specific coefficients calibrated separately for each suite. (e.g. \textsc{IllustrisTNG}, \textsc{Astrid}, \textsc{SIMBA}, or \textsc{Swift-EAGLE}).

\begin{table}[h!]
\centering
\caption{
Simulation-specific analytic forms of the $\Omega_{m}$ predictor at $z = 0$.
Each expression shares the same functional structure of Eq.~(\ref{z=0,eq}) but uses independently calibrated coefficients
for each simulation suite.}
\begin{tabular}{l c}
\hline\hline
\textbf{Simulation} & \textbf{Analytic Expression} \\[3pt]
\hline
\\[2pt]
\textsc{IllustrisTNG} &
$\displaystyle 
\Omega_{m}^{\mathrm{TNG}}
=
\ln\!\left[
  \sigma\!\left(
     \frac{Z_\star}{M_\star}
    \frac{\Omega_b V_{\mathrm{eff}}^{3}}{G H_0}
    \frac{\Omega_b}{5.68}
  \right)
  + 0.62
\right] -\frac{0.0024}{R_{\mathrm{compact}}}
$ \\[10pt]

\textsc{Astrid} &
$\displaystyle 
\Omega_{m}^{\mathrm{Astrid}}
=
\ln\!\left[
  \sigma\!\left(
     \frac{Z_\star}{M_\star}
    \frac{\Omega_b V_{\mathrm{eff}}^{3}}{G H_0}
    \frac{\Omega_b}{4.71}
  \right)
  + 0.61
\right]
-
\frac{0.005}{R_{\mathrm{compact}}}
$ \\[10pt]
\textsc{SIMBA} &
$\displaystyle 
\Omega_{m}^{\mathrm{SIMBA}}
=
\ln\!\left[
  \sigma\!\left(
     \frac{Z_\star}{M_\star}
    \frac{\Omega_b V_{\mathrm{eff}}^{3}}{G H_0}
    \frac{\Omega_b}{14.88}
  \right)
  + 0.62
\right] +\frac{0.0014}{R_{\mathrm{compact}}}
$ \\[10pt]
\textsc{Swift-EAGLE} &
$\displaystyle 
\Omega_{m}^{\mathrm{Swift-EAGLE}}
=
\ln\!\left[
  \sigma\!\left(
     \frac{Z_\star}{M_\star}
    \frac{\Omega_b V_{\mathrm{eff}}^{3}}{G H_0}
    \frac{\Omega_b}{8.58}
  \right)
  + 0.66
\right] -\frac{0.0035}{R_{\mathrm{compact}}}
$ \\[10pt]

\hline
\end{tabular}
\label{tab:omega_m_eqs}
\end{table}

Importantly, the analytic structure in Eq.~(\ref{z=0,eq}) is not retrained on any other simulation suite data.
The symbolic form of the predictor is learned exclusively from the \textsc{IllustrisTNG-LH} dataset and is subsequently
applied to \textsc{SIMBA}, \textsc{ASTRID}, and \textsc{Swift-EAGLE} without modification to its functional form.
When transferring the equation to other simulation suites, we only refit the numerical coefficients
$\{k_{\mathrm{sim}},\,c_0^{\mathrm{sim}},\,a_0^{\mathrm{sim}}\}$ to account for differences in subgrid physics,
while keeping the latent-variable structure fixed.

In practice, we find that recovering an equally compact analytic relation with clear physical interpretation
when training directly on other simulation suites is substantially more difficult, owing to stronger degeneracies
among feedback parameters and increased stochasticity in baryonic responses.
The simulation-specific coefficients and resulting analytic expressions are summarized in Table~\ref{tab:omega_m_eqs}.

\vspace{2mm}

\begin{figure*}[h!]
    \centering
    % === Row 1: TNG and ASTRID ===
    \begin{minipage}{0.48\textwidth}
        \centering
        \includegraphics[width=\linewidth]{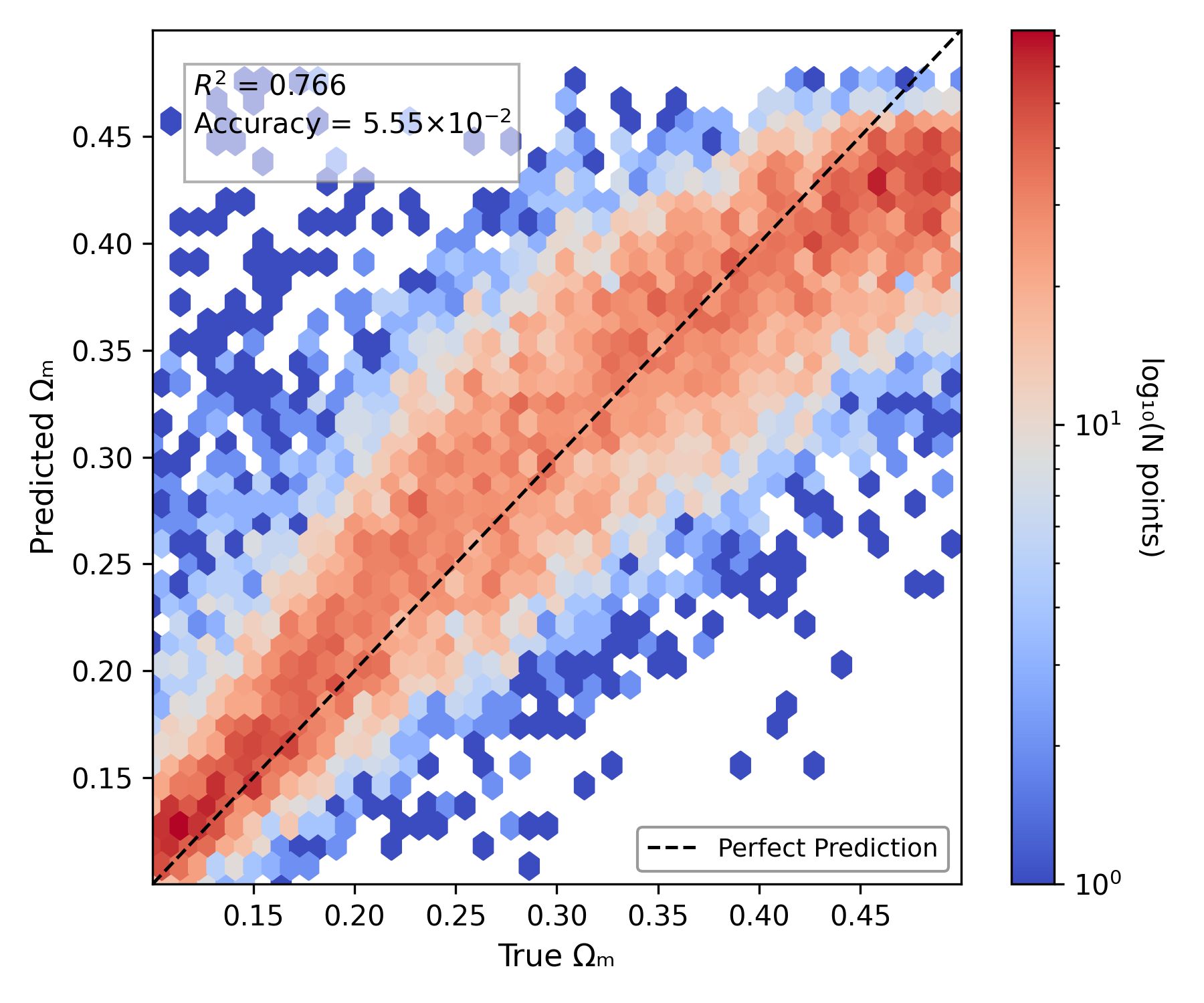}
        \vspace{4pt}
        \small\textbf{IllustrisTNG, $z=0$}
    \end{minipage}
    \hfill
    \begin{minipage}{0.48\textwidth}
        \centering
        \includegraphics[width=\linewidth]{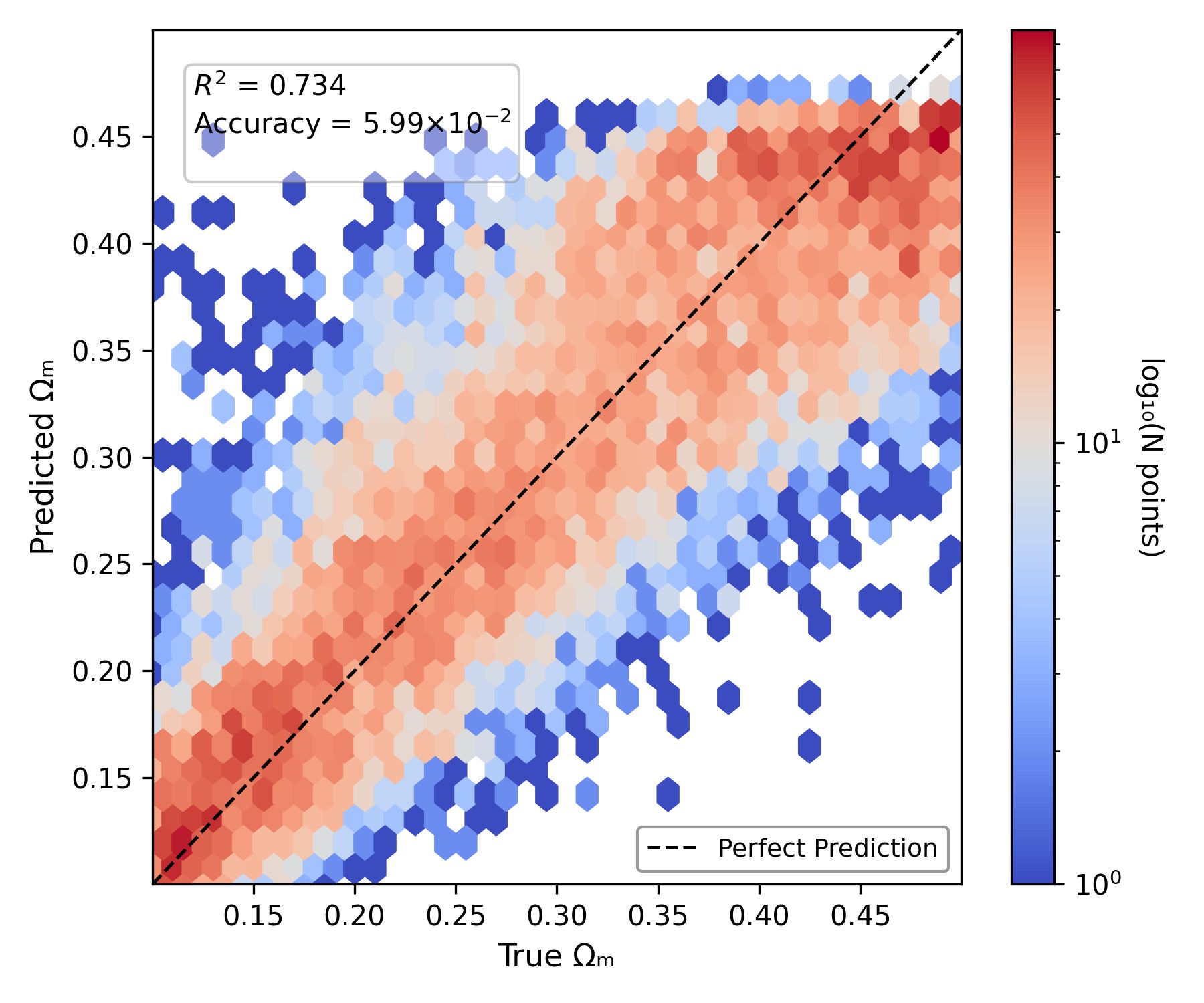}
        \vspace{4pt}
        \small\textbf{ASTRID, $z=0$}
    \end{minipage}

    \vspace{10pt}

    % === Row 2: SIMBA and SWIFT ===
    \begin{minipage}{0.48\textwidth}
        \centering
        \includegraphics[width=\linewidth]{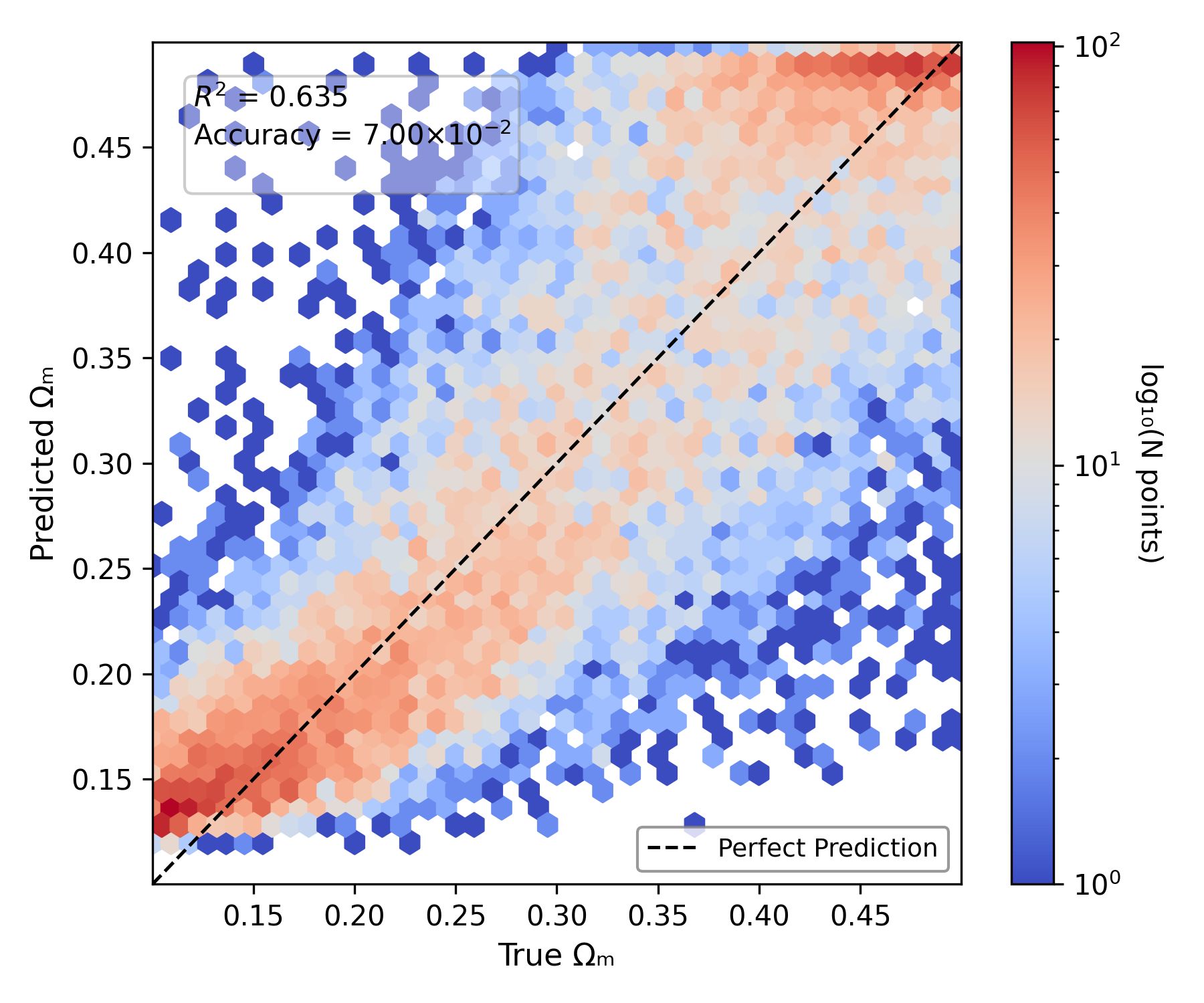}
        \vspace{4pt}
        \small\textbf{SIMBA, $z=0$}
    \end{minipage}
    \hfill
    \begin{minipage}{0.48\textwidth}
        \centering
        \includegraphics[width=\linewidth]{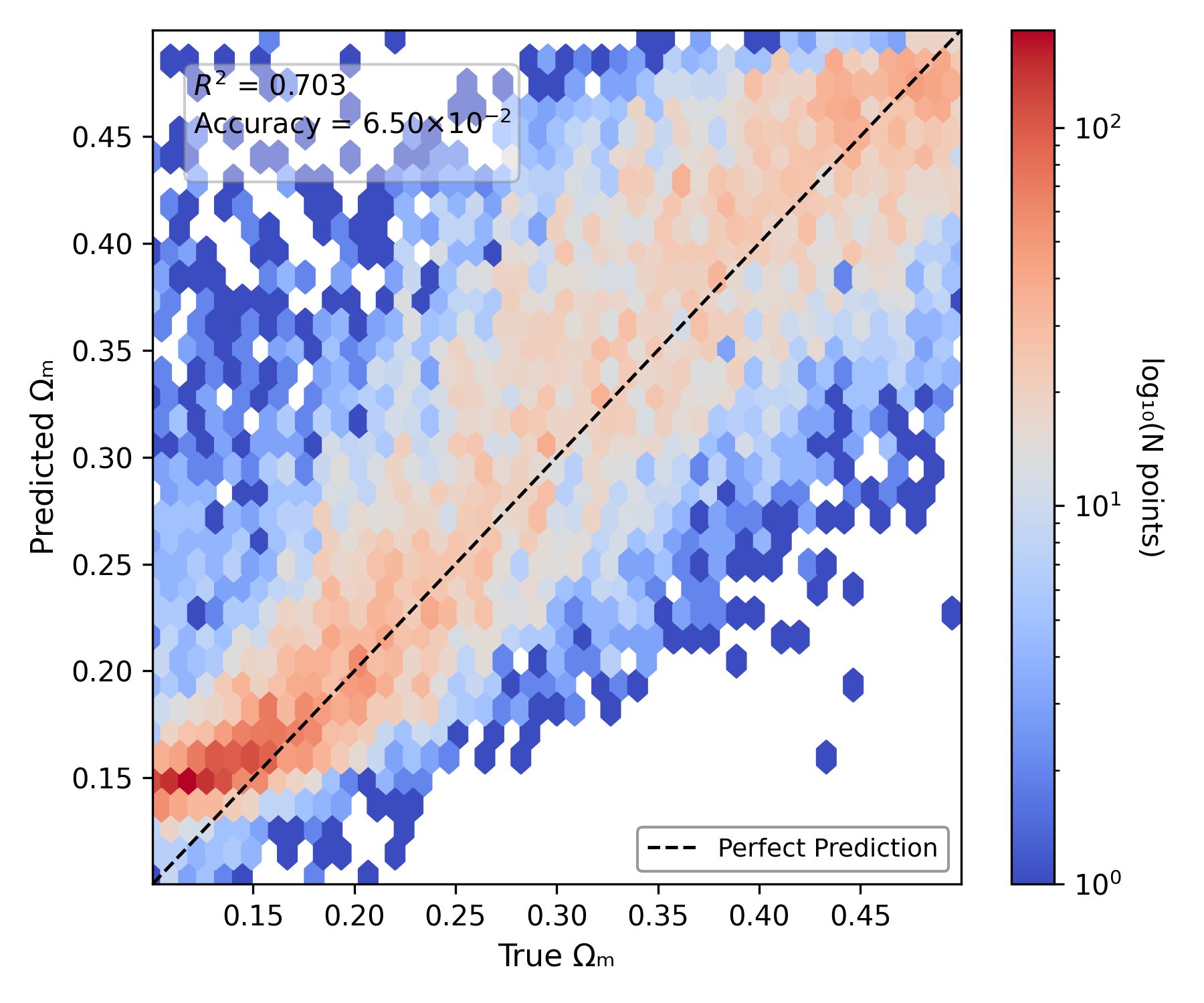}
        \vspace{4pt}
        \small\textbf{Swift-EAGLE, $z=0$}
    \end{minipage}

    \vspace{10pt}

    \caption{
    \textbf{Performance of the analytic $\Omega_m$ predictor across four simulation suites at $z=0$.}
    Each panel compares predicted versus true $\Omega_m$ for galaxies from the \textsc{IllustrisTNG}, \textsc{ASTRID}, \textsc{SIMBA}, and \textsc{Swift-EAGLE} simulations.
    Points are binned using a hexagonal grid and colored by the base-10 logarithm of the number of galaxies in each bin.
    The dashed line denotes perfect prediction, and the quoted $R^2$, and accuracy values summarize performance in linear space.
    }
    \label{fig:omega_m_fit_z0}
\end{figure*}

Figure~\ref{fig:omega_m_fit_z0} shows the performance of the analytic
$\Omega_m$ predictor at $z = 0$ for the
\textsc{IllustrisTNG}, \textsc{ASTRID}, \textsc{SIMBA}, and \textsc{Swift-EAGLE}
simulation suites.
Each panel compares predicted and true values using galaxies drawn from
1,000 independent realizations per suite, with 20 randomly selected galaxies
per simulation; for \textsc{IllustrisTNG}, we sample a different set of galaxies than those used for training.
Colors indicate the logarithmic number density in each hexagonal bin.

The model accurately recovers the underlying cosmological parameter
in both \textsc{IllustrisTNG} and \textsc{ASTRID},
achieving coefficients of determination of
$R^2 = 0.77$ and $R^2 = 0.73$, respectively.
The corresponding accuracies are $5.6 \times 10^{-2}$ and $6.0 \times 10^{-2}$.
Results for \textsc{SIMBA} and \textsc{Swift-EAGLE} exhibit systematically
larger scatter and degraded statistical performance,
reflecting differences in subgrid feedback implementations.
These cross-suite comparisons therefore probe physical consistency rather than
model overfitting, indicating that the extracted functional form captures
a genuine coupling between galaxy properties and the matter density $\Omega_m$.
At the high-$\Omega_m$ end, a mild underprediction is visible,
consistent with the saturation behavior of the logarithmic--sigmoid mapping

We note that the machine-learning models in previous work \citep{Villaescusa-Navarro:2022twv} were trained using flexible neural networks on large samples of galaxies drawn jointly from both the \textsc{IllustrisTNG} and \textsc{SIMBA} suites of the CAMELS simulations. In contrast, the symbolic-regression framework adopted here is intentionally constrained to identify a compact analytic relation from a controlled subset of galaxies, and is trained exclusively on galaxies from a single simulation suite (\textsc{IllustrisTNG}) at $z=0$, before being applied and tested across different simulation suites. Neural networks optimize predictive accuracy through high functional flexibility, whereas our approach prioritizes physical interpretability and robustness across simulation suites, and the differences in performance should be interpreted accordingly.

\subsection{Analytic predictor for \(\Omega_{m}\) at high redshift}
\label{sec:LH-higher_redshift}

For high redshift galaxies, we adopt an analytic form in which the predictor includes two $(1+z)$ power--law scalings, 
fitted across redshift snapshots $z = 0.5,\,1.0,\,1.5,\,2.0,$ and $3.0$ for \textsc{IllustrisTNG} and \textsc{ASTRID} datasets: 
\begin{equation}
\Omega_{m}^{\mathrm{sim}}(z)
=
\ln\!\left[
  \sigma\!\left(
    \frac{Z_\star}{M_\star}
    \frac{\Omega_b V_{\mathrm{eff}}^{3}}{G H_0}
    \frac{\Omega_b}
         {k_{\mathrm{sim}}}
    \frac{1}{(1+z)^{\alpha_k}}
  \right)
  + c_0^{\mathrm{sim}}
\right]
- \frac{a_0^{\mathrm{sim}}\,(1+z)^{\alpha_\gamma}}
        {R_{\mathrm{compact}}}\,.
\label{eq:Omega_m_sim_z}
\end{equation}
At each redshift, galaxies are independently identified from the corresponding simulation snapshot using the same selection criteria.

\begin{figure*}[h]
\centering
\setlength{\tabcolsep}{4pt}

% ---------- (a) IllustrisTNG ----------
\begin{tabular}{c c c}
\includegraphics[width=0.33\textwidth]{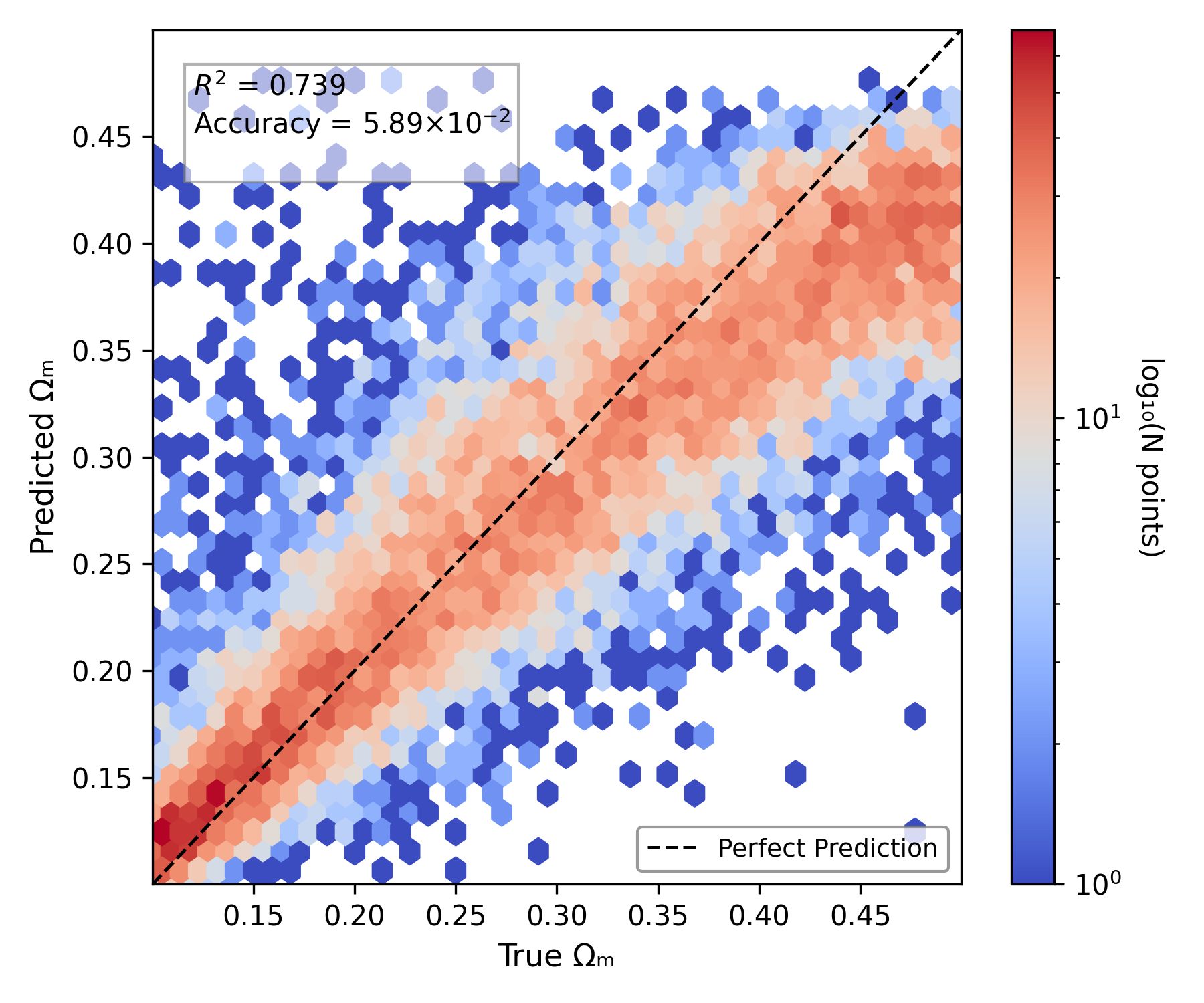} &
\includegraphics[width=0.33\textwidth]{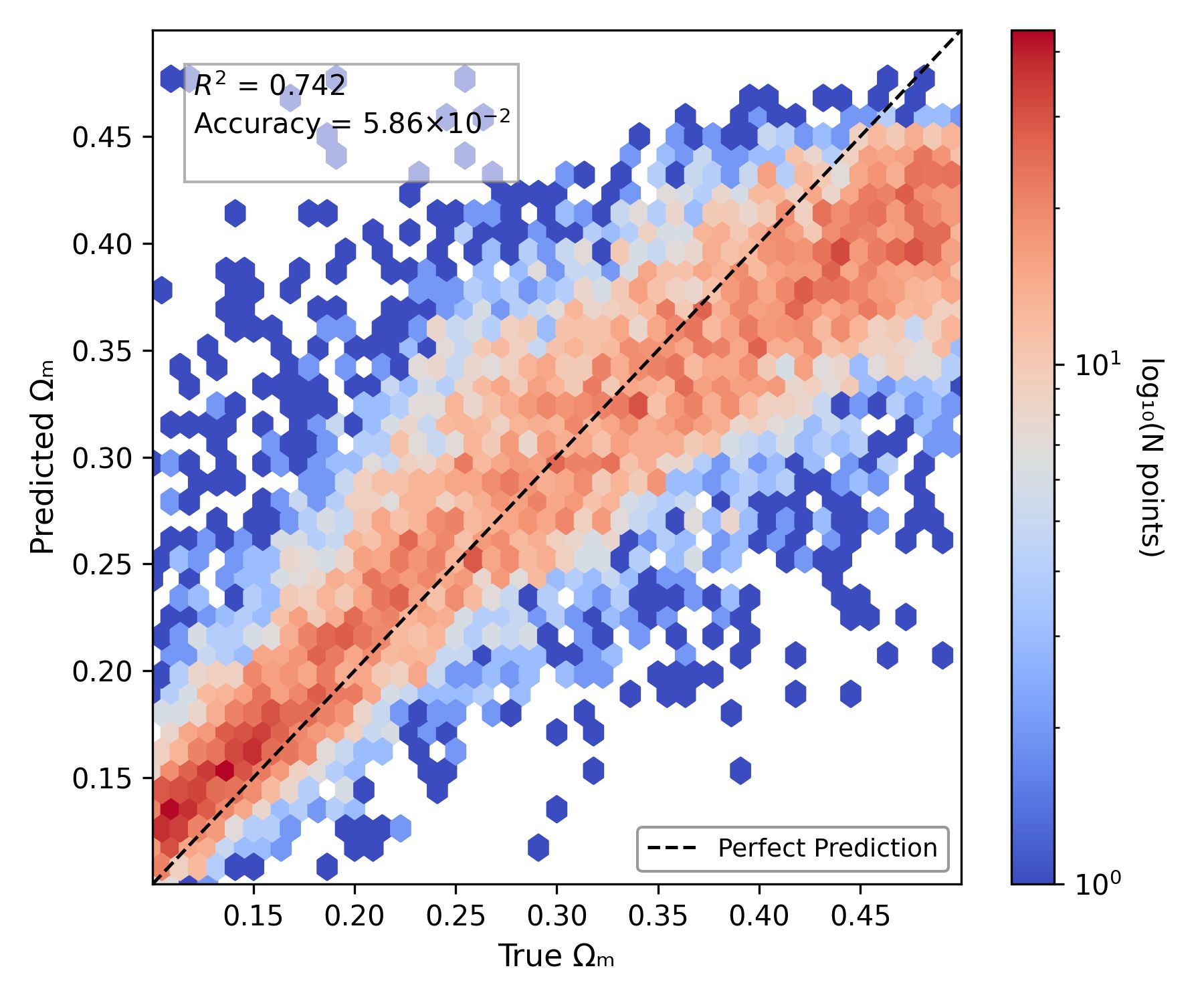} &
\includegraphics[width=0.33\textwidth]{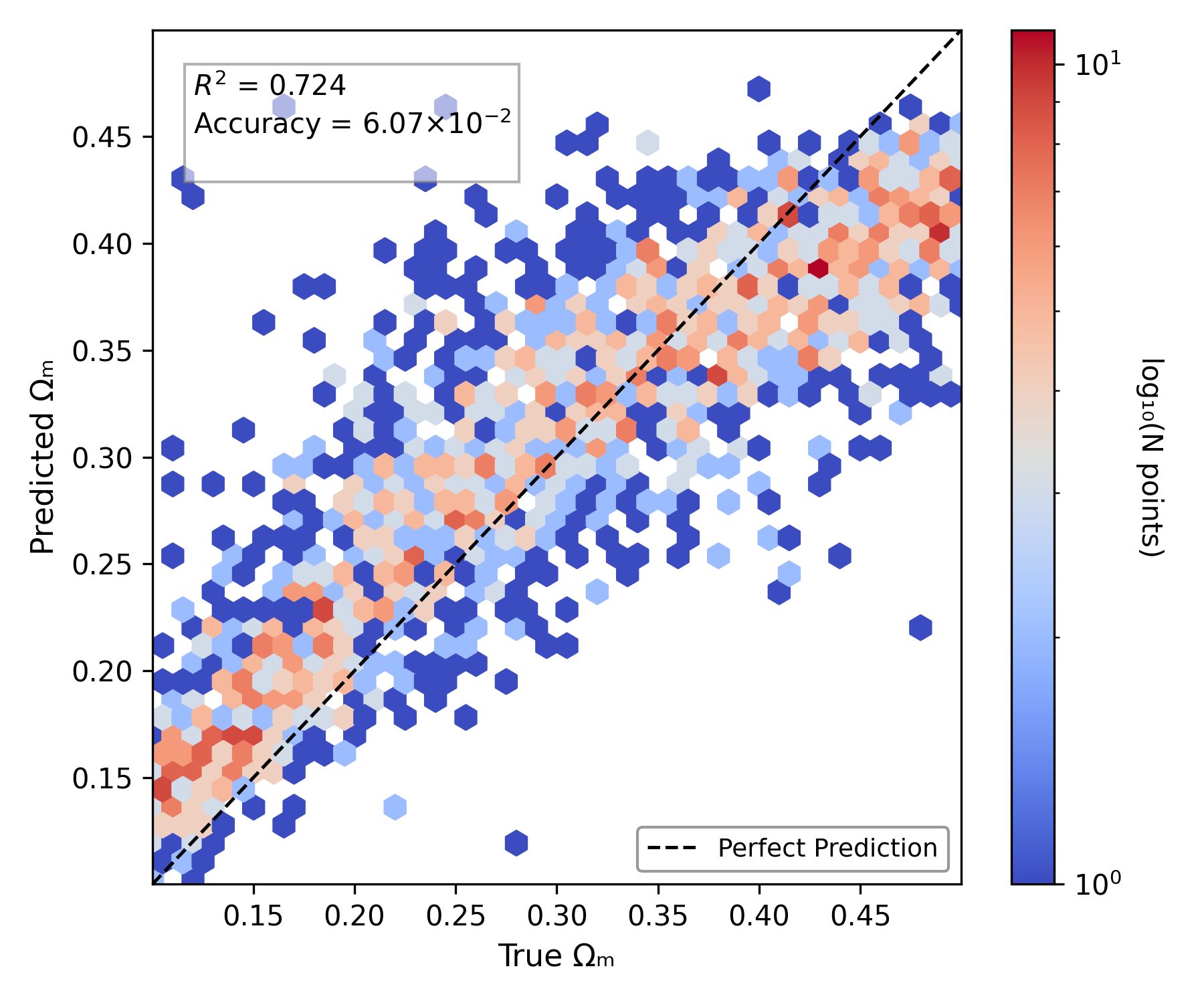}
\end{tabular}

\vspace{0.25em}

{\footnotesize \textbf{(a) IllustrisTNG at $z=1,2,3$.}}

\vspace{0.8em}

% ---------- (b) ASTRID ----------
\begin{tabular}{c c c}
\includegraphics[width=0.33\textwidth]{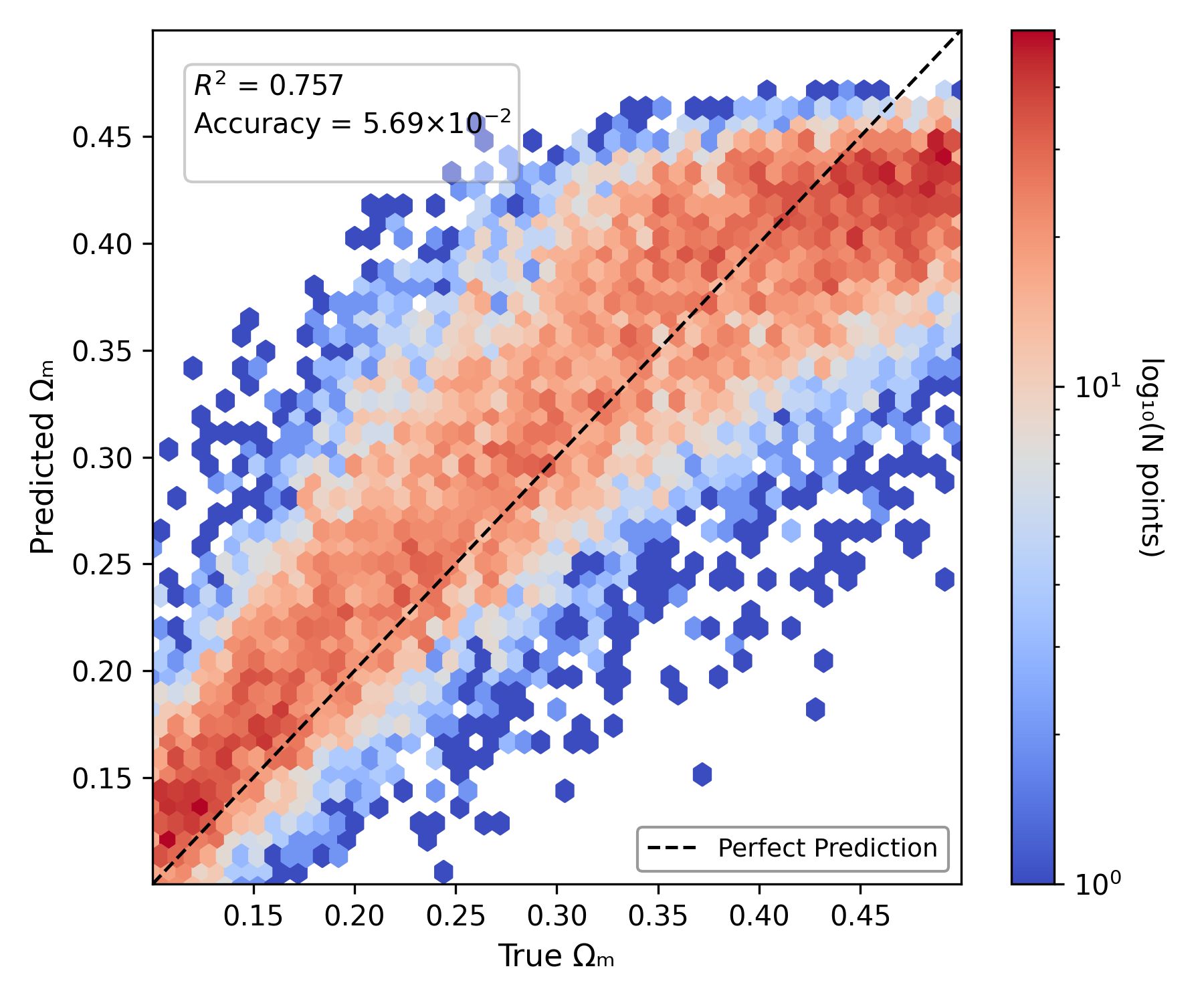} &
\includegraphics[width=0.33\textwidth]{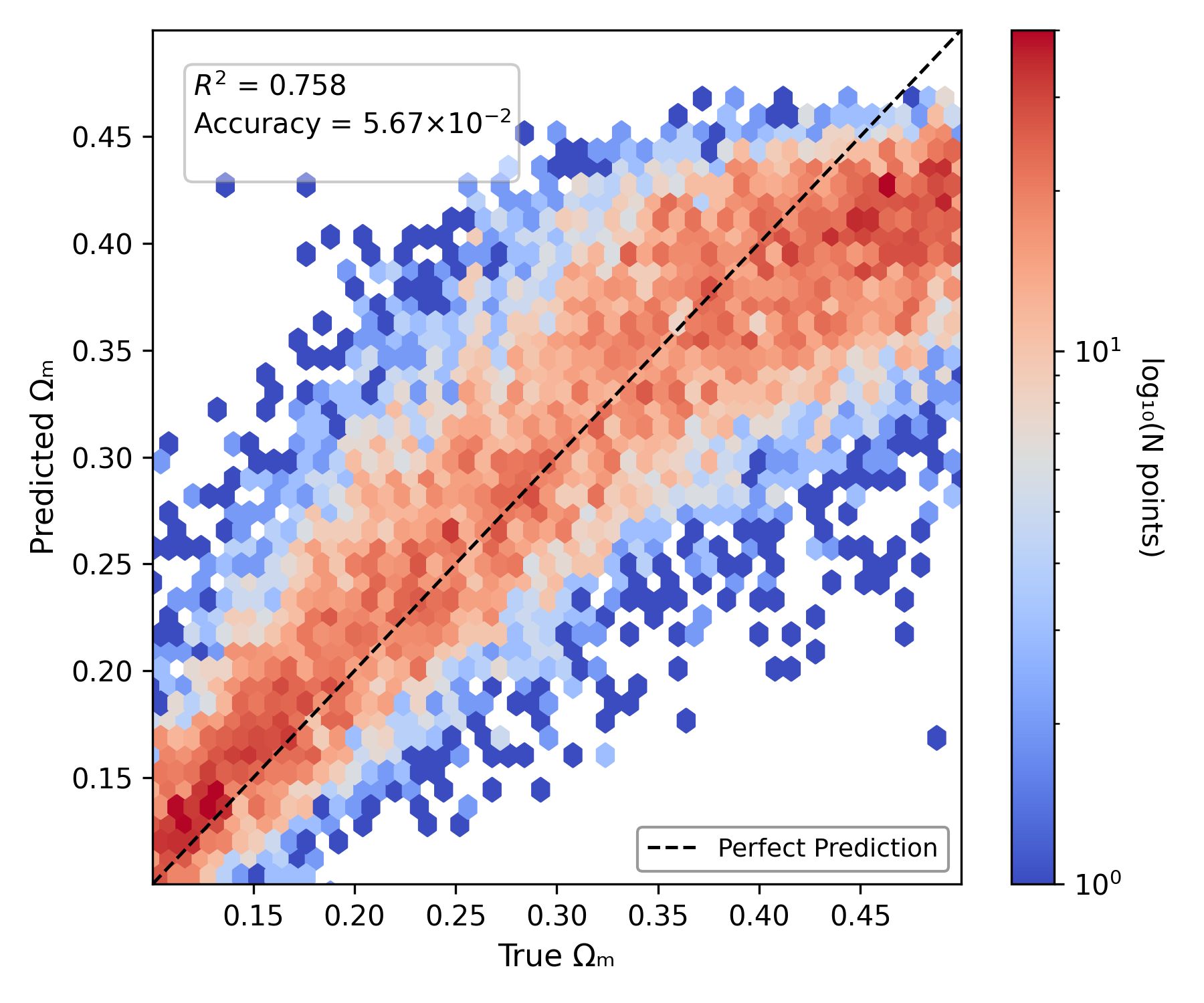} &
\includegraphics[width=0.33\textwidth]{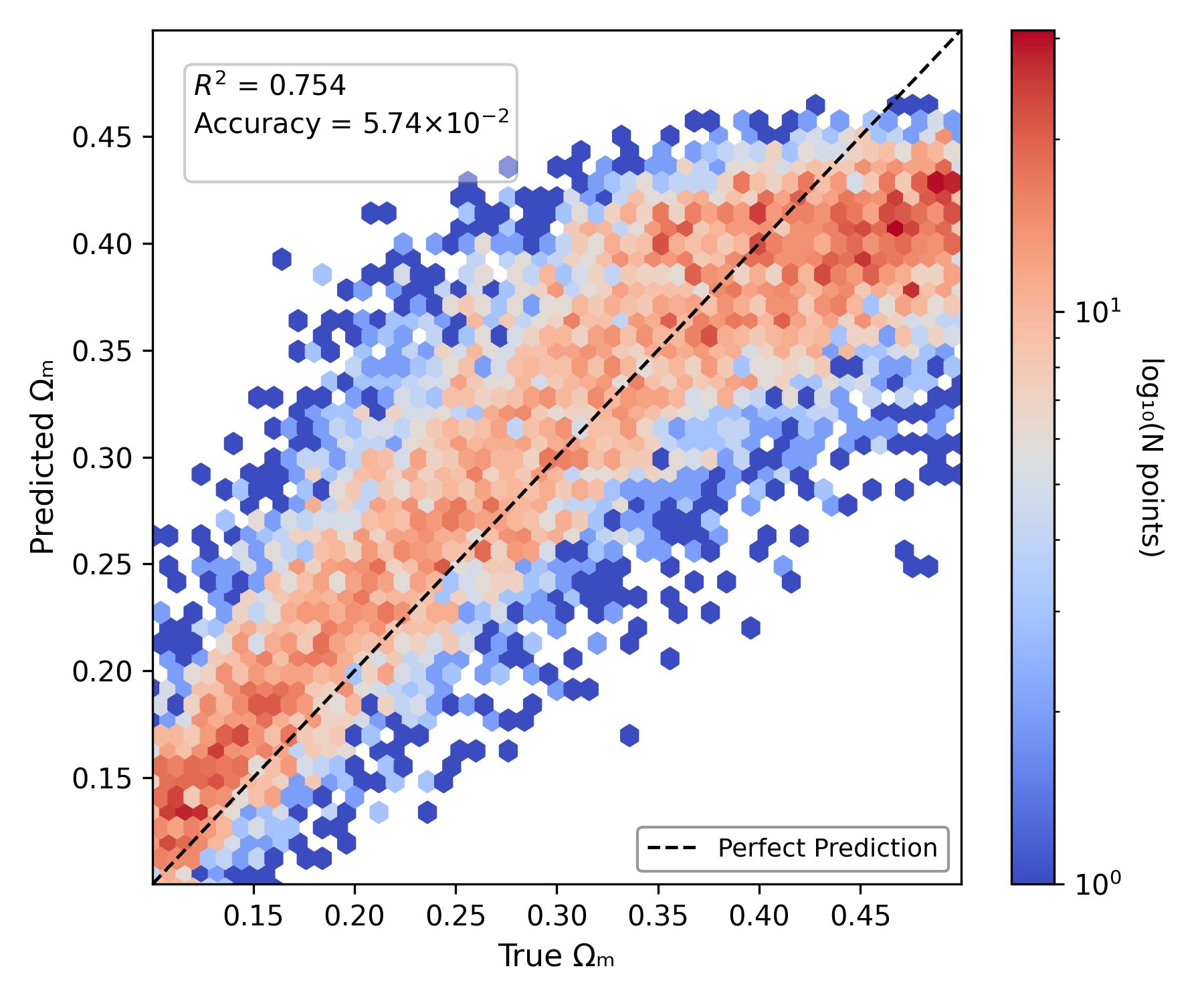}
\end{tabular}

\vspace{0.25em}

{\footnotesize \textbf{(b) ASTRID at $z=1,2,3$.}}

\caption{
\textbf{Performance of the analytic $\Omega_m$ predictor across different redshifts.}
Each panel compares predicted and true $\Omega_m$ values.
Top row: \textsc{IllustrisTNG}. Bottom row: \textsc{ASTRID}.
Colors represent the logarithmic number density per hexagonal bin.
The dashed red line indicates perfect prediction.
}
\label{fig:performance_z}
\end{figure*}

The best--fit power--law exponents are 
$\alpha_k = 0.45$ and $\alpha_\gamma = 0.33$.
Although we tested a third $(1+z)$ scaling applied to the $c^{\mathrm{sim}}_0$ term, its fitted response remained consistent with $(1+z)^0$, 
indicating that the offset component does not evolve significantly with redshift.
We present the full redshift evolution of the $\Omega_m$ predictor
for both \textsc{ IllustrisTNG} and \textsc{Astrid} simulations simulation suites in Figure~\ref{fig:performance_z}.
In contrast, applying the same redshift parameterization to the
\textsc{SIMBA} and \textsc{Swift-EAGLE} simulations does not yield
comparably stable performance across redshift.
This suggests that while the underlying latent structure is present in all
four models at $z=0$, its redshift transportability depends on the details of the baryonic physics implementation.
We therefore restrict presentation of the redshift-dependent fits to
\textsc{IllustrisTNG} and \textsc{ASTRID} here.

Our results show that for both simulation suites, the correlation between predicted and true $\Omega_m$ remains strong ($R^2\simeq0.74$–0.76) with comparable accuracy ($\simeq0.056- 0.06$), indicating that the redshift-dependent power-law scaling in Eq.~\ref{eq:Omega_m_sim_z} effectively captures the evolution of the baryon–cosmology coupling.
At $z=3$, the \textsc{IllustrisTNG} panels contain fewer data points due to the smaller number of galaxies per simulation at early cosmic times, while \textsc{ASTRID} suite retains higher sampling density.
Despite this difference, 
the consistency observed between the \textsc{IllustrisTNG} and \textsc{ASTRID} simulations in both redshift scaling and predictive performance, extending to higher redshift, supports a physical interpretation of the analytic relation rather than one driven by simulation-specific tuning. 
In contrast, its degraded performance at higher redshift in the \textsc{SIMBA} and \textsc{Swift-EAGLE} models suggests that additional modifications may be necessary to achieve robustness across diverse subgrid physics implementations.

We note, however, that the analytic structure itself does not appear to break down in the \textsc{SIMBA} and \textsc{Swift-EAGLE} simulations at higher redshifts. Instead, the power--law exponents describing the redshift evolution of the coefficients, $\alpha_k$ and $\alpha_\gamma$, which provide a consistent description for the \textsc{IllustrisTNG} and \textsc{Astrid} suites, do not remain universal when applied to the \textsc{SIMBA} and \textsc{Swift-EAGLE} datasets. In these cases, the relation continues to provide reasonable predictive performance when the coefficients are recalibrated separately for each simulation, indicating that the underlying functional dependence remains valid but that the effective normalization and scaling depend on the specific subgrid physics model. Training symbolic regression across multiple snapshots ($z=0$--$3$) while including redshift as an input variable was also explored, but the algorithm showed little sensitivity to this parameter, likely because redshift is constant within each snapshot while galaxy properties dominate the variance of the training data.

\section{Physical Interpretation}
\label{sec:LH-physics}
In this section, we explore the physical basis of our analytic predictor and provide a conceptual framework that links its mathematical form to the underlying astrophysical processes.

Before interpreting the physical meaning of the equations themselves, we begin by examining how feedback processes shape the stellar mass–metallicity relation (MZR) 
~\citep{2008AJ....135.1877E,1979A&A....80..155L,Tremonti:2004et,Koppen:2006ji} in the simulations through Subsection~\ref{sec:feedbackregulation}. 
As shown later, this provides the foundation for why a single-galaxy tracer could capture cosmological information across diverse galaxy populations and simulation models.

\subsection{Feedback Regulation of the Mass–Metallicity Relation Across Simulation Suites}
\label{sec:feedbackregulation}

\begin{figure*}[h]
\centering

% --- IllustrisTNG z = 0 ---
\begin{minipage}{0.49\textwidth}
    \centering
    \includegraphics[width=\linewidth]{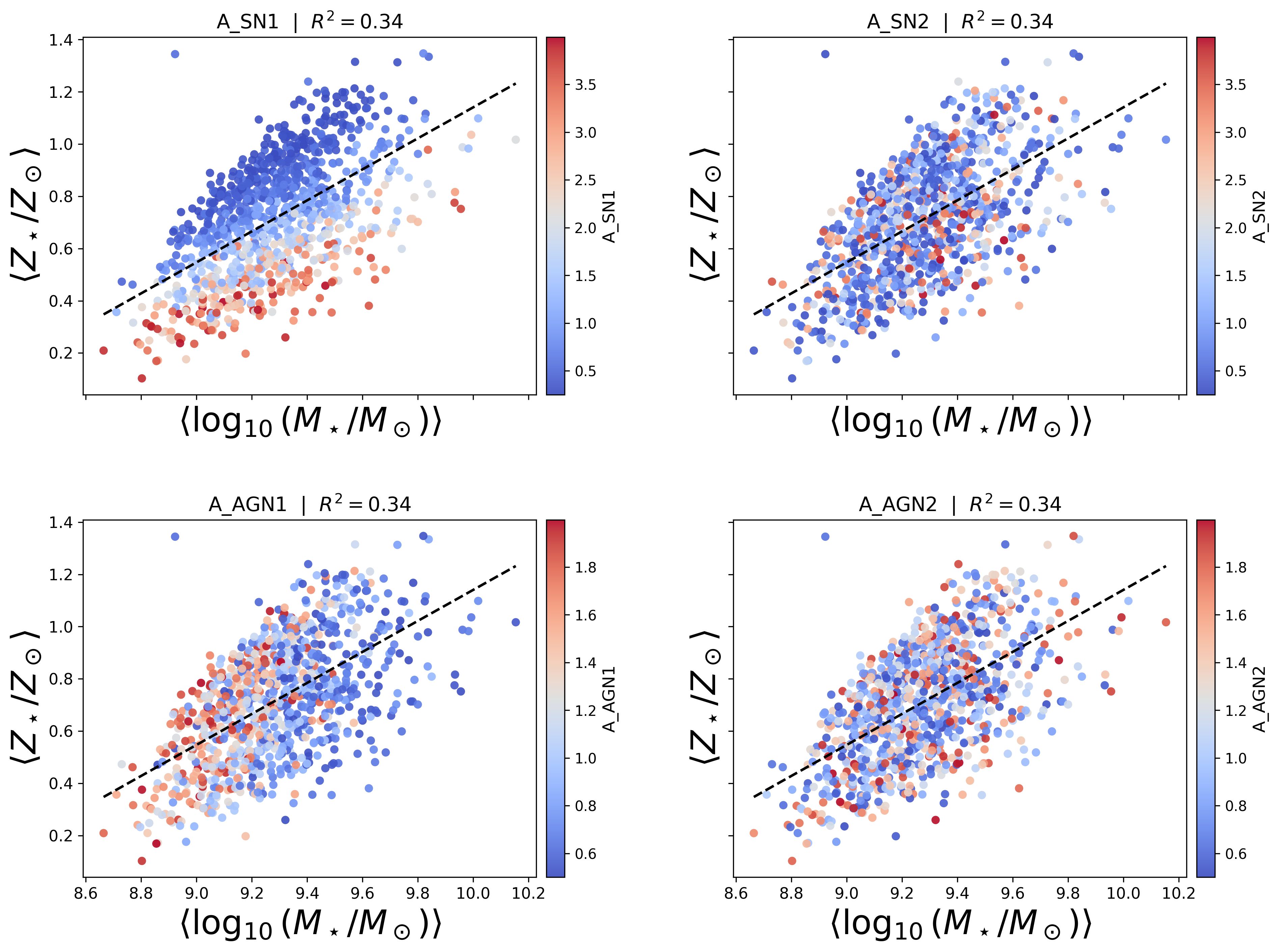}
    \vspace{3pt}
    \textbf{IllustrisTNG, $z=0$}
\end{minipage}
\hfill
% --- Astrid z = 0 ---
\begin{minipage}{0.49\textwidth}
    \centering
    \includegraphics[width=\linewidth]{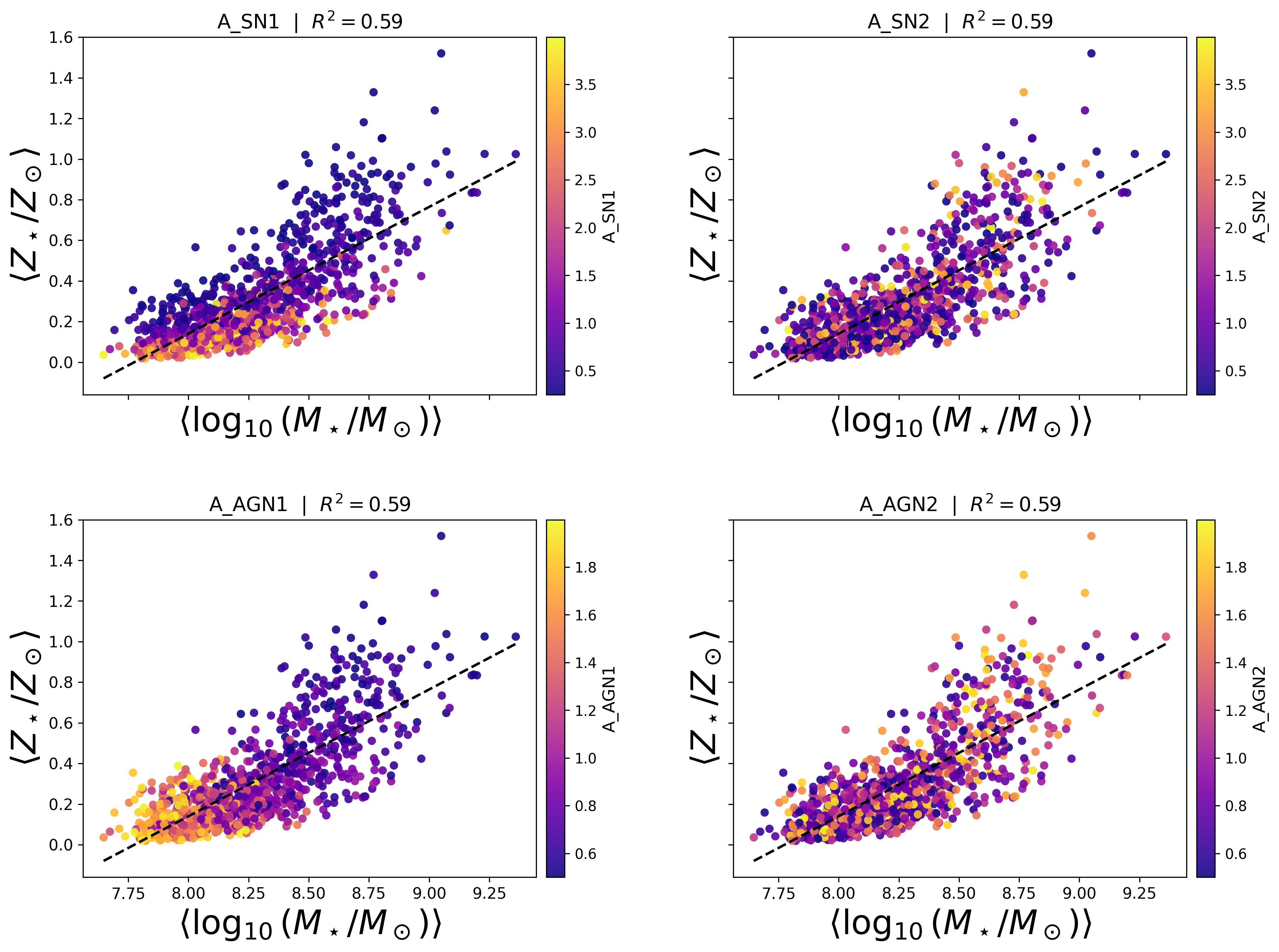}
    \vspace{3pt}
    \textbf{Astrid, $z=0$}
\end{minipage}

\vspace{0.7em}

\caption{
Stellar mass--metallicity relations (\textit{MZR}) at redshift $z=0$ for the 
\textsc{IllustrisTNG} (left) and \textsc{Astrid} (right) simulation suites. 
Each point represents an individual galaxy colored by its feedback parameters 
$A_{\mathrm{SN1}}, A_{\mathrm{SN2}}, A_{\mathrm{AGN1}},$ and $A_{\mathrm{AGN2}}$.
The dashed black lines show the best--fit power--law trends in 
$\log M_\star$--$Z_\star$ space, with annotated $R^2$ values quantifying 
the correlation strength. 
These relations illustrate that both simulations exhibit similar MZR slopes and 
feedback‐driven metallicity modulation, providing a baseline comparison 
for the analytic $\Omega_m$ tracer model.
}
\label{fig:MZ_feedback_z0}
\end{figure*}

Within the \textsc{CAMELS-LH} simulations, four feedback parameters are systematically varied: 
SN feedback  $A_{\mathrm{SN1}}$ and $A_{\mathrm{SN2}}$, 
and AGN feedback  $A_{\mathrm{AGN1}}$ and $A_{\mathrm{AGN2}}$. 
Figure~\ref{fig:MZ_feedback_z0} shows the resulting $\log M_\star$–$Z_\star$ MZRs at redshift $z=0$ for both the \textsc{IllustrisTNG} and \textsc{Astrid} simulations, highlighting how variations in feedback alter the metallicity normalization and slope. 
Results at higher redshifts ($z = 1$, $2$, $3$) are provided in Appendix \ref{sec:FeedbackonMZRathighredshift} (see Fig. \ref{fig:MZ_feedback_evolution}) for completeness.

In the \textsc{IllustrisTNG} suite,
at redshift $z=0$, galaxies with stronger SN feedback 
($A_{\mathrm{SN1}}$ and $A_{\mathrm{SN2}}$; red points) tend to exhibit lower metallicities 
at fixed stellar mass, indicating that enhanced supernova activity 
efficiently removes enriched gas from the interstellar medium and suppresses 
metal retention \citep{Torrey_2019,Dav__2011}. In contrast, the AGN feedback parameters 
($A_{\mathrm{AGN1}}$ and $A_{\mathrm{AGN2}}$) show weaker or more scattered correlations.
Parameter $A_{\mathrm{SN1}}$ emerges as the dominant driver of the MZR, controlling both the normalization and scatter of the relation. 
The secondary SN mode ($A_{\mathrm{SN2}}$) and the AGN parameters 
($A_{\mathrm{AGN1}}$, $A_{\mathrm{AGN2}}$) display weaker or more dispersed correlations, 
indicating a comparatively minor impact on the resulting metallicity.
The redshift evolution of this relation reveals a progressive tightening of the MZR, with the coefficient of determination increasing from $R^2 \simeq 0.3$ at $z=0$ to $R^2 \simeq 0.77$ by $z=3$ (see Appendix~\ref{sec:FeedbackonMZRathighredshift}, Figure~\ref{fig:MZ_feedback_evolution}).
At higher redshifts, galaxies exhibit steeper MZRs and more coherent color gradients: 
stronger feedback corresponds to systematically lower metallicities, reflecting the 
dominant role of feedback in regulating early baryon cycling and chemical enrichment. 
As cosmic time progresses, recycling and self-regulation processes 
diminish this dependence, leading to larger scatter at low redshift.

For the \textsc{Astrid} suite, at redshift $z=0$, SN and AGN feedback exhibit a more balanced influence on the MZR compared to \textsc{IllustrisTNG}. 
The parameter $A_{\mathrm{SN1}}$ remains a key driver of metallicity suppression, 
but the AGN parameters $A_{\mathrm{AGN1}}$ and $A_{\mathrm{AGN2}}$ show comparable metallicity gradients, 
indicating that AGN-driven outflows play a prominent role in regulating galactic enrichment. 
The overall correlations are tighter ($R^2 \simeq 0.6$) and span a broader stellar-mass range, 
signifying a more coherent coupling between feedback and star formation efficiency as shown in Figure~\ref{fig:MZ_feedback_evolution} right column.
Across redshift, the MZR slope and scatter remain relatively stable up to $z \simeq 3$, 
with both SN and AGN feedback maintaining clear anti-correlations with metallicity. 
Unlike \textsc{IllustrisTNG}, where the $A_{\mathrm{SN1}}$ mode dominates the enrichment history, 
feedback in \textsc{Astrid} acts in a more integrated and self-consistent fashion: 
supernovae and AGN processes operate concurrently to regulate baryon cycling and chemical enrichment, 
producing a smoother MZR evolution and a persistent balance between feedback strength and metallicity buildup.

In contrast, the \textsc{SIMBA} and \textsc{Swift-EAGLE} suites exhibit weaker and less coherent mass–metallicity relations. In \textsc{SIMBA}, variations in both SN and AGN feedback parameters primarily increase the scatter of the MZR rather than produce systematic metallicity offsets at fixed stellar mass, indicating a reduced sensitivity of stellar metallicity to feedback strength. \textsc{Swift-EAGLE} occupies an intermediate regime: feedback-driven metallicity gradients remain visible, but only over a restricted stellar-mass range, and they do not persist across the full galaxy population, limiting the effectiveness of metallicity as a global tracer of baryon-processing efficiency.

\subsection{Physical meaning of the Latent Variable}
\label{sec:latent variable}

The contrasting feedback behaviors observed in our results
motivate a closer examination of how metallicity enters our analytic formulations (Equations \ref{z=0,eq} and \ref{eq:Omega_m_sim_z}). 
From the structure of the analytic formula, we identify that the latent variable $x$
inside the sigmoid operator $\sigma (x)$ naturally encodes the stellar 
\textit{mass--metallicity relation (MZR)}—denoted as term~$(\mathrm{I})$:
\[
x \equiv
\underbrace{\frac{Z_\star}{M_\star}}_{(\mathrm{I})}
\underbrace{\frac{\Omega_b V_{\mathrm{eff}}^{3}}{G H_0}}_{(\mathrm{II})}
\underbrace{\frac{\Omega_b}{k_{\rm sim}}}_{(\mathrm{III})}\,.
\]
Isolating the MZR term reveals that the remaining components of $x$ 
carry distinct and complementary physical interpretations.

Inside term~$(\mathrm{II})$, the factor $V_{\mathrm{eff}}^{3}/(G H_0)$ has the same dimensional form as a mass term, corresponding to the characteristic mass enclosed by a system with dynamical velocity $V_{\mathrm{eff}}$ in a Universe expanding at rate $H_0$.
Multiplication by the cosmic baryon fraction $\Omega_b$ projects this total mass into its baryonic component, under the assumption that galaxies inherit the cosmological baryon fraction.
Although this assumption implies that the local baryon-to-total-matter ratio matches the cosmological mean, feedback and gas dynamics may lead to departures from this idealized value.

Term~$(\mathrm{III})$ functions as a \textit{baryon--cosmology coupling coefficient} 
that encapsulates the influence of feedback regulation within the latent variable. 
Here, $\Omega_b$ represents the universal baryon supply, while the scaling factor $1/k_{\rm sim}$ 
adjusts this baseline to reflect the \textit{effective participation efficiency}—the fraction 
of baryons that remain gravitationally bound and actively engaged in the 
star‐formation and enrichment cycle. 
In galaxies with stronger feedback, energetic outflows driven by SN or AGN 
expel metal‐rich gas and decrease baryon retention, effectively increasing $k_{\rm sim}$ (\textsc{IllustrisTNG})
and lowering the contribution of baryons to subsequent enrichment. 
Conversely, weaker feedback allows gas to remain confined within the potential well, 
reducing $k_{\rm sim}$ (\textsc{Astrid}) and enhancing metallicity buildup. 
Term~$(\mathrm{III})$ therefore serves as a feedback‐regulation coefficient, 
linking local baryonic processes to the global baryon fraction $\Omega_b$ 
and mediating the connection between feedback physics and the cosmological 
sensitivity encoded in the analytic predictor.

Altogether, the latent variable $x$ reflects the interplay between a baryonic mass scale predicted by galaxy dynamics and cosmology, and the subsequent modification of this scale by feedback-driven baryon retention and enrichment. A related baryon–cosmology connection has previously been explored at the scale of galaxy clusters, where \cite{White:1993wm} argued that the baryon fraction within gravitationally bound systems can constrain the cosmic matter density through the approximate relation $f_b \equiv M_{\rm baryon}/M_{\rm total} \approx \Omega_b / \Omega_m$. Here, we extend this principle to galaxy scales by encoding, within the latent variable, the coupling between the cosmological baryon supply and the nonlinear thermodynamic and chemical evolution of baryons within galaxy potentials.

The sigmoid operator $\sigma(x)$  acting on the latent variable further distills this baryonic information. This imposes physically motivated saturation at both low and high baryon-processing capacity, reflecting the diminishing sensitivity of enrichment to variations in gas supply or feedback once the system becomes strongly metal-poor or metal-rich. 
The subsequent logarithm then restores the dynamic range in the high-capacity regime, yielding a smoothly varying, monotonic mapping between the processed baryonic content of a galaxy and the underlying cosmology. The additive constant in the argument of the logarithm introduces an enrichment floor, ensuring numerical stability at the same time. 

Finally, the small corrective term proportional to $R_{\rm compact}^{-1}$ encodes a subtle structural dependence, introducing a mild sensitivity to galaxy morphology and its evolution with redshift. Although this term appears with a small coefficient in the analytic expression, it improves the predictive accuracy of the relation by capturing systematic residual trends. In the results, removing the term decreases the coefficient of determination by approximately 11 percent, indicating that while the dominant contribution to the relation arises from the baryon-binding and metallicity term, the compactness correction provides a secondary structural adjustment that improves the overall accuracy. The amplitude of this correction also exhibits a mild redshift dependence through the factor $(1+z)^{\alpha_\gamma}$ with $\alpha_\gamma \simeq 0.33$, reflecting the evolving influence of galaxy structural properties with cosmic time.

\vspace{2mm}

With the full physical interpretation of the analytic predictor in hand, a natural question arises: \emph{why is its behavior effectively captured by a mass–metallicity relation}?
Our reasonable assumption is that, in a regime where astrophysical degrees of freedom are only minimally perturbed, the baryon cycle does not explore the full richness of feedback-regulated behavior. Indeed, in the CAMELS-LH simulation runs — where the cosmological parameters, including $\Omega_b$ and $H_0$, are held fixed and only a limited subset of astrophysical variations are restricted—the baryon cycle remains sufficiently regular that this integrated behavior is well approximated by a clean mass–metallicity–like dependence. As a result, the stellar mass–metallicity relation serves as the dominant channel through which $\Omega_m$ imprints on individual galaxies.

A clearer view of the mathematical structure underlying this behavior emerges from examining power–law transformations of the analytic relation in Eq.~\ref{eq:Omega_m_sim_z}, which provide complementary insight into how the latent variables combine to yield the observed scaling. 
The derivation of these transformations, together with a visualization of the resulting galaxy–cosmology manifold that captures how individual galaxies populate the latent predictor space, are presented in Appendix~\ref{sec:Appendix_power-law} and Appendix~\ref{sec:gal_manifold} respectively.

To reinforce the robustness of this interpretation, we extend our analysis in Section \ref{sec:SB28tests} to the CAMELS-TNG-SB28 simulation suites, in which variations in cosmology and a wider astrophysical parameter space introduce richer baryonic pathways. This broader setting allows us to assess how the underlying physical mechanisms adapt under more complex conditions.

\vspace{-7pt}

\section{Generalization Across extended cosmology and feedback variations}
\label{sec:SB28tests}

\vspace{2mm}

The CAMELS project primarily uses sets of hydrodynamic simulations in which six cosmological and astrophysical parameters are varied across 1{,}000 realizations for each of the \textsc{ASTRID}, \textsc{IllustrisTNG}, \textsc{SWIFT-EAGLE}, and \textsc{SIMBA} models. These original datasets, referred to as the \textbf{CAMELS-LH} sets, (detailed in ~\cite{Villaescusa-Navarro:2022twv}) provide a controlled framework for exploring the impact of a small number of dominant parameters on large-scale structure and galaxy formation. However, while this design captures the leading-order sources of variation, it does not span the full range of uncertainty present in modern cosmological and subgrid models. Many additional components of astrophysical modeling, including prescriptions for stellar evolution, galactic winds, and feedback from active galactic nuclei, are not explicitly sampled in the CAMELS-LH configuration, despite their potentially significant impact on galaxy observables and cosmological inference.
To address this limitation, CAMELS introduced a series of extended simulation sets designed to explore substantially higher-dimensional parameter spaces. One such extension is the \textbf{TNG-SB28} simulation suite \citep{CAMELS:2023wqd}, which expands the \textsc{IllustrisTNG} model by simultaneously varying 28 independent parameters (see more details in Section ~\ref{sec:style}).

Significantly for this work, the SB28 dataset allows both $\Omega_b$ and the Hubble parameter $H_0$ to vary self-consistently alongside astrophysical parameters for the first time within the \textsc{IllustrisTNG} framework. This feature enables direct tests of whether galaxy-scale properties remain informative about cosmology when the background expansion rate and baryon content are no longer fixed, providing a uniquely stringent environment for evaluating analytic models of cosmological inference from galaxy observables.

In Section~\ref{sec:SB28:result}, we present the results of applying symbolic regression to the TNG-SB28 simulations, focusing on how the analytic form of the predictor adapts in the presence of a substantially expanded parameter space. In the following subsection~\ref{sec:gas regulae}, we interpret these results within the framework of the gas regulator model, providing a physical explanation for the emergent structure and its dependence on baryonic processes.

\subsection{TNG-SB28 Set Result}
\label{sec:SB28:result}

We apply the same symbolic regression framework (see Section~\ref{sec:ML-SR}) and galaxy-scale feature set (see Section~\ref{sec:dimensionless_par_construction}) used for the CAMELS-LH datasets to the TNG-SB28 dataset. In particular, we retain the same physically motivated dimensionless parameters, including the core structural parameter $S_{\rm core}$, the stellar mass fraction $f_\star$, and the stellar metallicity $Z_\star$. In addition, to account for the explicit variation of the background cosmology in TNG-SB28, we extend the analytic formulation to include a dependence on the Hubble parameter through the ratio $H_0 / H_{0,\rm fid}$, where 
\(
H_{0,\rm fid} = 0.067~\mathrm{km\,s^{-1}\,kpc^{-1}},
\)
denotes the fiducial value adopted in the CAMELS-LH simulations.

We recover an analytic form for $\Omega_m$ in the TNG-SB28 dataset that is closely related to our fiducial cosmology result in Eq.~\ref{z=0,eq} at redshift $z=0$, differing by small modifications in the fitted coefficients and functional normalization. The resulting expression is summarized in Table~\ref{tab:omega_m_equation_sb}.

\begin{table}[h]
\centering
\caption{Analytic $\Omega_m$ predictor for the TNG-SB28 simulation suite.}
\begin{tabular}{c}
\hline\hline

$\displaystyle
\Omega_m
= 0.47\Bigg[
\ln \sigma\!\left(
\frac{\Omega_b}{34.96}
\frac{\Omega_b\,V_{\rm eff}^{3}}{G\,H_0}
\frac{E(Z_\star,f_\star)}{M_\star}
\right)
+ 0.89
\Bigg]
+ 0.07\!\left(\frac{H_0}{H_{0,\mathrm{fid}}}-1\right)
$
\\[12pt]

$\displaystyle
E(Z_\star,f_\star)
=
\frac{1}{
1 + \dfrac{1}{Z_\star} + \dfrac{0.24}{f_\star}
}
$
\\

\hline
\end{tabular}
\label{tab:omega_m_equation_sb}
\end{table}

For clarity, we write the model in the following general form:
\begin{equation}
    \Omega_m
= B_0\Bigg[
\ln \sigma\!\left(x
\right)
+ c_0
\Bigg]
+ a_0\!\left(\frac{H_0}{H_{0,\mathrm{fid}}}-1\right),
\label{eq:sb28-eq}
\end{equation}
with latent variable
\[
x
= 
\left(\frac{\Omega_b}{k}\right)
\left(\frac{\Omega_b\,V_{\rm eff}^{3}}{G\,H_0}\right)
\left(\frac{E(Z_\star,f_\star)}{M_\star}\right),
\]
and fixed numerical values
\(
k = 34.96,\;
c_0 = 0.89,\;
a_0 = 0.07,\;
H_{0,\mathrm{fid}} = 0.067,\;
B_0 = 0.47,\;
\delta_0 = 0.24.
\)

\vspace{1mm}

At higher redshift ($z = 1.0$, $2.0$, $3.0$), the equation keeps the same general structure, adding a redshift term dependence of \(\alpha_{k}\) = 0.5 following: 
\begin{equation}
    \Omega_m(z)
= B_0\Bigg[
\ln \sigma\!\left(
\frac{x}{(1+z)^{\alpha_k}}
\right)
+ c_0
\Bigg]
+ a_0\!\left(\frac{H_0}{H_{0,\mathrm{fid}}}-1\right) .
\end{equation}

\begin{figure*}[h]
\centering
\setlength{\tabcolsep}{6pt}   % control horizontal spacing
\renewcommand{\arraystretch}{1.0}

{\setlength{\tabcolsep}{2pt}   % tighten horizontal spacing
\renewcommand{\arraystretch}{0.9} % tighten vertical spacing
\begin{tabular}{cc}
    \includegraphics[width=0.51\textwidth]{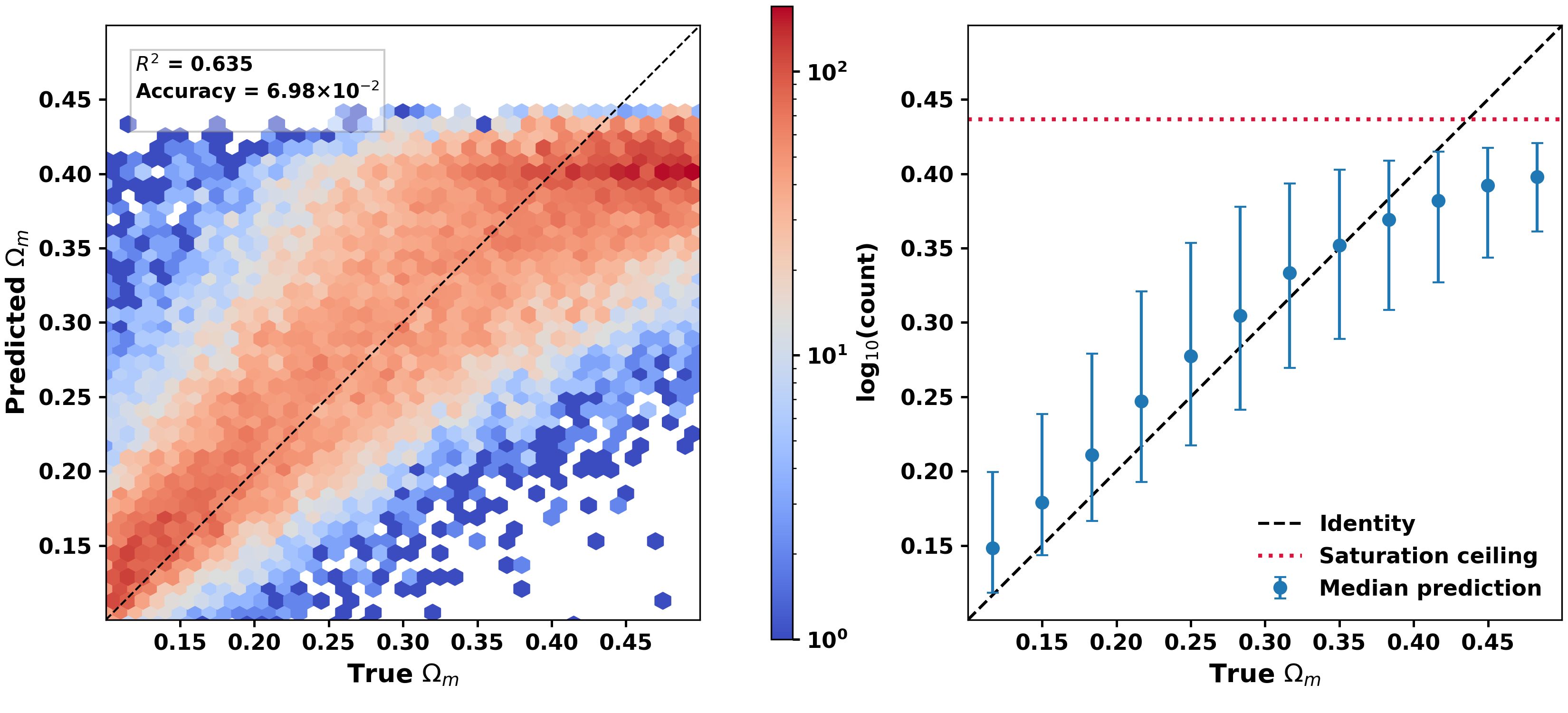} &
    \includegraphics[width=0.51\textwidth]{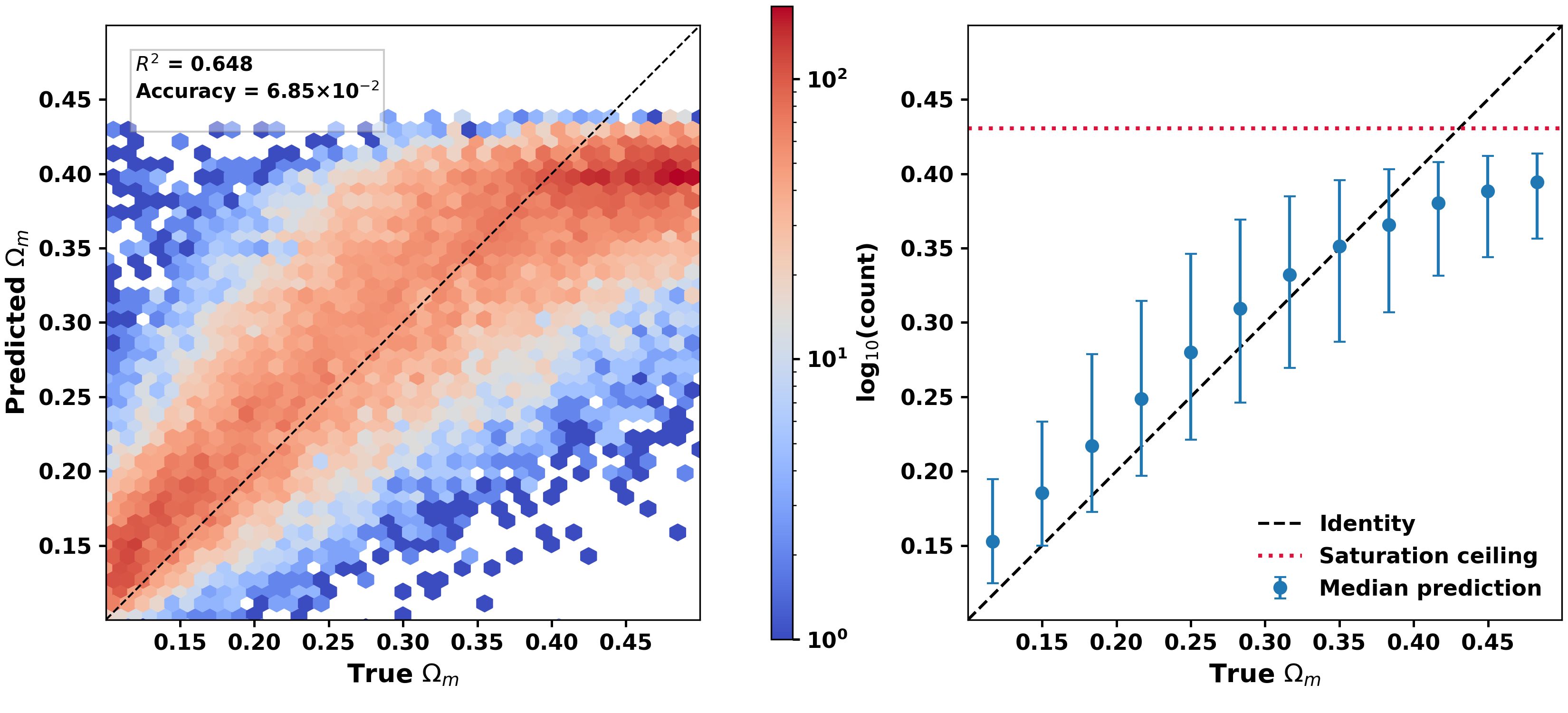} \\[-4pt]
    \small $z=0$ & \small $z=1$ \\[-2pt]
    \includegraphics[width=0.51\textwidth]{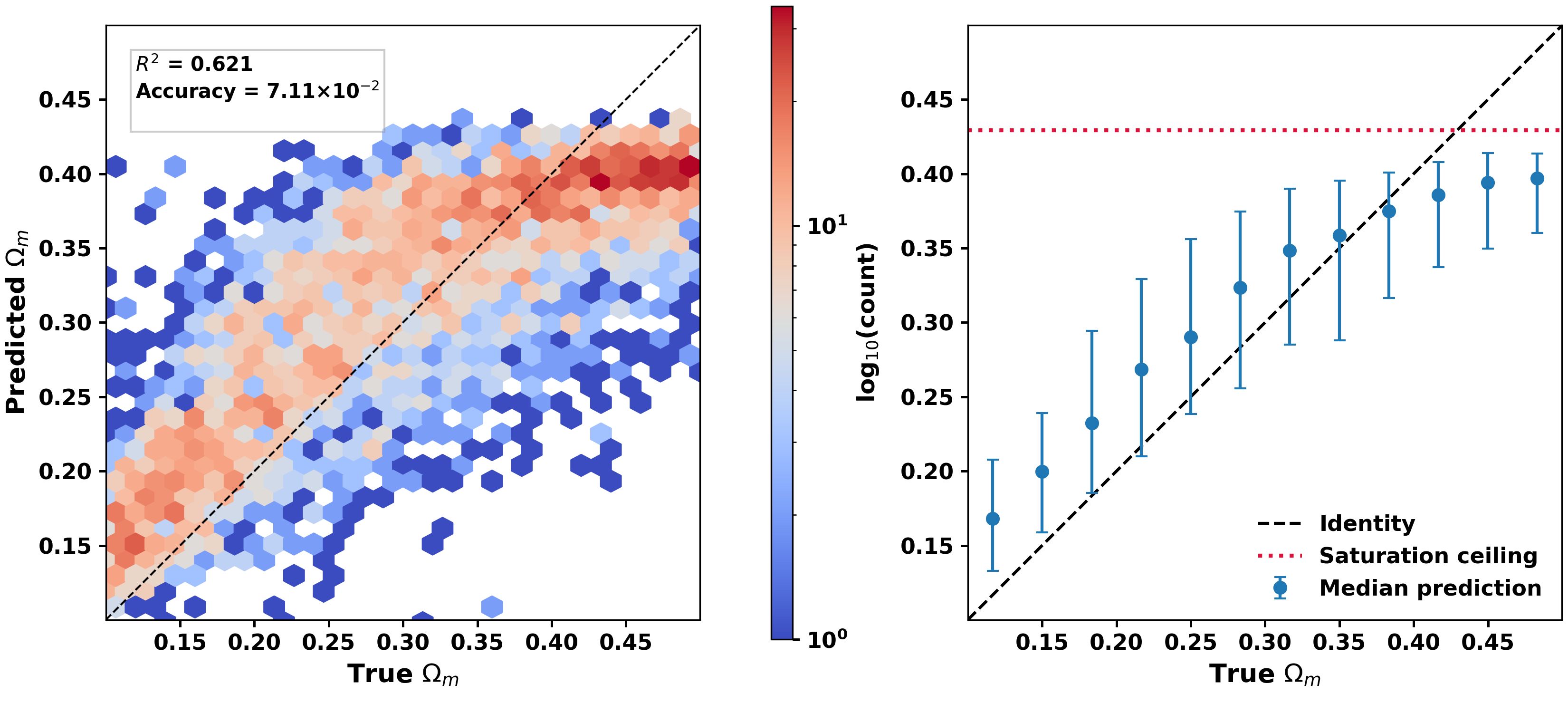} &
    \includegraphics[width=0.51\textwidth]{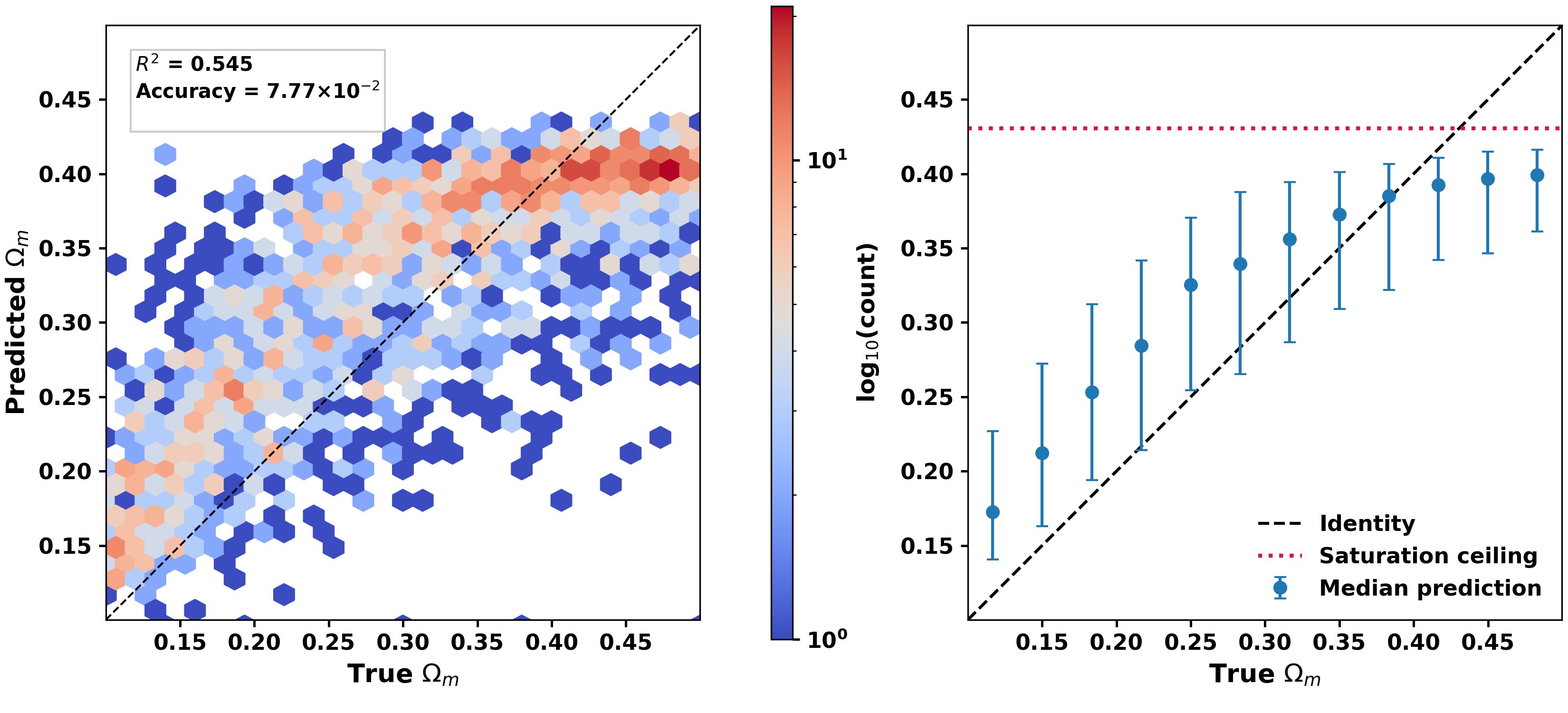} \\[-4pt]
    \small $z=2$ & \small $z=3$
\end{tabular}
}

\vspace{6pt}

\caption{ Performance of the analytic $\Omega_m$ predictor on the TNG-SB28 dataset for different redshifts.
Each panel contains a combined diagnostic: a hexbin density plot (left) comparing true versus predicted $\Omega_m$ and a calibration curve (right) showing the median prediction in bins of the true value with 16--84th percentile bands.
Results are shown for redshifts $z = 0.0, 1.0, 2.0, 3.0$ (top-left to bottom-right).
A saturation regime appears at high $\Omega_m$ (indicated by doted horizontal red lines, of $\Omega_m$ $\simeq$ 0.43) in all cases, indicating a limit beyond which galaxy properties no longer carry independent cosmological information under the SB28 parameter variations.
}
\label{fig:sb28_performance}
\end{figure*}

We evaluate the performance of the redshift-dependent model through one-to-one comparisons between predicted and true values of $\Omega_m$ at each redshift, as shown in Fig.~\ref{fig:sb28_performance}. At all redshifts, the hexbin density plots exhibit a strong monotonic correlation between predictions and ground truth, demonstrating that the latent structure successfully captures a systematic relationship between galaxy properties and cosmology, even under the extreme parameter variations of SB28. This confirms that the equation does not lose sensitivity to cosmology in a general sense.
However, the calibration curves reveal a physical limitation of the analytic form in the high-$\Omega_m$ regime: beyond a characteristic threshold $\Omega_m$ $\simeq$ 0.43, the median predicted value flattens and becomes insensitive to further increases in the matter density. 

This saturation does not necessarily imply a failure of the latent variable itself, but may instead reflect an intrinsic information ceiling in the TNG–SB28 dataset. Because SB28 explores a high-dimensional parameter space in which 28 astrophysical and cosmological parameters are varied simultaneously, it introduces strong degeneracies between feedback physics, enrichment history, and cosmology. In this regime, changes in cosmology may be increasingly offset by baryonic processes, producing convergence in galaxy-scale observables across distinct models. The observed saturation is therefore likely a manifestation of this extreme degeneracy, rather than a direct limitation of the analytic form alone.
Alternatively, it may reflect limitations of the symbolic-regression framework, including restricted functional flexibility or operator bias, which could prevent the analytic form from fully capturing residual cosmological dependence.
A direct way to test this possibility would be to apply more flexible numerical models, such as neural networks. Exploratory analyzes using more flexible machine-learning models on the dataset show qualitatively degradation across the $\Omega_m$ range, suggesting that this behavior is not obviously alleviated by increased model flexibility. While such approaches prioritize predictive accuracy, our objective here is to extract a compact, physically interpretable relation, and we do not pursue further optimization in that direction.

Overall, these results demonstrate that the latent variable structure robustly encodes the galaxy--cosmology connection across a wide range of cosmological backgrounds, while simultaneously suggesting a regime in which galaxy properties no longer provide unique cosmological information. In the following subsections, we turn to a physical interpretation of the analytic form itself, examining the origin of its functional structure, including the emergence of the characteristic knee behavior and the role of each component of the latent variable.

\subsection{Physical Interpretation via Gas Regulation}
\label{sec:gas regulae}

Although the overall structure of the latent variable remains closely aligned with the form identified in the CAMELS-LH datasets (see Equations \ref{z=0,eq} and \ref{eq:sb28-eq}), the internal organization of the equation changes in an important way. 
In our previous result, Eq. \ref{z=0,eq}, the dependence inside the sigmoid is largely captured by a clean mass–metallicity relation (MZR). In contrast, within the SB28 regime,
as the parameter space expands to include simultaneous variations in cosmology and a far broader set of feedback parameters, the equation reorganizes itself such that the dominant contribution is an efficiency-like factor that depends jointly on stellar metallicity and stellar mass fraction.

A central feature that emerges from the SB28 analysis is the appearance of an explicit retention-like term in the latent variable, represented by the function $E(Z_\star, f_\star)$.
This shift preserves the functional architecture of the latent variable while refining its physical content: rather than expressing baryonic regulation only through global empirical correlations, the model introduces an explicit term that modulates how efficiently baryons are retained and converted into stars.

\subsubsection{Gas–Regulator Framework}

Motivated by this connection, we now reinterpret the latent structure in the context of the gas--regulator framework ~\citep{2013ApJ...772..119L,2012MNRAS.421...98D} and show that the functional form of $E(Z_\star,f_\star)$ can be understood as an effective retention efficiency emerging from the steady-state behavior of the baryonic cycle.

Let $M_g$ be the gas mass, ${\rm SFR}$ the star-formation rate, $\dot M_{\rm in}$ the inflow rate,
$\eta$ the mass-loading factor (outflow rate $=\eta\,{\rm SFR}$), $R$ the instantaneous recycling fraction,
$y$ the net metal yield, and $Z$ the gas-phase metallicity (ISM). The gas-regulator model is then defined by the following continuity equations:
\begin{align}
\frac{dM_g}{dt} &= \dot M_{\rm in} - (1-R+\eta)\,{\rm SFR}, \label{eq:mg}\\
\frac{d(ZM_g)}{dt} &= Z_{\rm in}\,\dot M_{\rm in} + y(1-R)\,{\rm SFR}
- Z(1-R+\eta)\,{\rm SFR}. \label{eq:metal}
\end{align}
Eq.~\ref{eq:metal} describes the metal–mass continuity equation for the interstellar medium (ISM), written as the time derivative of the gas–phase metal mass. Each term admits a direct physical interpretation, in the way that defining
\(M_{Z,{\rm gas}}\equiv Z\,M_g\), we have
\begin{itemize}
  \item \textbf{Inflow metals:} \(+\;Z_{\rm in}\,\dot M_{\rm in}\)
  \item \textbf{Fresh production:} \(+\;y(1-R)\,{\rm SFR}\)
  \item \textbf{Lock-up in stars:} \(-\;Z(1-R)\,{\rm SFR}\)
  \item \textbf{Outflows (winds):} \(-\;Z\,\eta\,{\rm SFR}\) .
\end{itemize}

In the self-regulated regime, $dM_g/dt \approx 0$ and $d(ZM_g)/dt \approx 0$.
Using Eq.~\eqref{eq:mg} gives $\dot M_{\rm in} = (1-R+\eta)\,{\rm SFR}$, then
Eq.~\eqref{eq:metal} reduces to
\[
0 = Z_{\rm in}(1-R+\eta)\,{\rm SFR} + y(1-R)\,{\rm SFR} - Z(1-R+\eta)\,{\rm SFR}.
\]
Dividing by ${\rm SFR}$ and solving for $Z$ yields the standard equilibrium metallicity:
\begin{equation}
Z_{\rm eq} = Z_{\rm in} + \frac{y(1-R)}{(1-R)+\eta}.
\label{eq:Zeq_general}
\end{equation}

\vspace{2mm}

Neglecting the additive baseline from inflow metallicity ($Z_{\rm in}\!\approx\!0$),
so that the mass- and feedback-dependent term dominates, the equilibrium metallicity
can be written in terms of the \emph{retention fraction}:
\begin{equation}
\mathcal{R} \;\equiv\; \frac{1-R}{(1-R)+\eta},
\qquad
Z_{\rm eq} = y\,\mathcal{R}.
\label{eq:retention_general}
\end{equation}

Typically $R \sim 0.3$–$0.5$ for standard IMFs \citep{AudouzeTinsley1976}, and the combination
$y(1-R)$ is often absorbed into an effective yield $y_{\rm eff}$.
For the common analytic simplification, one may  set $R=0$ explicitly or, equivalently, absorb $(1-R)$ into the net metal yield $y$ and work with $y_{\rm eff}$:
\begin{equation}
Z_{\rm eq} = \frac{y}{1+\eta},
\qquad
\mathcal{R} = \frac{1}{1+\eta}.
\label{eq:R0_retention}
\end{equation}
Setting $R=0$ in the denominator of 
Eq. \eqref{eq:R0_retention} does not materially alter the gas–cycling budget.
In practice, $R$ contributes only a small and nearly constant term compared to 
$\eta$, which vary by orders of magnitude across the galaxy population.

\vspace{2mm}

\paragraph{Connection to the reduced cosmology–and–feedback variation regime case.}
Mapping the retention relation to our previous defined 
\[
E(Z_\star,f_\star)
=
\frac{1}{
\displaystyle
1 \;+\; \frac{1}{Z_\star}+\;\frac{\delta_0}{f_\star} \;
}.
\] 
Where the unity term in the denominator represents the regulator floor.
The $1/Z_\star$ term acts as a monotonic surrogate for chemical retention history, tracing the cumulative impact of metal loss and feedback-driven outflows in the system.
$\delta_0/{f_\star}$ term captures an additional structural channel associated with baryon conversion efficiency, which in simulations correlates with variations in halo growth and feedback regulation. We therefore interpret this term as an effective structural channel that modifies the mass loading factor from an idealized single-parameter regulator case.
When cosmology is fixed, baryon conversion efficiency and chemical enrichment correlate strongly with galaxy mass/structure properties and $E\!\propto\!Z_\star$
in the metal-poor regime; the latent variable reduces to the familiar
i.e.\ MZR-like proxy. When $\Omega_b$ or $H_0$ vary, $E(Z_\star,f_\star)$ remains robust since it
explicitly encodes structural baryon-processing effects ($\propto \delta_0/f_\star$) and chemical retention ($\propto 1/Z_\star$). 

\vspace{2mm}

\begin{figure*}[h]
    \centering
    \begin{minipage}{0.48\textwidth}
        \centering
        \includegraphics[width=\linewidth]{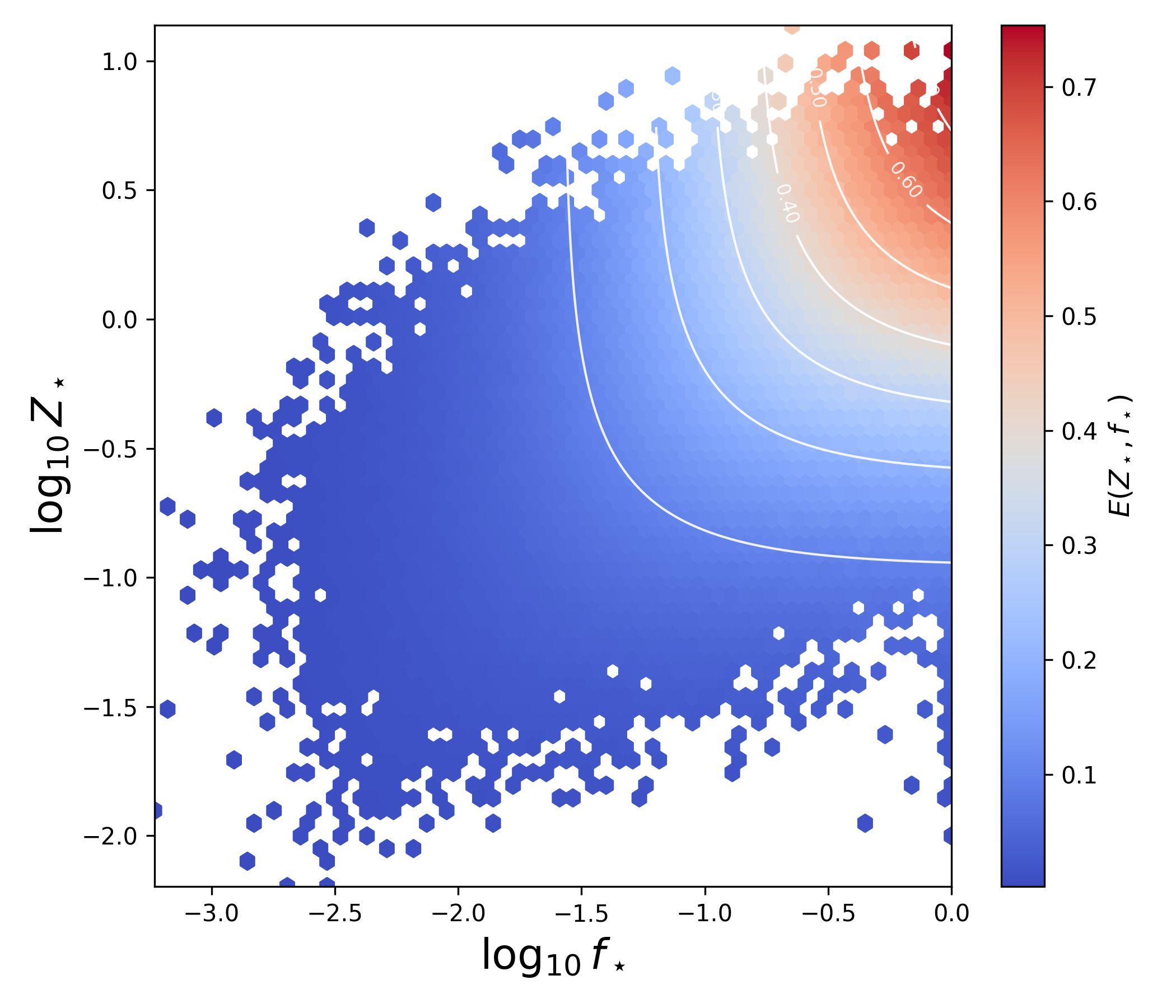}
        {\small (a)}
    \end{minipage}
    \hfill
    \begin{minipage}{0.48\textwidth}
        \centering
        \includegraphics[width=\linewidth]{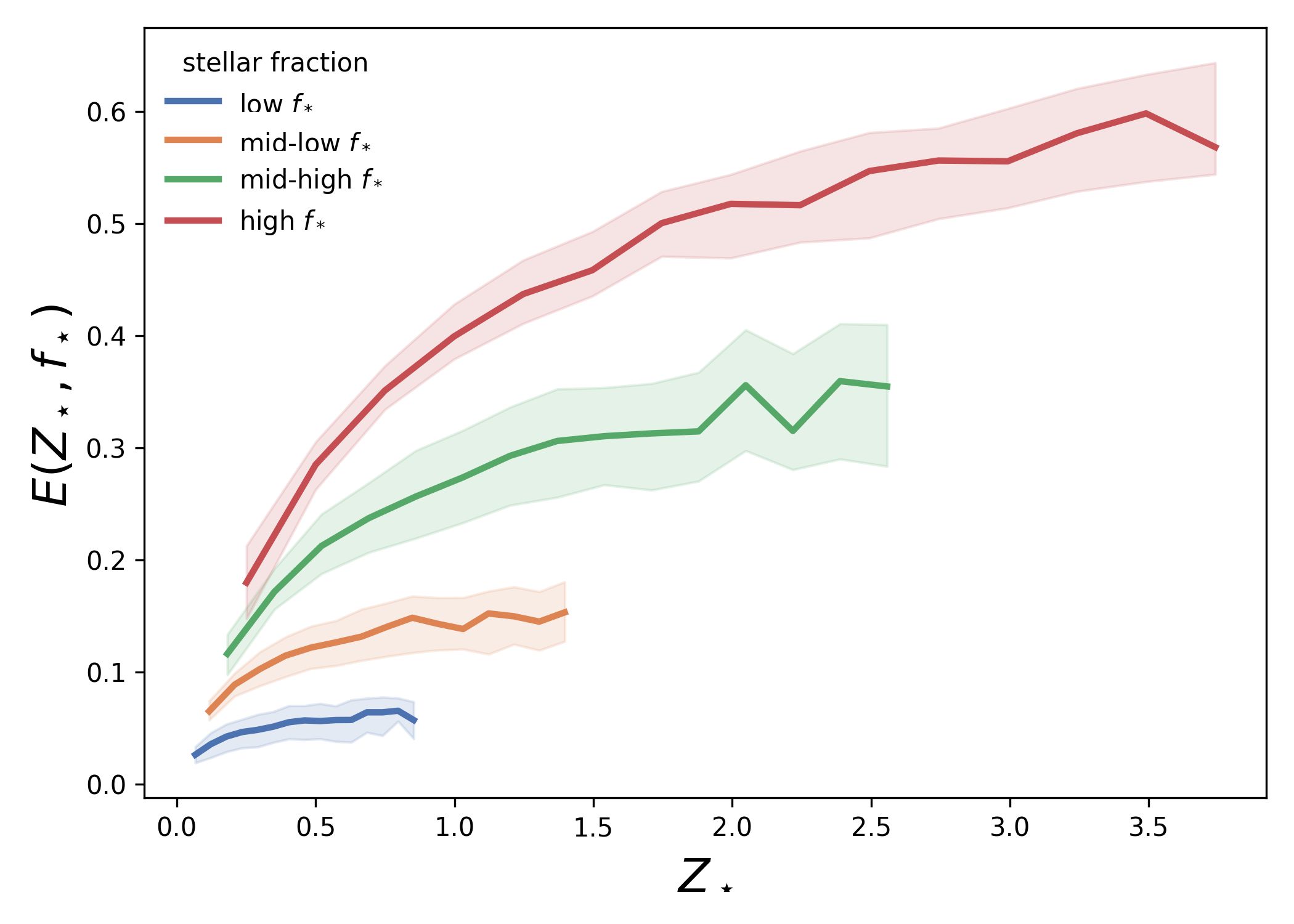}
        {\small (b)}
    \end{minipage}

    \caption{
   Baryonic retention–enrichment efficiency at redshift \(z=0\) across the stellar fraction–metallicity plane.
    \emph{(a)} The efficiency field 
    \(
    E(Z_\star,f_\star) = \bigl(1 + Z_\star^{-1} + \delta_0/f_\star \bigr)^{-1}
    \)
    evaluated over the galaxy population and shown in the $(\log f_\star, \log Z_\star)$ plane.
    White contours mark constant efficiency.
    \emph{(b)} Median efficiency as a function of stellar metallicity for quartiles in stellar fraction.
    At fixed $Z_\star$, galaxies with higher $f_\star$ are systematically more efficient, while at fixed $f_\star$ the efficiency increases with metallicity and saturates at high $Z_\star$.
    }
    \label{fig:efficiency_mzr}
\end{figure*}

In Figure~\ref{fig:efficiency_mzr}, we illustrate how the function $E(Z_\star,f_\star)$ maps onto the joint stellar metallicity–stellar fraction space for TNG-SB28 dataset. The resulting structure exhibits behavior consistent with baryonic retention and enrichment efficiency, with smoothly varying trends across both dimensions. At redshift \(z=0\), panel (a) shows the efficiency field evaluated across the $(f_\star, Z_\star)$ plane, revealing a smooth, monotonic structure in which efficiency increases toward higher stellar fraction and higher metallicity. The contours of constant $E$ are consistent with trends expected from gas-regulated systems, indicating that galaxies with both efficient star formation and enriched stellar populations retain baryons more effectively.
Panel~(b) further clarifies this behavior by showing the median efficiency as a function of stellar metallicity in four bins of stellar mass fraction, corresponding to the 0--25\%, 25--50\%, 50--75\%, and 75--100\% percentiles.
At fixed metallicity, galaxies with larger $f_\star$ consistently exhibit higher efficiency, while at fixed $f_\star$ the efficiency increases with $Z_\star$ and approaches saturation at high metallicity. This pattern is consistent with expectations from gas-regulator models, in which enrichment reflects cumulative star-formation processing while the stellar fraction traces the integrated conversion of gas into stars.

Importantly, the appearance of the retention term $E(Z_\star,f_\star)$ in the SB28 expression (see Table \ref{tab:omega_m_equation_sb}) does not represent a departure from the latent structure (Eq.~\ref{z=0,eq}) identified in the fiducial CAMELS-LH datasets. Rather, it should be viewed as a refinement of the same underlying relationship, in which the mass–metallicity relation is no longer treated as a single empirical scaling but is instead resolved into a more explicit, non-linear representation of baryonic processing. In this sense, the SB28 formulation preserves the core architecture of the model while extending it into a higher-dimensional parameter space, allowing the latent variable to respond more flexibly to variations in enrichment history and star-formation efficiency.

Overall, the analytic expression introduced here retains the same overall architecture as the 
fiducial cosmology predictor in Eq.~\ref{z=0,eq}: an effective enrichment term, an outer 
logarithmic--sigmoid operator, and a secondary correction term.
Two controlled modifications distinguish the new formulation. 
First, the normalization constant outside the logarithmic--sigmoid operator 
sets a uniform enrichment baseline and ensures that the latent variable maintains 
a consistent monotonic scaling across the full dynamic range.
Second, the structural correction used in the fiducial cosmology model, previously expressed through 
$R_{\rm compact}^{-1}$, is replaced by a direct dependence on the Hubble parameter. 
Once cosmology is allowed to vary, differences in expansion rate become a more relevant 
source of secondary variation than galaxy compactness, and the predictor reorganizes 
accordingly to incorporate this dependence.

\section{Summary}\label{sec:summary}

Modern cosmology has achieved remarkable precision through probes such as the cosmic microwave background~\citep{1965ApJ...142..419P}, baryon acoustic oscillations~\citep{1970ApJ...162..815P,Sunyaev:1970eu,2017MNRAS.470.2617A}, and large-scale structure statistics~\citep{1980lssu.book.....P,SDSS:2003eyi}. In parallel, galaxy formation studies have focused on understanding how nonlinear baryonic processes—gas accretion, star formation, feedback, and chemical enrichment—shape individual galaxies and their observable properties~\citep{1978MNRAS.183..341W,1991ApJ...379...52W}.
Motivated by these developments, large suites of cosmological hydrodynamic simulations have been developed to model galaxy formation within a controlled cosmological framework, enabling cosmological parameters and baryonic feedback processes to be varied systematically.
More recently, machine learning analyzes of these simulations have shown that cosmological parameters can be inferred directly from individual galaxy properties \citep{Villaescusa-Navarro:2022twv, Lue:2025zqk}.

In this paper we have demonstrated a unified picture in which the baryonic evolution of 
individual galaxies encodes information about the matter density parameter $\Omega_m$ 
through a physically grounded analytic framework that serves as a 
single-galaxy cosmological tracer.
Focusing first on the reduced cosmology–and–feedback variation regime of the CAMELS project (the CAMELS–LH datasets), which span multiple independent simulation suites—\textsc{IllustrisTNG}, \textsc{SIMBA}, \textsc{ASTRID}, and \textsc{SWIFT–EAGLE}—we find that this analytic structure (see Eq.~\ref{z=0,eq}) is remarkably stable despite substantial differences in numerical implementation. Moreover, this structure (see Eq.~\ref{eq:Omega_m_sim_z}) remains robust across redshift when augmented by a simple redshift-dependent rescaling, yielding a unified form applicable from low to intermediate redshifts.
In the extended cosmology–and–feedback variation suite (the CAMELS–SB28 datasets), a closely related, physically motivated analytic form (see Eq.~\ref{eq:sb28-eq}) remains robust across redshift, with the latent variable reorganizing while preserving the same underlying physical structure.

In the reduced cosmology-and-feedback variation regime, the latent variable reduces to a 
clean mass–metallicity-like dependence, reflecting the regularity of the baryon cycle when feedback variations are modest. In the extended cosmology–and–feedback variation suite, the latent variable reorganizes into the explicit 
retention function $E(Z_\star, f_\star)$, which captures the combined effects of chemical retention history and structural baryon-processing efficiency associated with stellar mass buildup and feedback regulation. Despite these differences in functional form, both expressions share 
the same underlying architecture and rely on the same physical ingredients, demonstrating 
that the imprint of $\Omega_m$ is encoded through a stable and interpretable combination 
of enrichment, baryon retention, and gravitational depth.

Our theoretical premise is that the background cosmology is approximately imprinted in 
each galaxy through the way baryons are accumulated, processed, and enriched across 
cosmic time. The formation history, metal production, and retention efficiency of a galaxy 
are all shaped by the depth of its potential well and the cosmological supply of baryons, 
linking local baryonic evolution to global cosmological parameters. By expressing these 
processes through an analytically tractable latent variable, and by demonstrating its 
validity across both fixed-cosmology and varying-cosmology regimes. We show that, in addition to the well-established constraints from large-scale 
clustering, the matter density parameter $\Omega_m$ leaves a measurable imprint on the 
internal baryon cycle of individual galaxies. This provides a surprising and 
complementary small-scale avenue for cosmological inference.

Our analysis is performed within a cosmological simulation framework, but the physical mechanisms driving the tracer are generic and suggest the possibility of observational validation. The existence of a single-galaxy cosmological tracer offers a complementary avenue for probing $\Omega_m$, extending cosmological inference beyond traditional large-scale structure measurements. Together, these results highlight a physically 
grounded connection between galaxy-scale baryon cycling and the matter density of the Universe, opening a new window on the interplay between galaxy formation and cosmology and suggesting a promising direction for future theoretical and observational exploration.

Looking ahead, new generations of cosmological simulations with varied subgrid physics, 
alternative feedback prescriptions, and tighter physical constraints will offer valuable 
opportunities to test and refine the functional form of this tracer. Evaluating its 
robustness across different numerical approaches and galaxy formation models will be 
crucial for assessing its universality and for clarifying the physical pathways through 
which cosmology imprints itself on galaxy-scale baryon cycling. Such efforts will help 
bridge the gap between simulated and observed galaxies, opening the way toward a broader 
and more predictive framework at the intersection of galaxy formation and cosmology.

\begin{acknowledgments}
We thank Adrian Bayer, Christopher Lovell, Daisuke Nagai, Oliver Philcox, Yacine Ali-Haimoud, and the whole CAMELS collaboration for insightful discussions and helpful comments that contributed to this work.
NSMS acknowledges partial support from the NSF CDSE grant AST-2408026 and the NASA TCAN grant 80NSSC24K0101.
\end{acknowledgments}

\appendix

\section{Feedback on MZR at high redshift} 
\label{sec:FeedbackonMZRathighredshift}

In this appendix, we present the feedback regulation of the Mass–Metallicity Relation at both simulation suites (\textsc{IllustrisTNG} and \textsc{Astrid}), similarly to what has been presented in Section \ref{sec:feedbackregulation}, but at higher redshifts ($z = 1.0$, $2.0$, $3.0$).

\begin{figure*}[h]
\centering

% ----------- z = 1 -----------
\noindent\makebox[3em][r]{\textbf{$z=1$}}\hspace{0.8em}%
\begin{minipage}{0.43\textwidth}
    \centering
    \includegraphics[width=\linewidth]{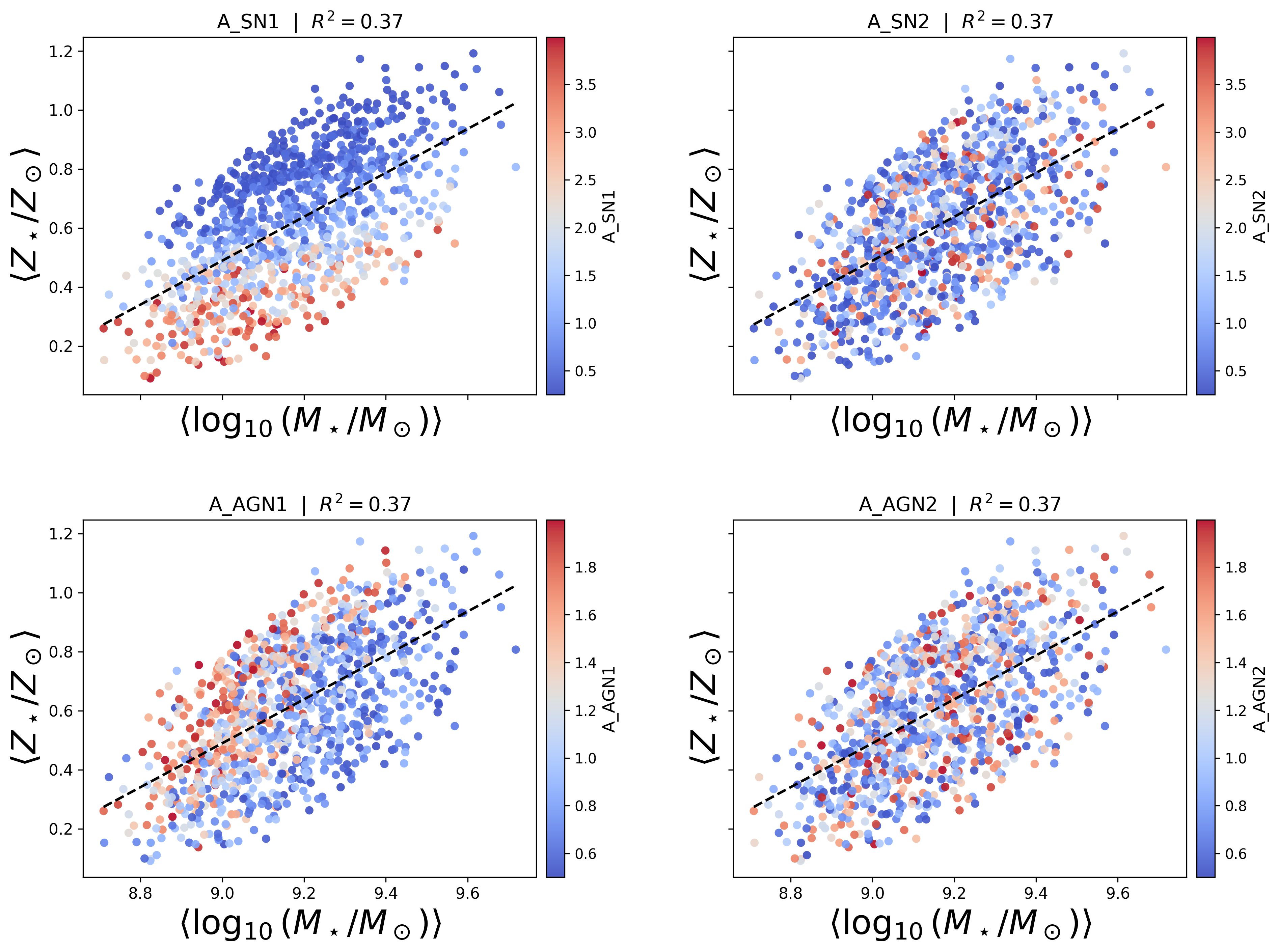}
\end{minipage}
\hfill
\begin{minipage}{0.43\textwidth}
    \centering
    \includegraphics[width=\linewidth]{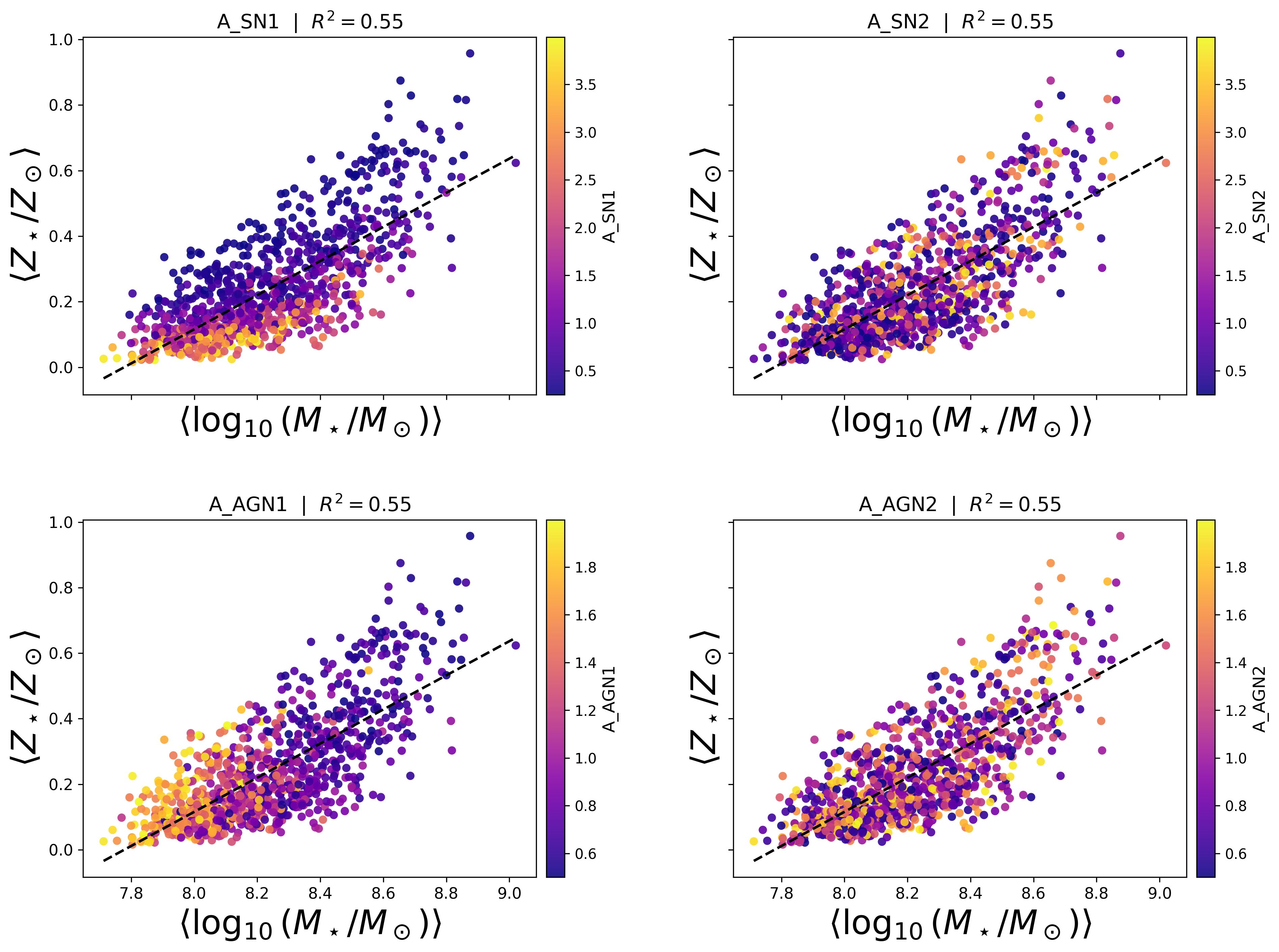}
\end{minipage}

\vspace{0.5em}

% ----------- z = 2 -----------
\noindent\makebox[3em][r]{\textbf{$z=2$}}\hspace{0.8em}%
\begin{minipage}{0.43\textwidth}
    \centering
    \includegraphics[width=\linewidth]{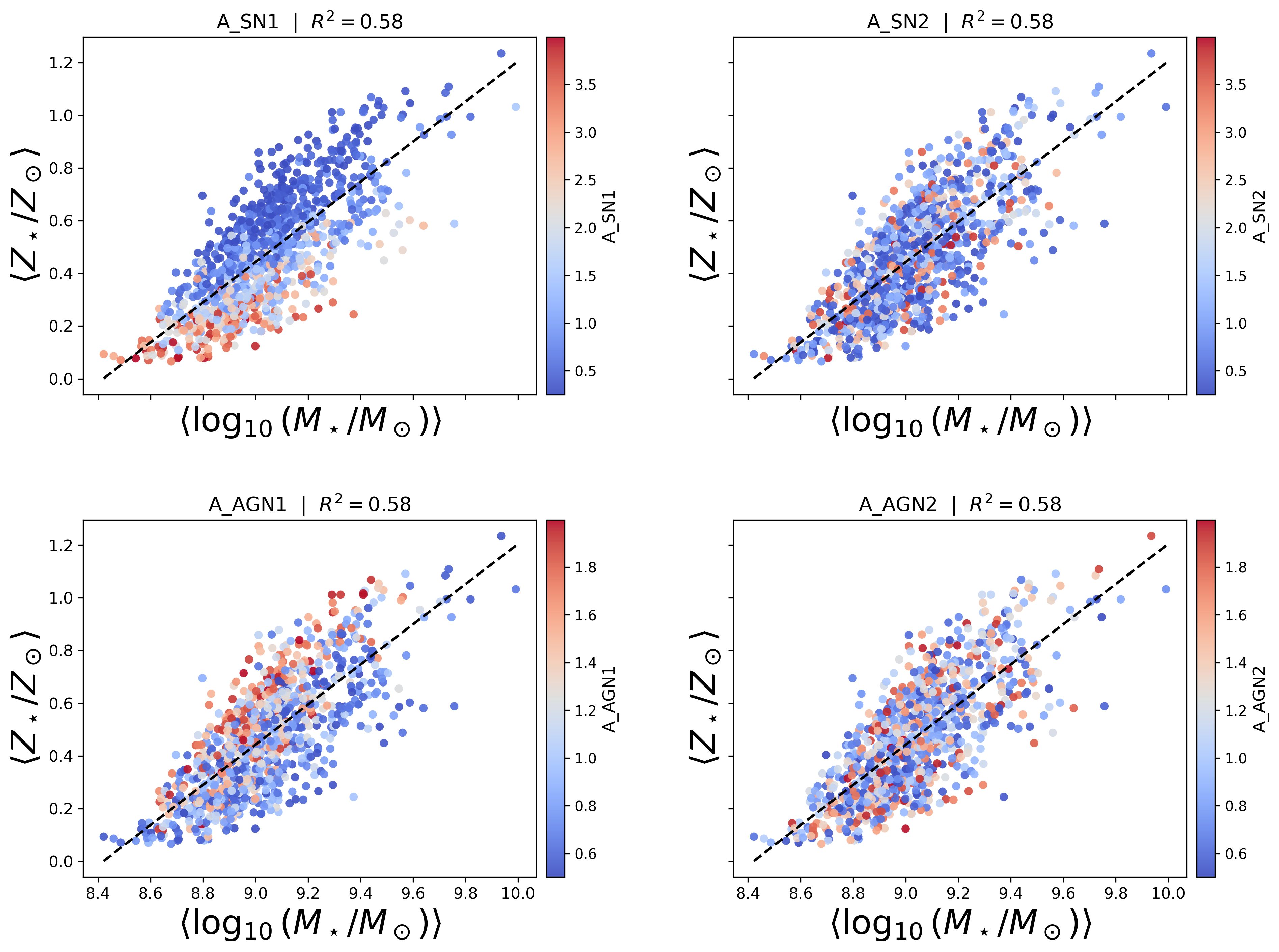}
\end{minipage}
\hfill
\begin{minipage}{0.43\textwidth}
    \centering
    \includegraphics[width=\linewidth]{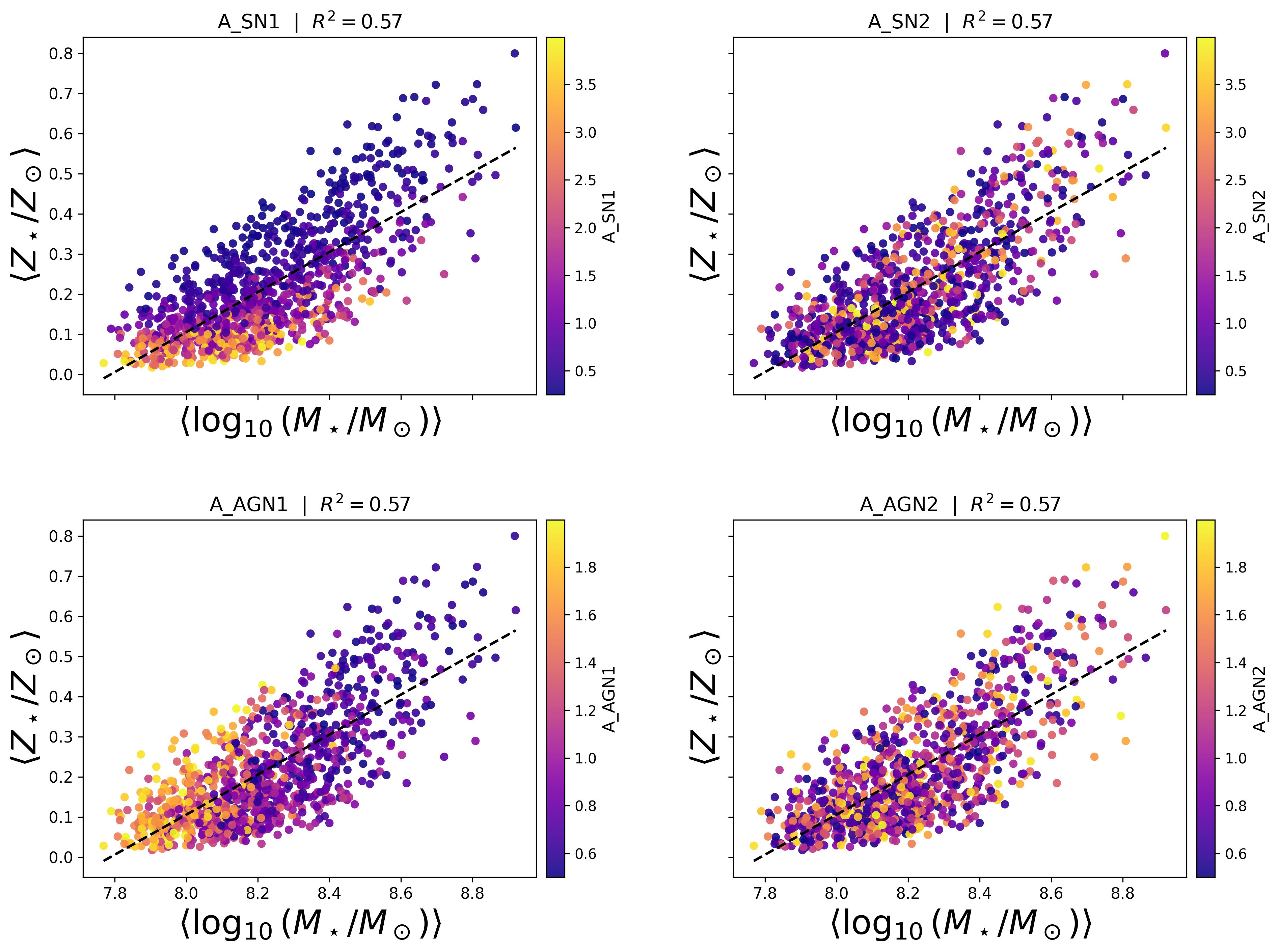}
\end{minipage}

\vspace{0.5em}

% ----------- z = 3 -----------
\noindent\makebox[3em][r]{\textbf{$z=3$}}\hspace{0.8em}%
\begin{minipage}{0.43\textwidth}
    \centering
    \includegraphics[width=\linewidth]{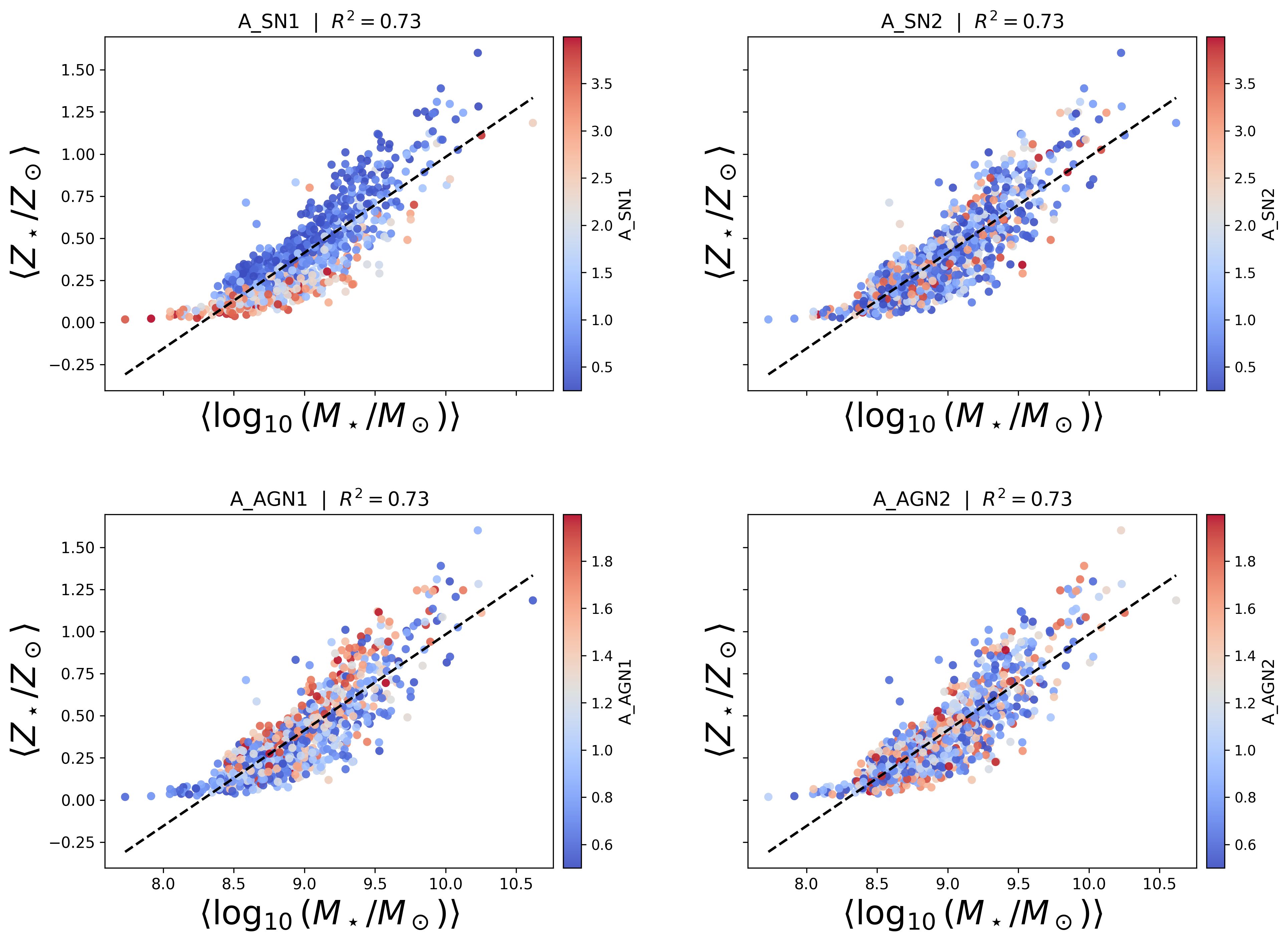}
\end{minipage}
\hfill
\begin{minipage}{0.43\textwidth}
    \centering
    \includegraphics[width=\linewidth]{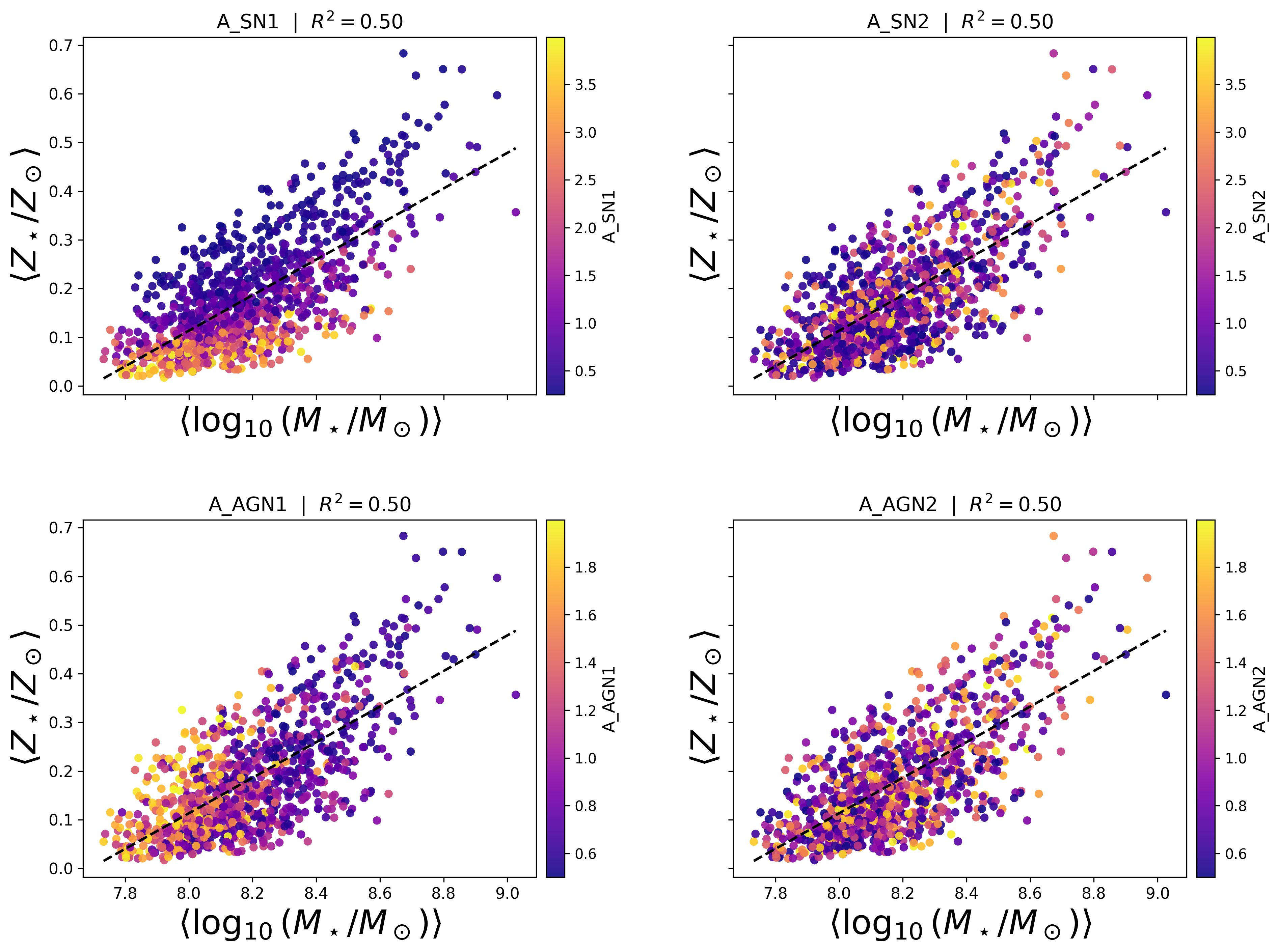}
\end{minipage}

\vspace{0.8em}

\caption{
Evolution of the stellar mass--metallicity relation (\textit{MZR}) across 
redshifts $z = 1.0,\ 2.0,\ 3.0$ for the 
\textsc{IllustrisTNG} (left column) and \textsc{Astrid} (right column) simulation suites. 
Each point represents a galaxy color--coded by its feedback parameters 
$A_{\mathrm{SN1}}, A_{\mathrm{SN2}}, A_{\mathrm{AGN1}},$ and $A_{\mathrm{AGN2}}$. 
Dashed lines indicate best--fit power--law trends in 
$\log M_\star$--$Z_\star$ space.
}
\label{fig:MZ_feedback_evolution}
\end{figure*}

\section{Power-Law Transformation}
\label{sec:Appendix_power-law}
The ln--sigmoid form from Equation~\ref{z=0,eq} captures the saturating behavior between metallicity enrichment 
and cosmological density, but it obscures how the slope of this relation evolves across redshift 
and feedback regimes. 
To better interpret the nonlinear structure of the original equation, 
we offer a format of power--law transformation to the analytic predictor.

We simplify the latent variable for mathematical analysis as:
\begin{equation}
x =
\left(
\frac{Z_\star}{M_\star}
\frac{\Omega_b V_{\mathrm{eff}}^{3}}{G H_0}
\frac{\Omega_b}{k_{\mathrm{sim}}}
\right)
= 
\frac{Z_\star}{k_{\mathrm{sim}}\,\varepsilon_\star},
\end{equation}
where $\varepsilon_\star$ represents the grouped parameters from the original relation for simplicity.  
To identify the \textit{knee point}---the location where $\Omega_{m}$ is most sensitive to metallicity---we impose the condition on Equation~\ref{eq:Omega_m_sim_z}:
\begin{equation}
f(x_k; c_0^{\mathrm{sim}})=\frac{d^{2}\Omega_{m}}{d(\ln Z_\star)^{2}} = 0.
\end{equation}
This yields the nonlinear equation,
\begin{equation}
f(x_k; c_0^{\mathrm{sim}})
= 
\frac{1}{s_k + c_0^{\mathrm{sim}}}
+ x_k
\frac{c_0^{\mathrm{sim}}(1 - 2s_k) - s_k^{2}}{(s_k + c_0^{\mathrm{sim}})^{2}}
= 0,
\quad
s_k = \sigma(x_k),
\end{equation}
where the subscript ``$k$'' denotes quantities evaluated at the \textit{knee point} of the relation, 
and $x_k$ is the positive root of $f(x_k; c_0^{\mathrm{sim}}) = 0$. 
The corresponding stellar metallicity at the knee is then given by
\begin{equation}
Z_{\star,k} = x_k\,k_{\mathrm{sim}}\,\varepsilon_\star,
\end{equation}
which defines the metallicity scale where the slope of the underlying ln–sigmoid relation 
reaches its maximum slope.

Evaluating the predictor at this point, we define $\mathbf{S}_k = \sigma(x_k)$ and obtain the slope and normalization at the knee:
\begin{align}
\mathbf{S}_k &=
\left.\frac{d\Omega_m}{d\ln Z_{\star}}\right|_{Z_{\star,k}}
= \frac{x_k\,s_k(1 - s_k)}{s_k + c_0^{\mathrm{sim}}}, 
\label{eq:Sk_def} \\[4pt]
\Omega_m^{k} &=
\Omega_m(Z_{\star,k})
= \ln\!\left(s_k + c_0^{\mathrm{sim}}\right)
- \frac{a_0^{\mathrm{sim}}}{R_{\mathrm{compact}}}.
\label{eq:Omk_def}
\end{align}

Around the knee, its local slope behaves as a rescaled version of itself with
respect to 
\vspace{-3pt} $Z_{\star}/Z_{\star,k}$,
\[
\frac{d\Omega_m}{d\ln Z_{\star}} \propto 
\left(\frac{Z_{\star}}{Z_{\star,k}}\right)^{\nu},
\]
\vspace{-5pt}
which represents a general power-law form.
To preserve the correct normalization at the knee, where
$\left.\tfrac{d\Omega_m}{d\ln Z_{\star}}\right|_{Z_{\star,k}} = \mathbf{S}_k$,
the proportionality can be written explicitly as
\begin{equation}
\frac{d\Omega_m}{d\ln Z_{\star}}
= \mathbf{S}_k
\left(\frac{Z_{\star}}{Z_{\star,k}}\right)^{\nu}.
\end{equation}
the value of $\Omega_m(Z_{\star})$ relative to the knee is obtained by integrating
with respect to $\ln Z_{\star}$:
\[
\Omega_m(Z_{\star}) - \Omega_m^k
= 
\int_{\ln Z_{\star,k}}^{\ln Z_{\star}}
\frac{d\Omega_m}{d\ln Z_{\star}'}\,d(\ln Z_{\star}')
= 
\mathbf{S}_k
\int_{\ln Z_{\star,k}}^{\ln Z_{\star}}
\left(\frac{Z_{\star}'}{Z_{\star,k}}\right)^{\nu} d(\ln Z_{\star}').
\]
Since $d(\ln Z_{\star}') = dZ_{\star}'/Z_{\star}'$, the integral evaluates to a simplified power-law expansion:
\begin{equation}
\Omega_m(Z_{\star})
\simeq
\Omega^k_m
+ \frac{\mathbf{S}_k}{\nu}
\left[
\left(\frac{Z_{\star}}{Z_{\star,k}}\right)^{\nu}
- 1
\right],
\label{power-law-expansion}
\end{equation}
The $1/\nu$ factor appears as a direct consequence of integration,
ensuring that the normalization at $Z_{\star} = Z_{\star,k}$ is preserved. 
The simulation specified formats of this power law transformation (see Eq. \ref{power-law-expansion}) are presented in Table ~\ref{tab:omega_m_powerlaw}.

\begin{table}[h!]
\centering
\begin{tabular}{l c}
\hline\hline
\textbf{Simulation} & \textbf{Power-Law Transformation} \\[3pt]
\hline
\\[2pt]
\textsc{IllustrisTNG} &
$\displaystyle
\Omega_{m}^{\mathrm{TNG}}(Z_{\star})
\simeq
\Omega_m^{k}
+ \frac{\mathbf{S}_k}{0.3}
\!\left[
\left( \frac{Z_{\star}}{Z_{\star,k}} \right)^{0.3}
- 1
\right]
$ \\[10pt]

\textsc{Astrid} &
$\displaystyle
\Omega_{m}^{\mathrm{ASTRID}}(Z_{\star})
\simeq
\Omega_m^{k}
+ \frac{\mathbf{S}_k}{0.02}
\!\left[
\left( \frac{Z_{\star}}{Z_{\star,k}} \right)^{0.02}
- 1
\right]
$ \\[9pt]
\hline
\end{tabular}
\caption{Power--law expansion of $\Omega_m(Z_{\star})$ around the metallicity knee for the IllustrisTNG and Astrid simulations.}
\label{tab:omega_m_powerlaw}
\end{table}

The knee metallicity $Z_{\star,k}$ marks the inflection point of the $\Omega_m$--$Z_{\star}$ relation, 
where the slope $\tfrac{d\Omega_m}{d\ln Z_{\star}}$ reaches its maximum and the sensitivity of $\Omega_m$ to metallicity is strongest. 
Physically, this point corresponds to the transition between cosmology--dominated and feedback--regulated regimes: 
at low metallicities ($Z_{\star}<Z_{\star,k}$), enrichment is limited by the global matter density, 
while at high metallicities ($Z_{\star}>Z_{\star,k}$), stellar and AGN feedback saturate the enrichment efficiency, 
weakening the dependence on $\Omega_m$.
The slope $\mathbf{S}_k$ quantifies the strength of the baryon--cosmology coupling at the transition point, 
representing how efficiently metallicity traces the cosmic matter field. 
The exponent $\nu$ governs how rapidly this coupling decays away from the knee and thus measures the 
elasticity of the $\Omega_m$--$Z_{\star}$ relation. 
A larger $\nu$ indicates that the cosmological imprint persists over a broader metallicity range, 
while a smaller $\nu$ implies rapid self--regulation by feedback and diminished cosmological sensitivity.

The contrasting values of $\nu$ can be directly traced to the redshift evolution of the mass--metallicity relation (MZR) in each simulation suite. 
In \textsc{IllustrisTNG}, SN and AGN feedback drive a progressively steeper MZR at higher redshift, 
indicating that the efficiency of metal enrichment remains strongly coupled to the evolving baryonic environment 
and the underlying cosmological matter density. 
This evolving enrichment pattern preserves a measurable cosmological imprint, 
reflected in the relatively large elasticity exponent ($\nu\simeq0.3$).  
In contrast, the \textsc{Astrid} simulations exhibit a remarkably stable MZR across cosmic time, 
suggesting that feedback processes regulate metallicity growth in a way that quickly reaches equilibrium 
and becomes largely insensitive to cosmological variations. 
As a result, the inferred $\Omega_m$--$Z_{\star}$ coupling remains weak and nearly constant with redshift, 
leading to the much smaller exponent ($\nu\simeq0.02$).  
This comparison highlights that $\nu$ effectively encodes how feedback prescriptions control 
the redshift evolution of the MZR, and thus how strongly galaxies preserve 
cosmological memory in their chemical enrichment histories.

\section{The Galaxy--Cosmology Manifold Visualization}
\label{sec:gal_manifold}

In this section we extend this concept by presenting a direct 
manifold visualization derived from our analytic predictor.

To visualize the intrinsic geometry of galaxy properties, 
we employ the \textit{Uniform Manifold Approximation and Projection} (UMAP) algorithm ~\citep{mcinnes2018umap} 
to embed galaxies from a high-dimensional property space into a two-dimensional manifold. 
The workflow proceeds as follows:

\begin{enumerate}
    \item \textbf{Feature definition.} 
    For each galaxy, we consider four baryonic observables:
    stellar mass $(M_{\star})$, stellar metallicity $(Z_{\star})$, 
    effective velocity $(V_{\mathrm{eff}})$, and structural compactness $(R_{\mathrm{compact}})$. 
    These quantities span a four-dimensional physical space that 
    encapsulates the galaxy’s internal regulation and dynamical state.

    \item \textbf{Normalization.} 
    Each feature is standardized using a \textsc{StandardScaler} 
    to ensure that all parameters contribute comparably to the metric distance. 
    This step removes biases due to differing physical units or dynamic ranges.

    \item \textbf{Local neighborhood graph.} 
    UMAP constructs a weighted graph that connects each galaxy 
    to its $n_{\mathrm{neighbors}}=30$ nearest neighbors in this 
    normalized property space, using Euclidean distance as the similarity metric.
    The resulting graph defines the local geometric relationships between galaxies.

    \item \textbf{Manifold learning.} 
    The algorithm optimizes a low-dimensional (2-D) representation 
    by minimizing the cross-entropy between the high- and low-dimensional 
    neighbor graphs. 
    This preserves the local topology: galaxies that are close in the 
    original parameter space remain close on the projected manifold.

    \item \textbf{Visualization.} 
    The manifold is color-mapped by derived or cosmological quantities.
\end{enumerate}

\begin{figure*}[h!]
    \centering
    \hspace*{-0.02\textwidth} 
    \setlength{\tabcolsep}{1.5pt} % small horizontal gap
    \renewcommand{\arraystretch}{0}
    \setlength{\arrayrulewidth}{0.2pt} % thin vertical lines

    \begin{tabular}{c|c|c|c|}
        %----------------------------------------------------------
        % Top labels
        %-----------------------------------------     %----------------------------------------------------------
        % Row: z = 0
        %----------------------------------------------------------
        \rotatebox{90}{\textbf{$z = 0$}} &
        \includegraphics[width=0.33\textwidth]{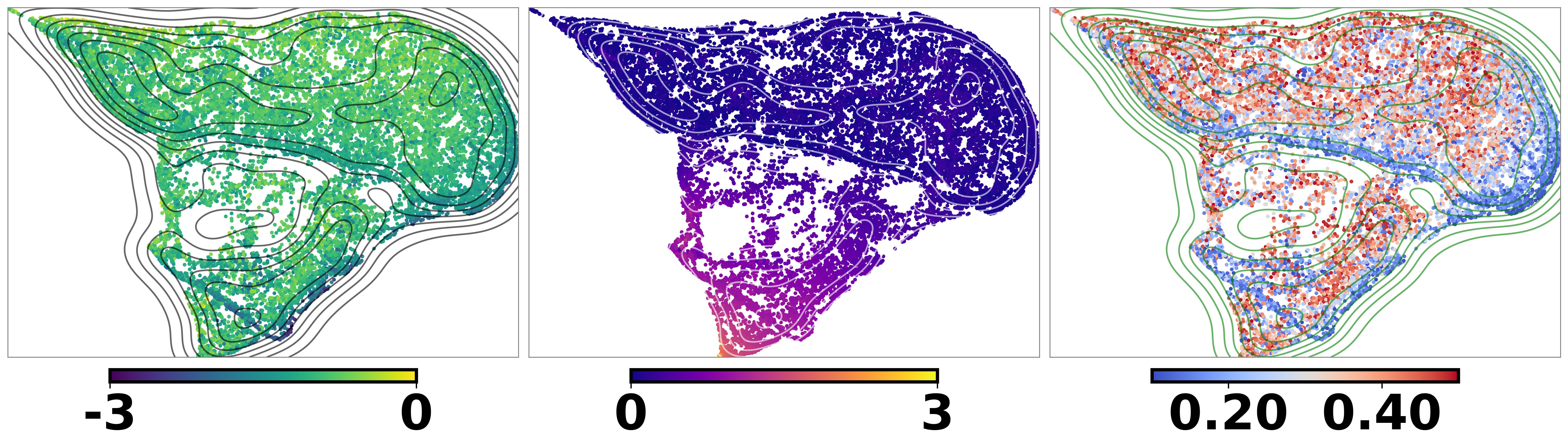} &
        \includegraphics[width=0.33\textwidth]{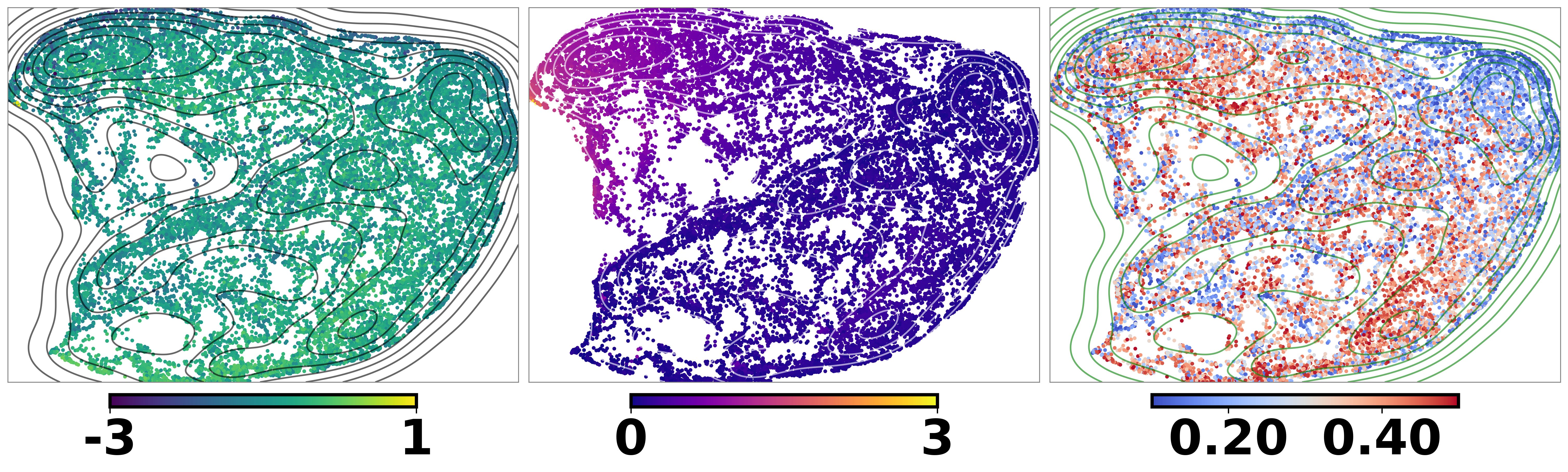} &
        \includegraphics[width=0.33\textwidth]{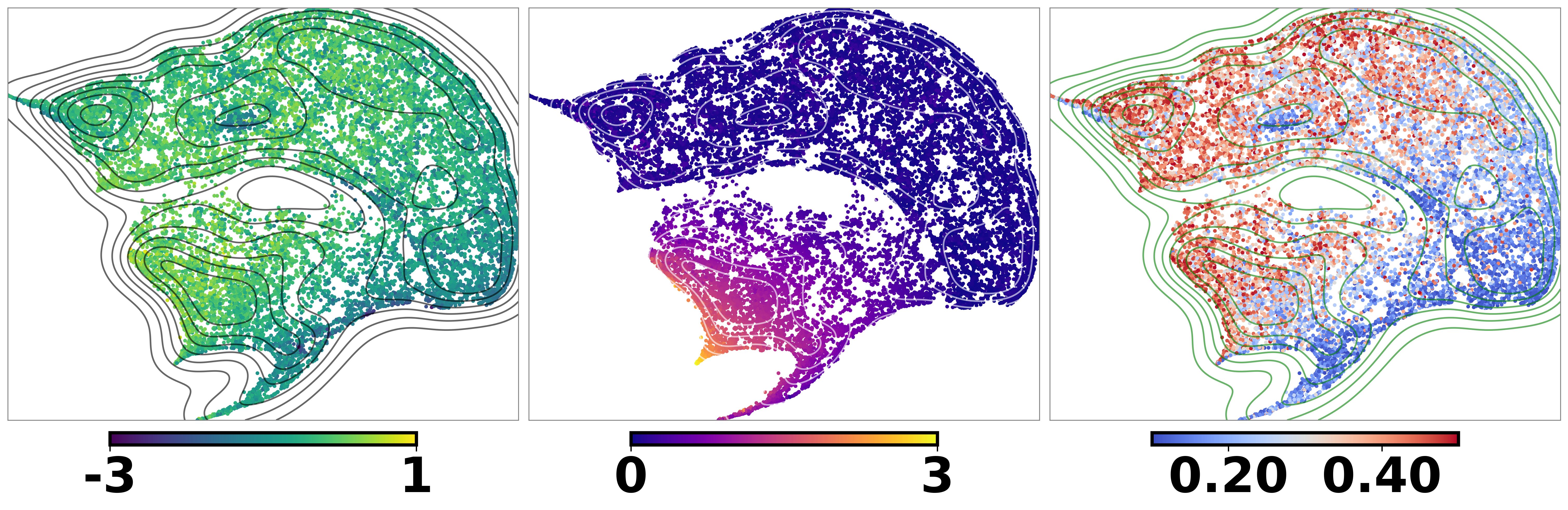} \\[4pt]

        %----------------------------------------------------------
        % Row: z = 1
        %----------------------------------------------------------
        \rotatebox{90}{\textbf{$z = 1$}} &
        \includegraphics[width=0.33\textwidth]{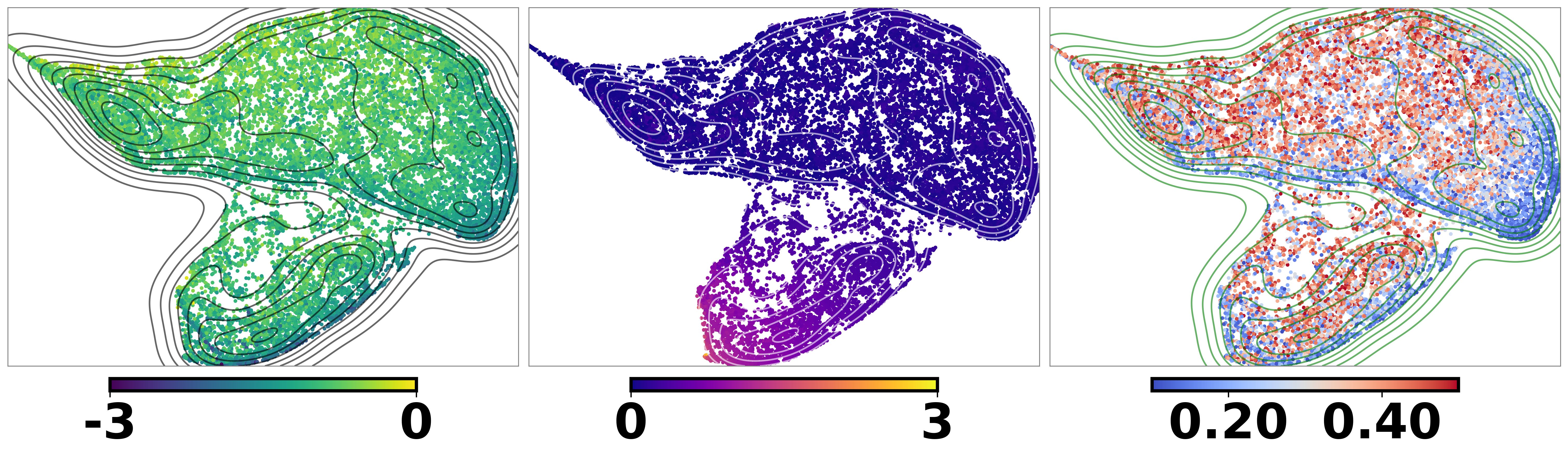} &
        \includegraphics[width=0.33\textwidth]{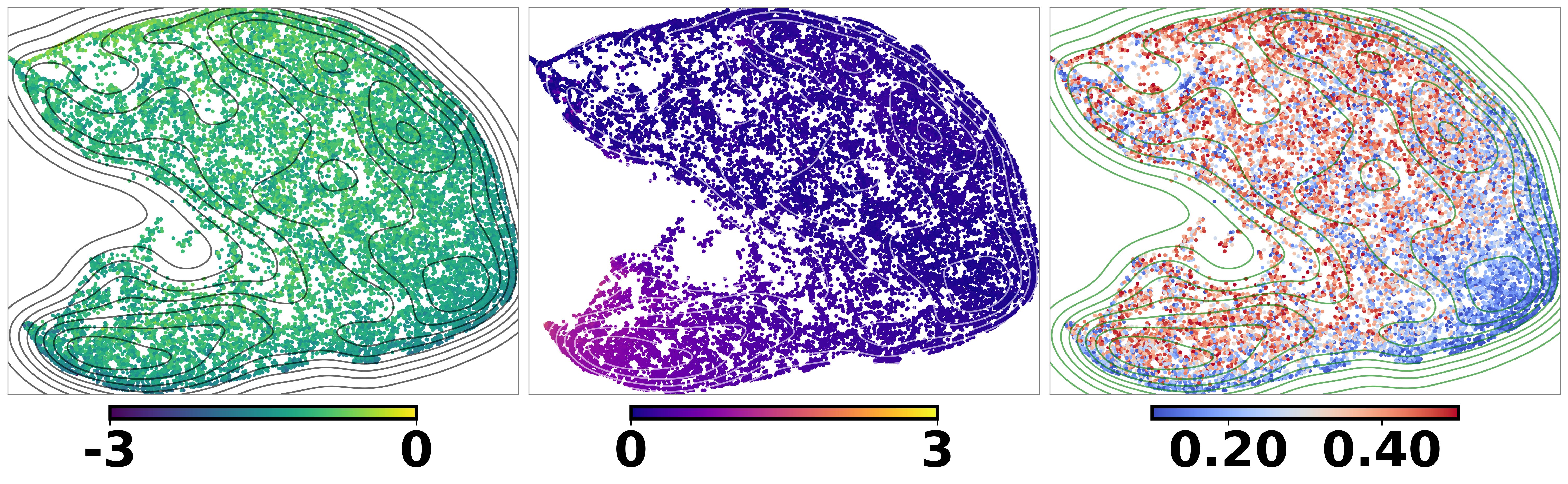} &
        \includegraphics[width=0.33\textwidth]{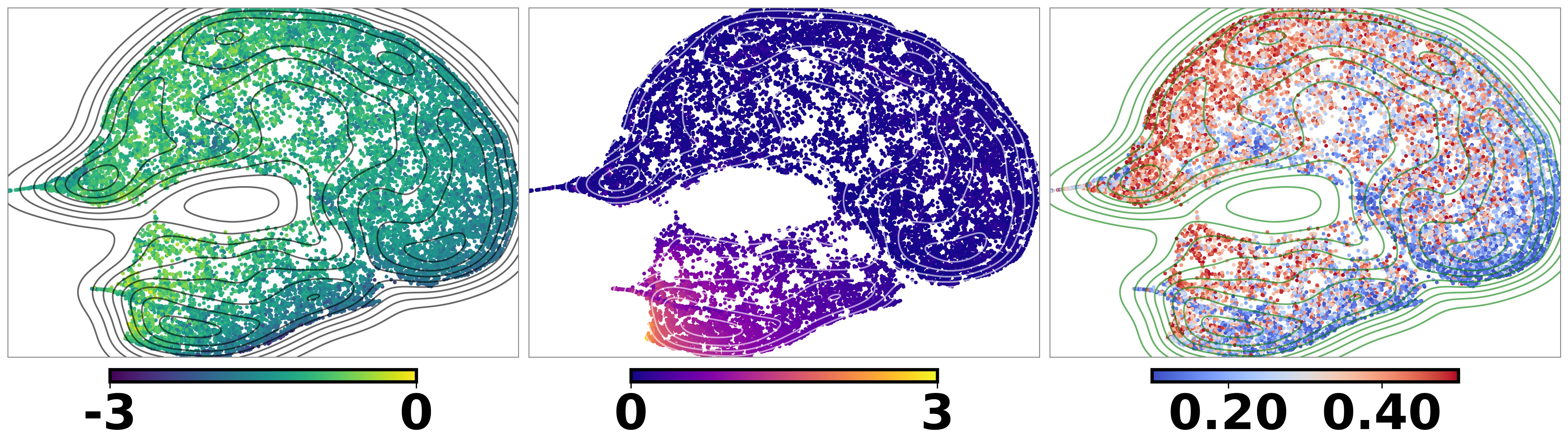} \\[4pt]

        %----------------------------------------------------------
        % Row: z = 2
        %----------------------------------------------------------
        \rotatebox{90}{\textbf{$z = 2$}} &
        \includegraphics[width=0.33\textwidth]{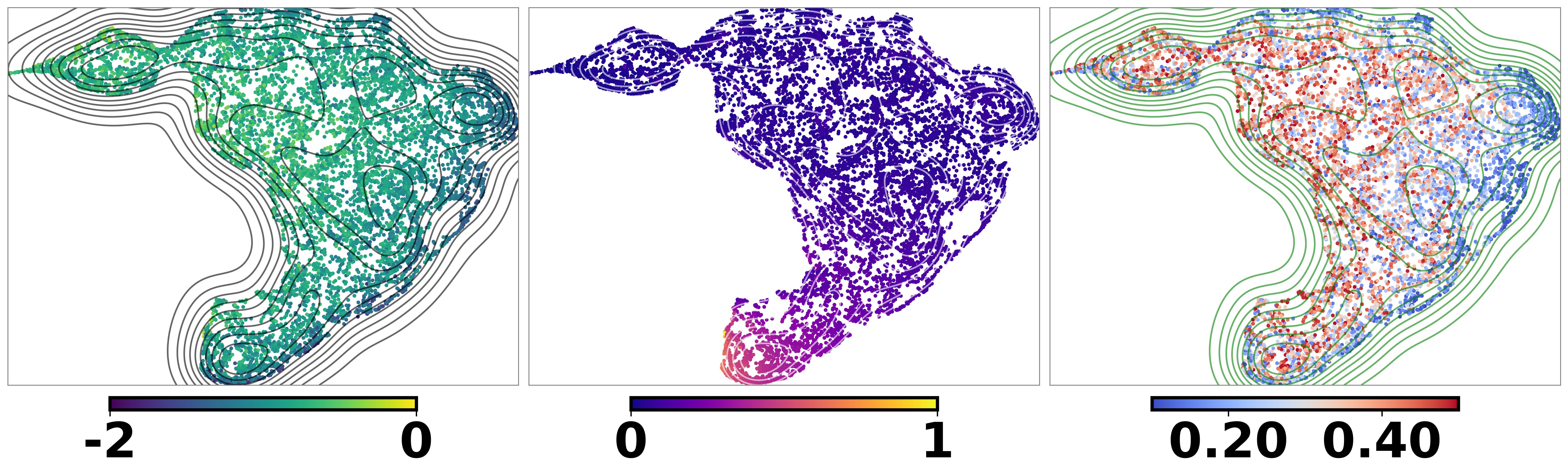} &
        \includegraphics[width=0.33\textwidth]{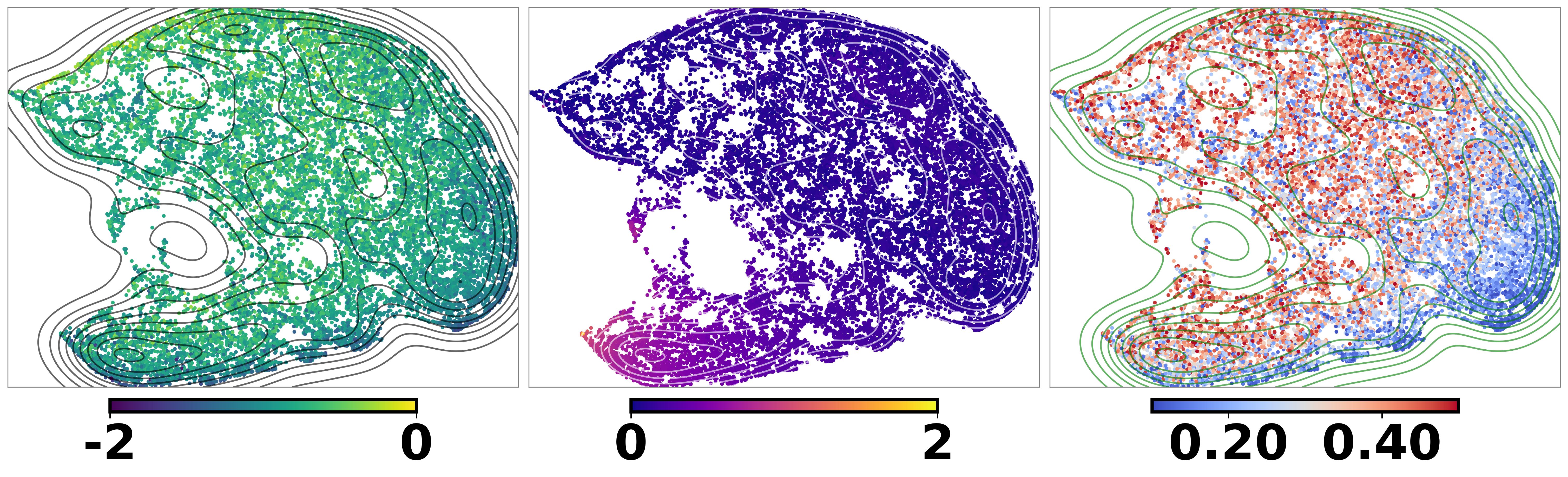} &
        \includegraphics[width=0.33\textwidth]{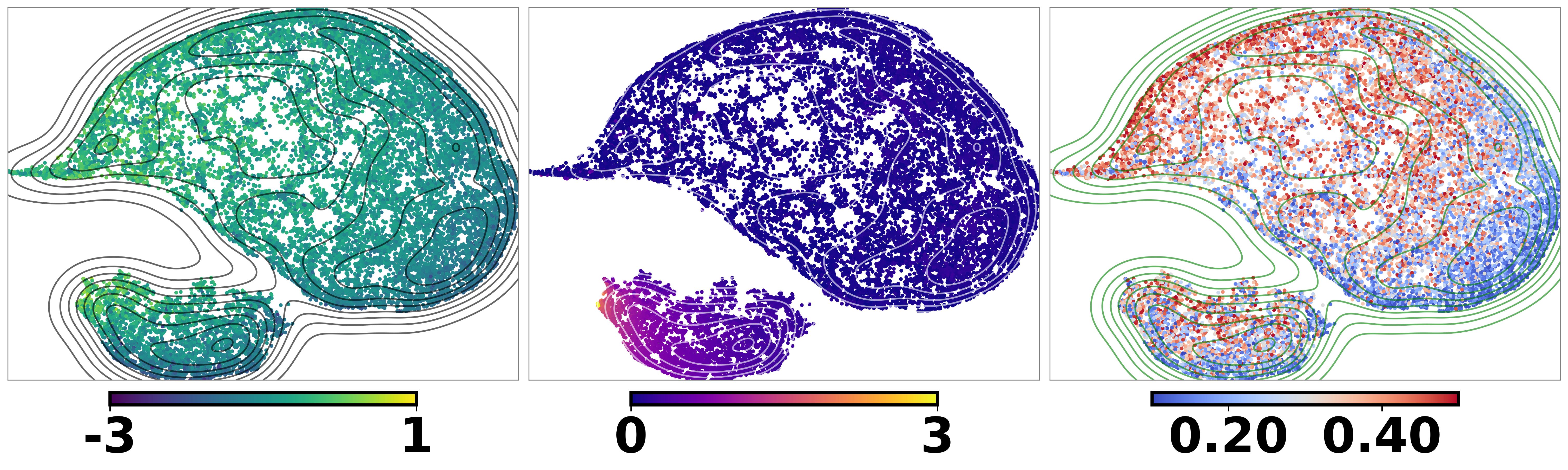} \\[4pt]

        %----------------------------------------------------------
        % Row: z = 3
        %----------------------------------------------------------
        \rotatebox{90}{\textbf{$z = 3$}} &
        \includegraphics[width=0.33\textwidth]{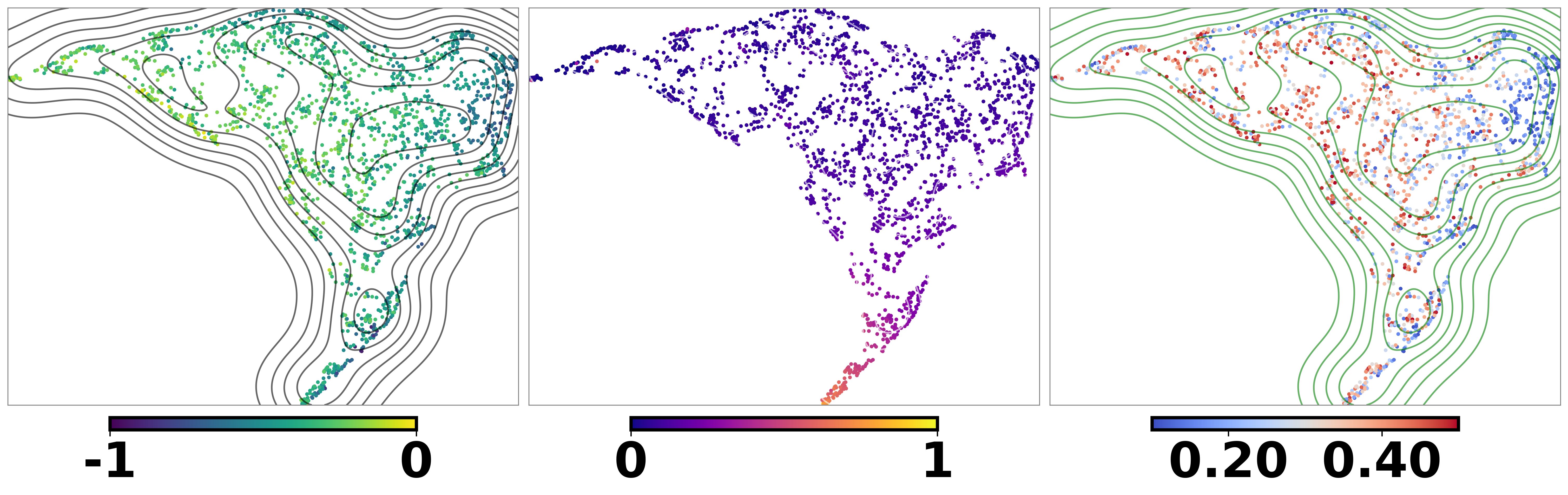} &
        \includegraphics[width=0.33\textwidth]{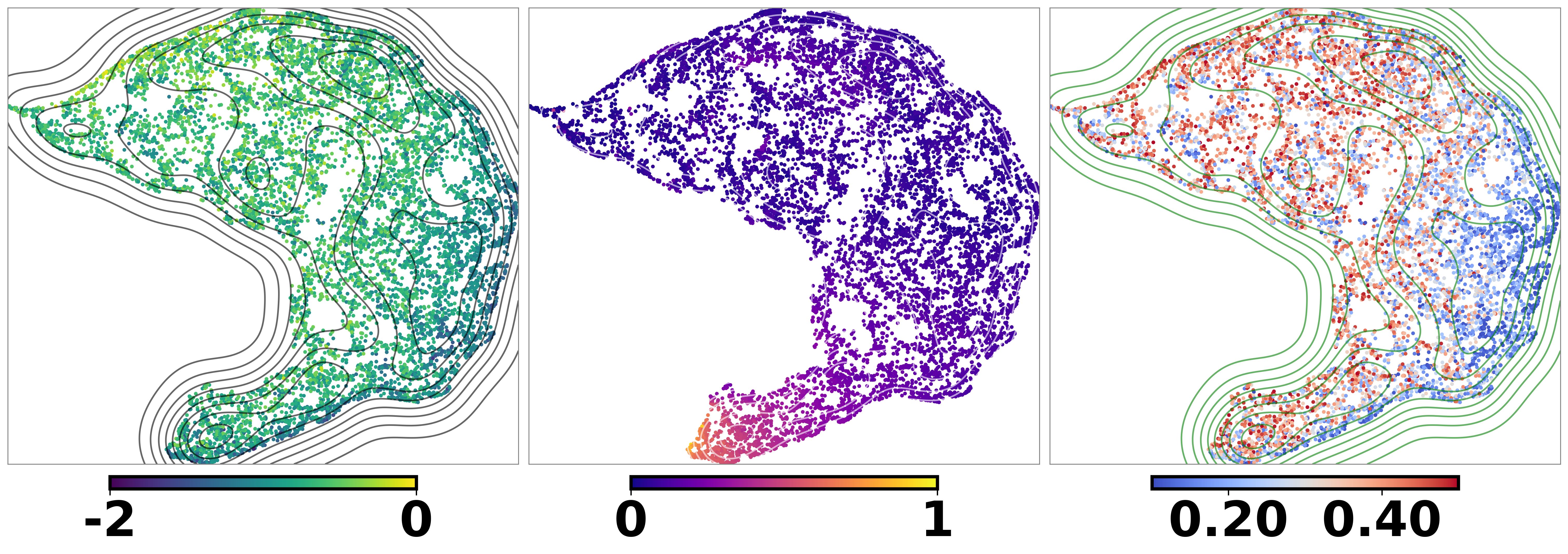} &
        \includegraphics[width=0.33\textwidth]{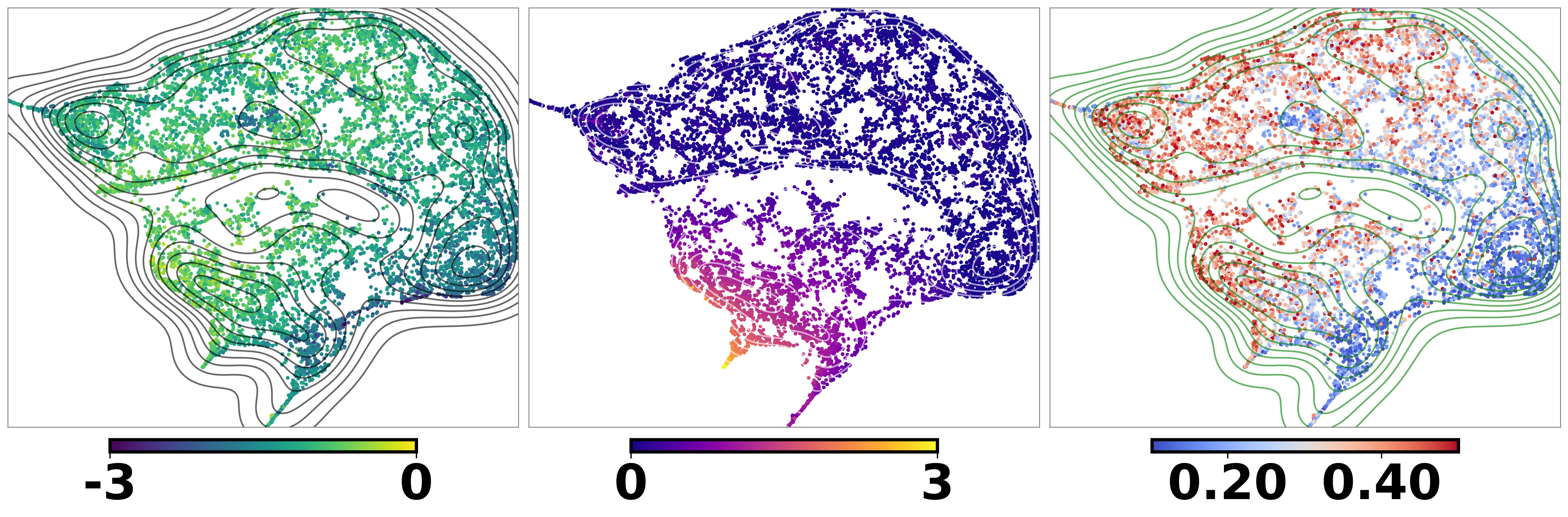} \\[2pt]
        \hline
        %----------------------------------------------------------
        % Bottom labels
        %----------------------------------------------------------
        & \small (a)\textsc{IllustrisTNG} & \small (b)~Astrid & \small (c)~SIMBA
    \end{tabular}

    \vspace{4pt}
    \caption{
        Evolution of the galaxy-property manifold across redshift and simulation suite.
        Each column corresponds to one CAMELS simulation suite 
        (\textsc{IllustrisTNG}, \textsc{Astrid}, \textsc{SIMBA}),
        and each row corresponds to a fixed redshift ($z = 0.0, 1.0, 2.0, 3.0$).
        Within each panel, colors encode:
        the \textbf{latent variable} log($x$) (left), 
        the \textbf{compactness ratio} $R_{\text{compact}}$ (middle), 
        and the corresponding \textbf{cosmological parameter} $\Omega_{m}$ (right).  
        Vertical dividers separate the three simulation columns for clarity,
        while redshift labels along the left indicate cosmic epoch.
        The manifold topology remains broadly consistent across simulations and redshifts,
        highlighting a universal low-dimensional relation among
        stellar mass, metallicity, kinematics, and structural compactness.
    }
    \label{fig:manifold_evolution}
\end{figure*}

The results of the three simulation suites’ galaxy-property manifolds across redshift ($z = 0.0, 1.0, 2.0, 3.0$) are presented in Fig.~\ref{fig:manifold_evolution}. 
Within each subpanel, every point on the manifold corresponds to an individual galaxy, while the color scale encodes different physical quantities: the latent variable $x$ (left, shown in logarithmic scale to better capture its variation), the compactness ratio $R_{\mathrm{compact}}$ (middle), and the inferred cosmological parameter $\Omega_m$ (right). 

The two-dimensional surface represents the nonlinear projection of the high-dimensional galaxy-property space obtained through the UMAP algorithm. 
The horizontal and vertical directions in each plot serve as abstract coordinates of this projected space, the smooth color gradients across the manifolds indicate correlated variations among the underlying baryonic, structural, and cosmological parameters.
Across all panels, the color gradients exhibit a coherent diagonal trend, 
indicating that enrichment efficiency and structural compactness vary jointly 
rather than independently. 
Galaxies with deeper potential wells are simultaneously more compact and 
more metal-enriched, reflecting a coupled baryon–feedback regulation process. 
The same diagonal orientation appears in the $\Omega_m$ panels, 
demonstrating that the cosmological sensitivity encoded by our analytic 
relation aligns with the latent direction of baryonic regulation within the 
manifold.

Remarkably, the orientation and continuity of these gradients remain stable 
across redshift and simulation suite. 
Although minor distortions are visible, the overall deformation of the manifold 
is very gentle—its topology and dominant gradient direction are largely 
preserved. 
Despite differing feedback and star-formation prescriptions in 
\textsc{IllustrisTNG}, \textsc{Astrid}, and \textsc{SIMBA}, 
the manifolds retain nearly identical geometries and color transitions. 
This persistence implies that galaxies from distinct physical models occupy a 
common, low-dimensional equilibrium surface where baryonic, structural, and 
cosmological dependencies co-evolve in a universal manner.

\bibliography{sample701}{}
\bibliographystyle{aasjournalv7}



\end{document}